\documentclass[twoside,symmetric,notoc,justified]{tufte-book}
\usepackage[danish,english]{babel} 
\usepackage{graphicx}
\usepackage{hyperref}
\usepackage{xcolor}

\newsavebox{\titleimage}
\savebox{\titleimage}{\includegraphics[height=7\baselineskip]{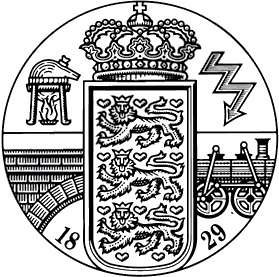}}

\usepackage{beramono}
\usepackage[strict]{changepage}

\publisher{%
\begin{tabular}{@{}c@{}}
\begin{minipage}[c]{0.7\linewidth}
\begin{flushleft}
Department of Applied Mathematics and Computer Science\\[2em]
Technical University of Denmark
\end{flushleft}
\end{minipage}%
\hfill 
\begin{minipage}[c]{0.3\linewidth}
\begin{flushright}
\usebox{\titleimage}
\end{flushright}
\end{minipage}
\end{tabular}
}

\usepackage{float}
\usepackage{booktabs}

\usepackage{graphicx}
\setkeys{Gin}{width=\linewidth,totalheight=\textheight,keepaspectratio}
\graphicspath{{graphics/}}

\usepackage{fancyvrb}
\usepackage{amsmath}
\usepackage{amssymb}
\fvset{fontsize=\normalsize}

\usepackage{xspace}
\usepackage[normalem]{ulem}

\newcommand{\monthyear}{%
  \ifcase\month\or January\or February\or March\or April\or May\or June\or
  July\or August\or September\or October\or November\or
  December\fi\space\number\year
}

\usepackage{units}
\usepackage{multirow}

\newcommand{\hlred}[1]{\textcolor{Maroon}{#1}}
\newcommand{\hangleft}[1]{\makebox[0pt][r]{#1}}

\providecommand{\XeLaTeX}{X\lower.5ex\hbox{\kern-0.15em\reflectbox{E}}\kern-0.1em\LaTeX}

\newcommand{\tuftebs}{\symbol{'134}}
\newcommand{\doccmddef}[2][]{%
  \hlred{\texttt{\tuftebs#2}}\label{cmd:#2}%
  \ifthenelse{\isempty{#1}}%
    {
      \index{#2 command@\protect\hangleft{\texttt{\tuftebs}}\texttt{#2}}
    }%
    {
      \index{#2 command@\protect\hangleft{\texttt{\tuftebs}}\texttt{#2} (\texttt{#1} package)}
      \index{#1 package@\texttt{#1} package}\index{packages!#1@\texttt{#1}}
    }%
}
\newcommand{\doccmd}[2][]{%
  \texttt{\tuftebs#2}%
  \ifthenelse{\isempty{#1}}%
    {
      \index{#2 command@\protect\hangleft{\texttt{\tuftebs}}\texttt{#2}}
    }%
    {
      \index{#2 command@\protect\hangleft{\texttt{\tuftebs}}\texttt{#2} (\texttt{#1} package)}
      \index{#1 package@\texttt{#1} package}\index{packages!#1@\texttt{#1}}
    }%
}

\usepackage{epigraph}
\usepackage{makeidx}
\makeindex

\newcommand{\autoforceversofloat}{
    \ifodd\thepage
        \textbf{Odd page - forcing verso float}
        \forceversofloat
    \else
        \textbf{Even page - no forcing}\\
    \fi
}

\usepackage{pdfpages}

\begin{document}

\frontmatter

\begin{titlepage}
    \centering 

    \vspace*{3cm} 

    {\Huge \textbf{Geometry \& Geography \\of\\ Complex Networks} \par}
    
    \vspace{2cm} 
    
    {\Large \textbf{Louis Boucherie} \par}

    \vfill 
    
    {\large 2025 \par} 

\end{titlepage}


\chapter{Abstract}

Complex systems\marginnote{\textit{\textbf{Keywords:} Network Science, Community Detection, Hierarchical Clustering, Markov Chain Monte Carlo, Network Embeddings, Graph Neural Networks, Geography, Human Mobility, Hyperbolic Space.}} are made up of many interacting components. Network science provides the tools to analyze and understand these interactions. Community detection is a key technique in network science for uncovering the structures that shape the behavior of these networks. This thesis introduces the Adaptive Cut, a novel method that improves clustering methods by employing multi-level cuts in hierarchical dendrograms. Overcoming the limitations of traditional single-level cuts-especially in unbalanced dendrograms-the Adaptive Cut provides a multi-level cut by optimizing a Markov chain Monte Carlo with simulated annealing. In addition, we propose the Balanceness score, an information-theoretic metric that quantifies dendrogram balance and predicts the benefits of multilevel cuts. Evaluations on over 200 real and synthetic networks show significant improvements in partition density and modularity.

In the second part, we explore network geometry using machine learning techniques like network embeddings and graph neural networks. Focusing on the Danish cohabitation network, we use methods such as node2vec and Infomap to study community structures and the interplay between higher-dimensional network geometry and geography. Our analysis shows that incorporating network geometry allows redefining administrative boundaries into non-contiguous regions that better reflect social and spatial dynamics. We also discuss the representation of hierarchical data in hyperbolic space through Poincaré maps, which can represent tree-like structures in low dimension.

In addition, we examine how geography constrains human mobility, an aspect often overlooked in scale-free characterizations of mobility. By incorporating geography via the pair distribution function from condensed matter physics, we separate geographic constraints from mobility choices. Analyzing datasets containing millions of individual movements, we identify a universal power law that spans five orders of magnitude, thereby bridging the divide between distance-based and opportunity-driven models of human mobility.

\tableofcontents

\chapter*{List of Publications}

This dissertation is based on the two following papers:
\begin{enumerate}
    \item[(i)] Boucherie, L., Ahn, Y-Y., \& Lehmann, S. Adaptive Cuts reveal multiscale complexity in networks arXiv preprint arXiv:2512.08741, 2025 \citep{boucherie2025adaptive}
    \item[(ii)] Boucherie, L., Maier, B. F., \& Lehmann, S. (2024). Decomposing geographical and universal aspects of human mobility. arXiv preprint arXiv:2405.08746. \citep{boucherie2024decomposing} \citep{boucherie2025decoupling}
\end{enumerate}
Other papers mentioned in the dissertation are:
\begin{enumerate}

    \item[(iii)] Nakis, N., Celikkanat, A., Boucherie, L., Djurhuus, C., Burmester, F., Holmelund, D. M., ... \& Mørup, M. (2023, April). Characterizing Polarization in Social Networks using the Signed Relational Latent Distance Model. In International Conference on Artificial Intelligence and Statistics (pp. 11489-11505). PMLR. \citep{nakis2023characterizing}
    \item[(iv)] Nakis, N., Celikkanat, A., Boucherie, L., Lehmann, S., \& Mørup, M. (2024, April). Time to Cite: Modeling Citation Networks using the Dynamic Impact Single-Event Embedding Model. In International Conference on Artificial Intelligence and Statistics (pp. 1882-1890). PMLR. \citep{nakis2024time}
    \item[(v)] Bhasker, N., Chung, H., Boucherie, L., Kim, V., Speidel, S., \& Weber, M. Contrastive Poincaré Maps for single-cell data analysis. In ICLR 2024 Workshop on Machine Learning for Genomics Explorations. \citep{bhasker2024contrastive} \citep{bhasker2025uncovering}
\end{enumerate}
Other papers not included in the dissertation are:
\begin{enumerate}

    \item[(vi)] Boucherie, L., Budnik, K. B. \& Panos, J. (2022). Looking at the evolution of macroprudential policy stance: A growth-at-risk experiment with a semi-structural model. ECB Occasional Paper, (2022/301). \citep{budnik2022looking} \citep{boucherie2022looking}
    \item[(vii)] Budnik, K. B., Boucherie, L., Borsuk, M., Dimitrov, I., Giraldo, G., Groß, J., ... \& Volk, M. (2022). Macroprudential Stress Test of the Euro Area Banking System Amid the Coronavirus (COVID-19) Pandemic. \citep{budnik2022macroprudential}
    \item[(viii)] Budnik, K., Boucherie, L., \& Panoš, J. (2024). Measuring the macroprudential policy stance in the euro area with a semi‐structural model. Economic Notes, 53(3), e12244. \citep{budnik2024measuring} 
    \item[(ix)] Budnik, K. B., Groß, J., Vagliano, G., Dimitrov, I., Lampe, M., Panos, J., Boucherie, L., \& Jancokova, M. (2023). BEAST: A model for the assessment of system-wide risks and macroprudential policies. \citep{budnik2023beast}
\end{enumerate}

\listoffigures

\listoftables

\chapter{Introduction}

This thesis is structured in reverse order from the title. It begins with complex networks, introducing the fundamentals of network science and gradually leading to community detection, which serves as the entry point for the first paper.  Prior to this, there is a discussion on heavy-tailed distributions and methods to estimate power-laws, a topic crucial for both complex networks and the final chapter of the thesis. The first paper introduces the Adaptive Cut algorithm, a method to improve partition metrics in community detection—namely, partition density for link clustering and modularity for the Louvain method. The algorithm optimizes a Markov chain Monte Carlo over hierarchical clustering dendrograms. This involves iteratively merging or splitting communities to improve the objective function, while allowing for stochastic acceptance of suboptimal steps to avoid becoming trapped in local optima.

The second chapter shifts focus towards the geometry of complex networks, specifically on graph representation learning. The aim here is to analyze low-dimensional embeddings of networks, capturing both structural and relational information for tasks such as node classification, link prediction, and clustering.  In this context, the cohabitation network of Denmark is introduced as a case study, this dataset is also central to the third chapter. Network science techniques from the earlier chapters are applied to this dataset, revealing interesting patterns, including the presence of non-contiguous administrative boundaries. This section is still a work in progress.

The final chapter examines the data-generating process of the co-habitation network, i.e. how individuals change addresses. This study is situated within the broader field of human mobility. The key insight lies in the interplay between geography and the geometry of the network, particularly in studying the distances involved in these moves. By accounting that mobility occurs on the physical structure of the built environment, the analysis reveals a power-law distribution that spans five orders of magnitude.

\mainmatter

\chapter{Complex Networks}

\newthought{Networks} form a fundamental part of contemporary society, influencing both daily interactions and broader global dynamics. We exist within a web of complex networks, spanning social relationships like family, friends, colleagues, and partners. Communication is facilitated through technological networks, such as telephones, the Internet, and the World Wide Web\footnote{The distinction lies in that the Internet refers to the underlying physical infrastructure of routers and servers, while the World Wide Web consists of the interconnections between web pages.}. The infrastructure supporting modern life is underpinned by networks that transport vital resources, such as electricity through power grids and water through underground piping systems. Our mobility is enabled by transportation networks, including roads, railways, and airways. On a biological level, networks serve as a cornerstone of human and natural systems, spanning scales from cellular interactions to expansive ecological networks.

The concept of networks has its roots in a long history, with a notable milestone being Leonhard Euler's solution to the Königsberg Bridge Problem in 1735. Euler's representation of this problem through nodes and links laid the groundwork for the development of graph theory\footnote{The distinction between graphs and networks is subtle: while a "graph" is an abstract mathematical construct, the term "network" often refers to a tangible physical system.}. Throughout the 20\textsuperscript{th} and early 21\textsuperscript{st} centuries, network science has experienced substantial growth, showing that networks can effectively model a wide range of systems beyond just physical connections.

\begin{figure*}[ht!]
\checkoddpage \ifoddpage \forcerectofloat \else \forceversofloat \fi
    \centering
    \includegraphics[width=\linewidth]{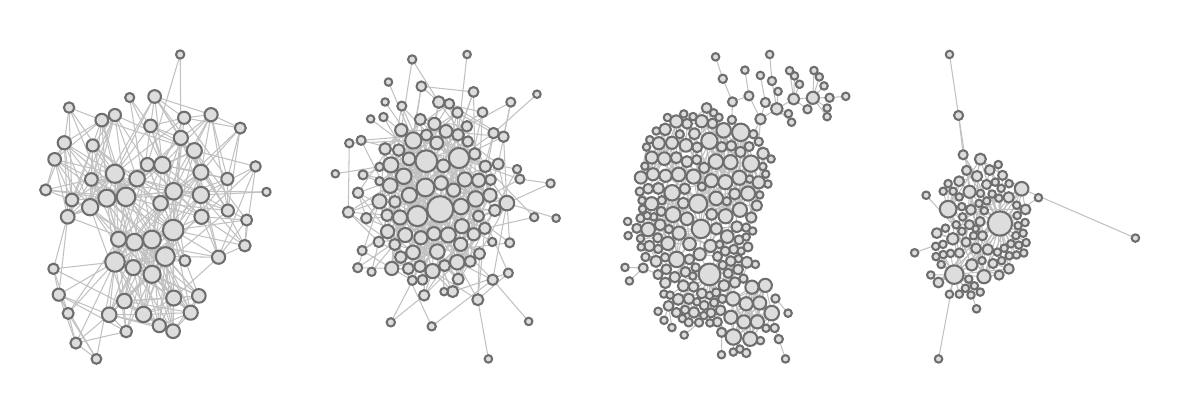}
    \caption[Examples of networks]{Network visualizations of four different systems. 
(A) Social network: representing multiplex social relationships among individuals at a research department at Aarhus University. \citep{Magnani2013}. 
(B) Informational network: A network of common adjective and noun adjacencies from the novel David Copperfield by Charles Dickens, showing linguistic connections between words \citep{Newman2006}.
(C) Technological network: The reduced Central Chilean power grid, illustrating the infrastructure and connections within the electrical system \citep{Kim2018}.
(D) Biological network: Vazquez \& Simberloff's plant-pollinator web, representing interactions between plant species and their pollinators, demonstrating ecological relationships \citep{Vazquez2003}.}
    \label{fig:4networks}
\end{figure*}

In recent years, network science has increasingly been applied to investigate and understand the properties of a variety of complex systems. This has facilitated the modeling of numerous real-world systems, which can generally be divided into four main categories: social, informational, technological, and biological.

\newthought{Social Networks} A social network consists of individuals and the pattern of interactions that occur between them \citep{wasserman1994social, scott2011sage,sekara2015dynamics}. The concept has been used to describe a variety of social dynamics, including friendships \citep{rapoport1961}, terrorist communications \citep{krebs2002mapping}, sexual contact patterns \citep{liljeros2001web, bearman2004chains}, business relations \citep{mariolis1975interlocking, davis2003small}, online interactions \citep{szell2010multirelational,szell2010measuring}, and intermarriages among prominent families \citep{padgett1993robust}. The foundation of social network analysis can be traced back to Jacob Moreno in the 1930s, who employed sociograms to examine relationships within small groups \citep{moreno1934sociometry}. Subsequent works by researchers like Davis et al. \citep{davis1941deep} and Rothlisberger and Dickson \citep{roethlisberger1939management} further developed the field. However, early social network research faced limitations due to the small sample sizes that resulted from the challenges of accurately mapping social ties. Seminal studies, such as Milgram's small-world experiment \citep{milgram1967small}, primarily utilized surveys, interviews, and self-reports, which were often subject to various biases and inaccuracies \citep{watts2007twenty, marsden1990network}.

The advent of the Internet has led to the availability of more comprehensive datasets, including scientific collaboration networks \citep{newman2001structure}, film actor networks \citep{amaral2000classes}, and telephone call networks \citep{aiello2000large}. Subsequent studies explored various modes of digital communication, such as email interactions \citep{ebel2002scale, newman2002email}, instant messaging \citep{leskovec2008planetary}, and online social platforms \citep{lewis2008tastes, aral2012identifying}, as well as mobile phone call patterns \citep{onnela2007structure, miritello2013time}. These advancements have significantly enhanced the capacity to analyze social networks in greater depth and over extended timeframes.

\newthought{Information Networks} represent the flow of knowledge between entities. A prime example is the network of scientific paper citations, where nodes correspond to individual papers and edges denote citations \citep{de1965networks}. Such citation networks are highly accessible and cover multiple decades, providing extensive datasets for research \citep{lotka1926frequency}. Another major instance of an information network is the World Wide Web, which consists of billions of web pages interconnected through hyperlinks \citep{barabasi2000scale, kleinberg1999authoritative}.

Further examples of information networks include patent citation networks \citep{jaffe2002patents} and the spread of mobile phone viruses \citep{wang2009network}. Moreover, the structure of cultural history has been explored through networks \citep{schich2014network}, and semantic word networks have been studied to understand the connections between words and meanings \citep{sigman2002global}.

\newthought{Technological Networks} are typically human-made and designed to distribute goods. A well-known example is the Internet, which is organized by the physical connections between routers and has been extensively mapped and analyzed \citep{faloutsos1999power}. Other examples include airline networks \citep{amaral2000classes}, road networks \citep{porta2006network, rosvall2005networks}, railways \citep{latora2002efficient, sen2003small}, and power grids \citep{watts1998collective}. Rivers and their drainage basin can also be considered distribution networks \citep{dodds1999geometry}. 

Technological networks are strongly influenced by spatial and geographical constraints that shape their structure \citep{yook2002modeling, daqing2011spatial}. The interplay of external factors influencing these spatially embedded networks remains an active area of research.

\newthought{Biological Networks} every organism exists within a complex web of biological networks. At the ecosystem level, species interactions form intricate webs \citep{baird1989seasonal, cohen1990community, dunne2002network}, which reveal the complexity of ecosystems. On a smaller scale, neural networks consist of interconnected neurons in the brain \citep{sporns2002human}, with progress being made in mapping them, as in the nematode C. elegans \citep{white1986structure}.
At the cellular level, gene regulatory networks allow organisms to adapt to environmental fluctuations. Proteins within cells form protein-protein interaction networks, which are crucial for cellular function and have been extensively studied \citep{uetz2000comprehensive, jeong2001lethality}. Additionally, metabolic processes are represented as networks of biochemical substrates and products, providing a framework to analyze cellular metabolism \citep{jeong2000large, ravasz2002hierarchical}. These different layers of biological networks offer valuable insights into the underlying complexity of living systems.

\begin{marginfigure}%
  \includegraphics[width=\linewidth]{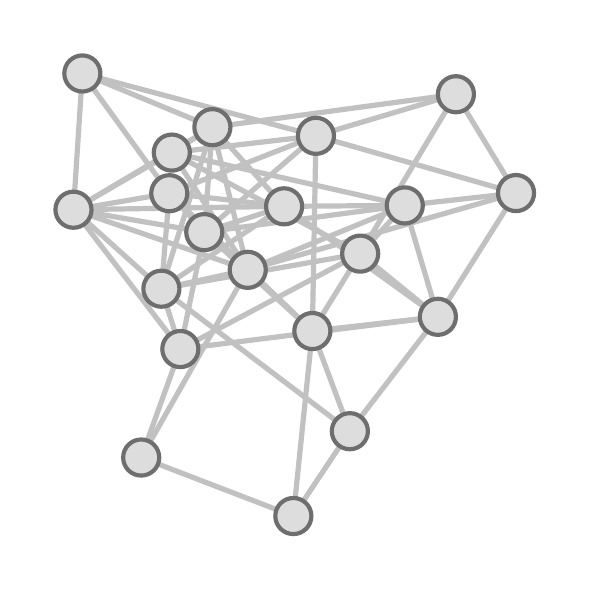}
  \caption[A network]{A visualization of an Erdős-Rényi random graph with 20 nodes and a connection probability of 0.3. The spring layout algorithm positions the nodes to reflect the network's structure.}
  \label{fig:network}
\end{marginfigure}

The networks described above differ significantly in their size, function, and nature, ranging from systems with just a few interconnected nodes to those involving billions of connected entities.

\section{Networks Definitions \& Properties}

\subsection{Adjacency Matrix}

A network is defined as a collection of entities and the connections between pairs of these entities\footnote{It is important to note that this definition focuses on connections between pairs of entities. In some contexts, interactions between three or more entities are considered, known as 'hypergraphs'; however, these are not relevant to the current discussion.}. In this context, the entities are referred to as ``nodes,'' while the connections between them are called ``links'' or ``edges.'' When a node \(i\) is connected to a node \(j\) by a link, node \(j\) is considered a ``neighbor'' of node \(i\) and vice versa. The set of all nodes connected to node \(i\) forms the neighbor set \(\mathcal{N}(i)\), also known as the neighborhood of \(i\).

\newthought{Adjacency matrix} since links only connect two nodes at a time, a network can be described using an adjacency matrix, a mathematical tool that records the presence or absence of connections between nodes. This matrix will be especially useful in later analyses (see Chapter \ref{sec:chapter2}). For a network with \(N\) nodes, the adjacency matrix \(A\) is an \(N \times N\) matrix. In the simplest case, where the network is both unweighted and undirected, the adjacency matrix is defined as follows,

\begin{marginfigure}%
  \includegraphics[width=\linewidth]{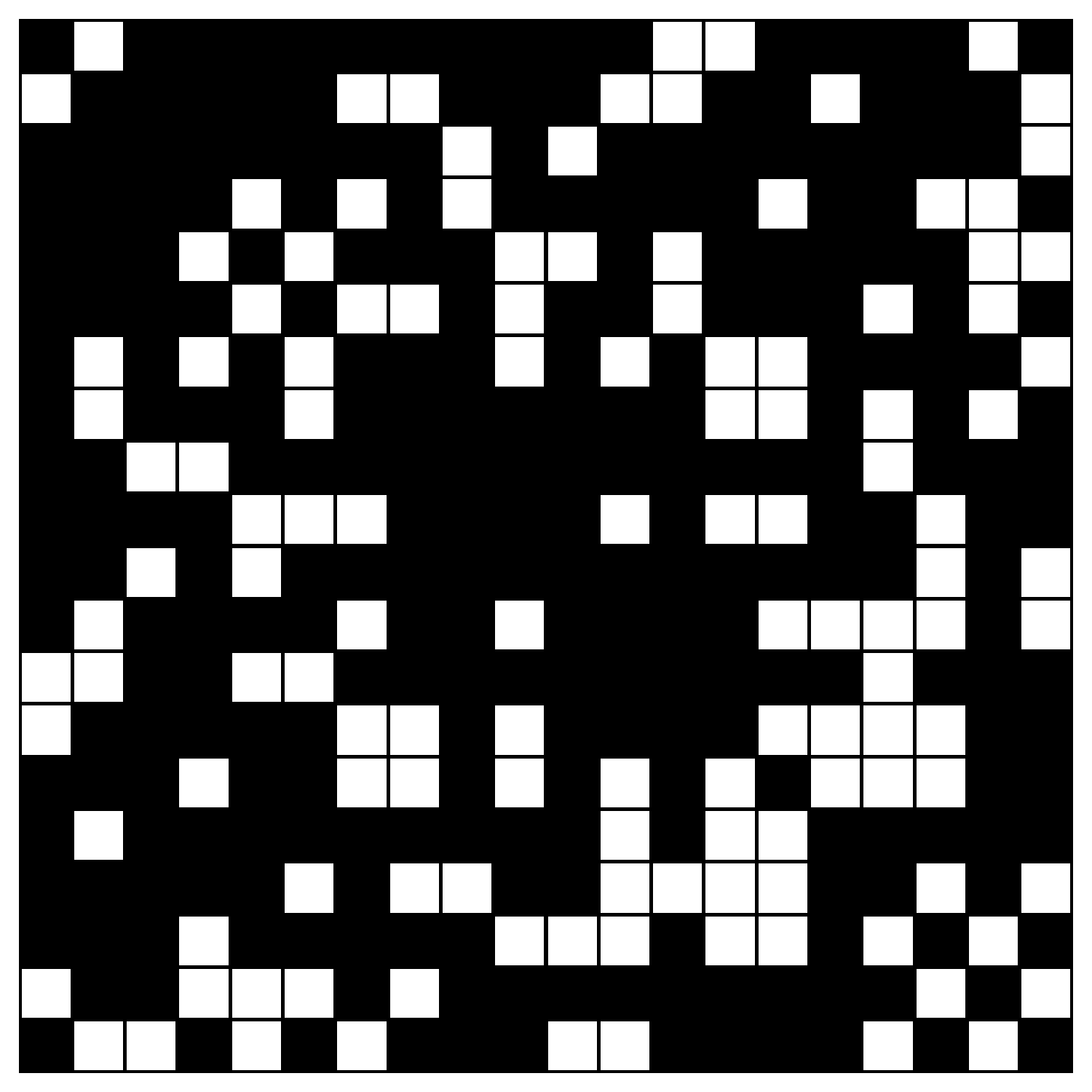}
  \caption[Adjacency matrix]{Adjacency matrix of the generated network. Each cell indicates the presence (white) or absence (black) of an edge between pairs of nodes.}
  \label{fig:adjacency}
\end{marginfigure}

\[
A_{ij} = \begin{cases} 
1 & \text{if } i \text{ is connected to } j, \\
0 & \text{otherwise}. 
\end{cases}
\]

\begin{marginfigure}%
  \includegraphics[width=\linewidth]{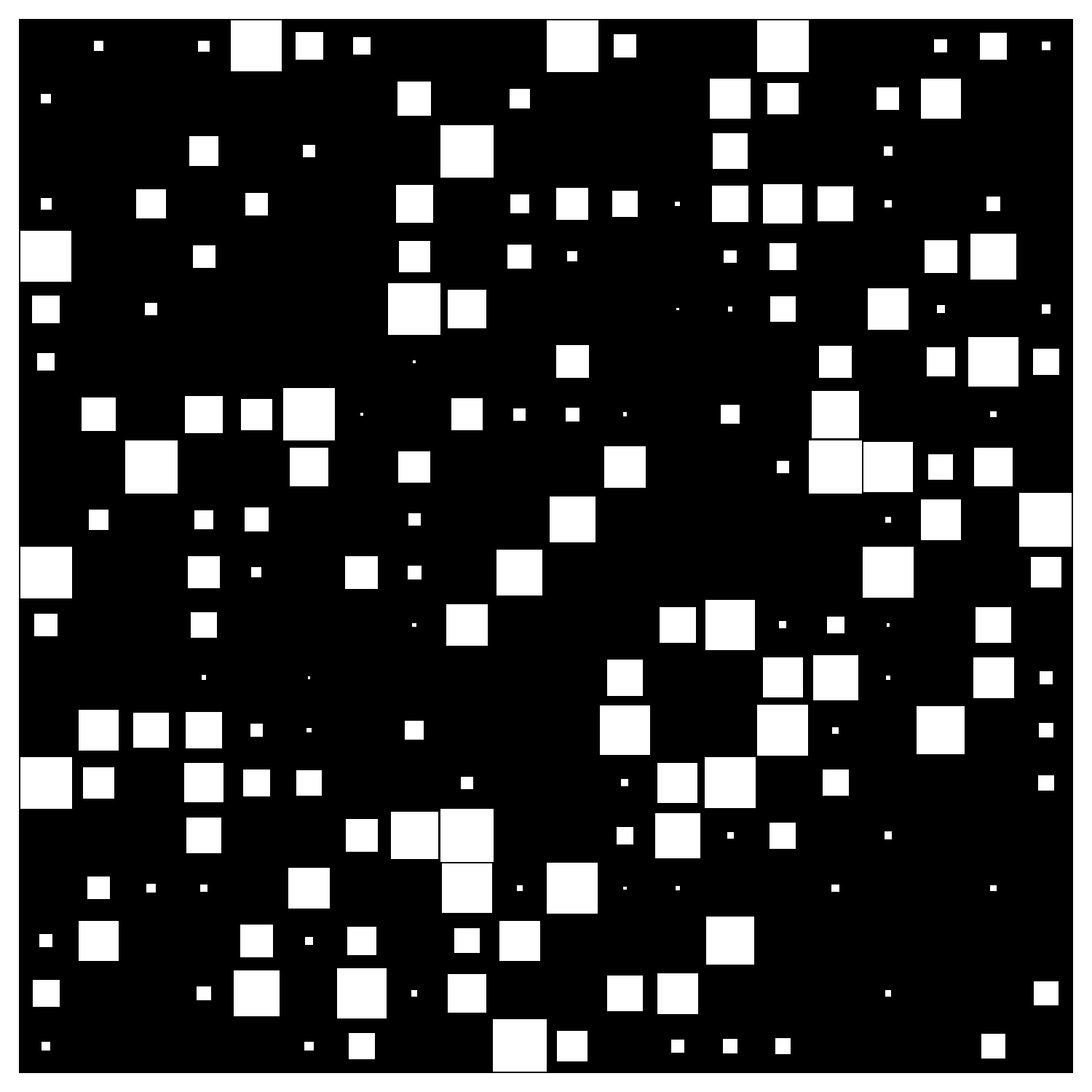}
  \caption[Weighted adjacency matrix]{Adjacency matrix of a weighted network, the size of the cell is proportional to the weight of the edges}
  \label{fig:adjacency=w}
\end{marginfigure}

\begin{marginfigure}%
  \includegraphics[width=\linewidth]{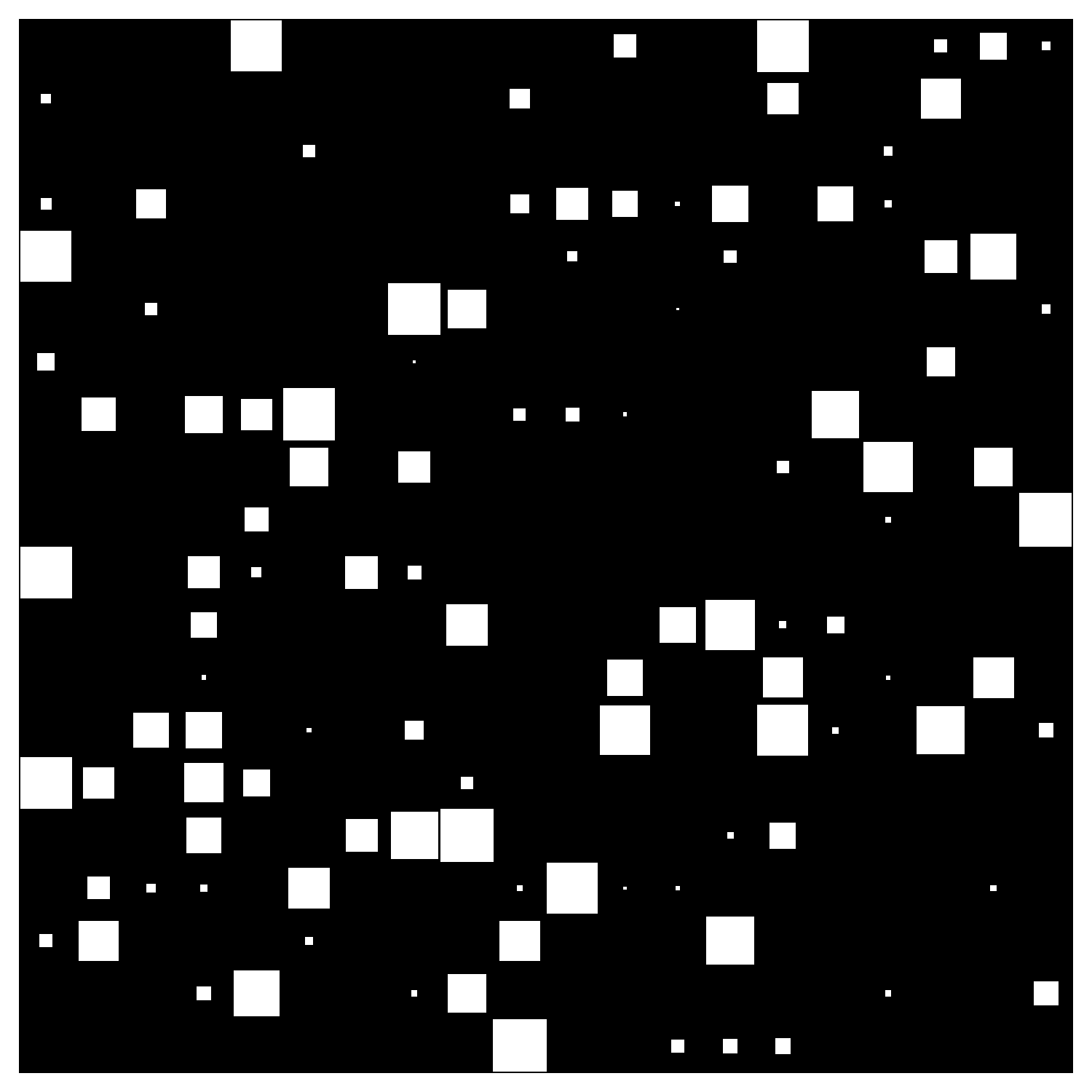}
  \caption[Directed and weighted adjacency matrix]{Adjacency matrix of a weighted and directed network, the adjacency matrix is not symmetric anymore}
  \label{fig:adjacency-wd}
\end{marginfigure}

In an undirected network, the existence of an edge \((i, j)\) implies that the edge is bidirectional, i.e., it connects node \(i\) to node \(j\) and vice versa. As a result, the adjacency matrix is symmetric, i.e., \(A = A^T\). The Figure \ref{fig:adjacency} shows the adjacency matrix of the network of Figure \ref{fig:network}. The total number of edges in the network can be determined by summing all the entries of 1 in the matrix \(A\).

\[
m = \frac{1}{2} \sum_{i=1}^{N} \sum_{j=1}^{N} A_{ji}.
\]

Usually empirical networks are directed and weighted, thus we also consider the case of directed and weighted networks. In weighted networks, for each pair of nodes \((i, j)\), the entry \(A_{ji}\) in the adjacency matrix reflects the strength of the interaction between node \(j\) and node \(i\). If \(A_{ij} = 0\), it indicates that there is no connection between \(i\) and \(j\). Conversely, if \(A_{ij} > 0\), a connection exists, and the weight is represented by \(A_{ij}\). Figure \ref{fig:adjacency=w} shows the adjacency matrix of a weighted network.

In a directed, weighted network, for each pair of nodes \((i, j)\), the entry \(A_{ij}\) in the adjacency matrix represents the weight of the interaction influencing node \(j\), where node \(i\) is the source. If \(A_{ij} = 0\), it means that node \(i\) cannot influence node \(j\), or that no flow can occur from \(i\) to \(j\). Conversely, if \(A_{ij} > 0\), node \(i\) can influence node \(j\). As a result, in directed networks, the adjacency matrices are typically asymmetric, i.e., \(A \neq A^T\). Figure \ref{fig:adjacency-wd} illustrates the adjacency matrix of a directed network.

\subsection{Node Degree \& Degree Distributions}

Nodes can play different roles depending on their characteristics and the specific analysis being performed. A basic characteristic of a node is the total number of its neighbors, which we will refer to as its degree.  

\begin{marginfigure}%
  \includegraphics[width=\linewidth]{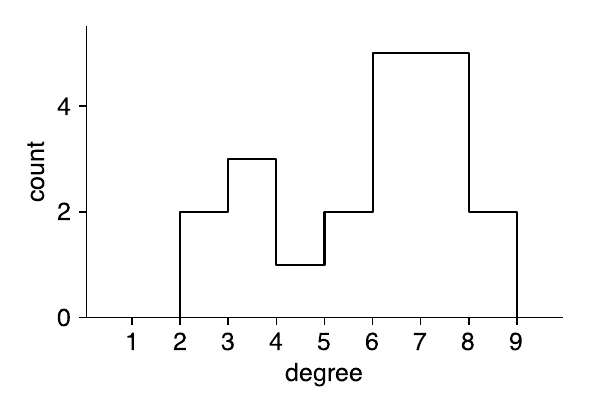}
  \caption[Degree distribution]{A histogram showing the distribution of node degrees within the network. The x-axis represents the degree (number of connections per node), and the y-axis represents the frequency of nodes with that degree. The plot provides insights into the connectivity patterns of the network. The top and right spines are removed for a cleaner appearance, and all text is presented in lowercase.}
  \label{fig:deg_dis}
\end{marginfigure}

The degree \( k \) of a node represents the number of links or edges connected to node \( i \). It can also be interpreted as the number of neighbors of \( i \), i.e. $k_i=|\mathcal{N}(i)|$. This value is computed by summing the entries in the corresponding row (or column) of the adjacency matrix\footnote{In an undirected network, the row and column sums are identical, leading to \( k_i = \sum_{j} A_{ji} = \sum_{j} A_{ij} \)}. In a directed network, however, these sums indicate the number of outgoing connections \( k_{\text{out}, i} = \sum_{j} A_{ij} \) and incoming connections \( k_{\text{in}, i} = \sum_{j} A_{ji} \). For an undirected network, the degree \( k_i \) of a node \( i \) is given by
\[
k_i = \sum_{j=1}^{N} A_{ji},
\]
which is also equal to the size of its neighborhood, \( k_i = |\mathcal{N}(i)| \). Since nodes generally do not have the same number of links the system is described by the degree distribution \( P(k) \), which gives the probability that a randomly chosen node has exactly \( k \) links. The Figure \ref{fig:deg_dis} shows the distribution of node degrees within the network of Figure \ref{fig:adjacency}.

In certain cases, it is sufficient to approximate the network by treating each node as an average node with a mean degree \( \langle k \rangle \), which is given by
\[
\langle k \rangle = \frac{2m}{N},
\]
where \( m \) is the total number of edges in a network of \( N \) nodes. The mean degree provides a measure of the network's density. When \( \langle k \rangle \) is much smaller than \( N \), the network is considered sparse, while an increase in the mean degree indicates a denser network. A network is called dense if, as \( N \) tends to infinity, the ratio \( \langle k \rangle / N \) remains greater than zero.

To gain a qualitative understanding of the degree distribution, we can examine its moments. The \( q \)\textsuperscript{th} moment of the degree distribution is defined as

\[
\langle k^q \rangle = \int k^q P(k) \, dk.
\]

In analytical calculations, it is often more convenient to assume that degrees can take any positive real values, hence the continuous definition. The lower moments offer important physical insights into the distribution:

\begin{itemize}
    \item \( q = 0 \): The zero\textsuperscript{th} moment \( \langle k^0 \rangle \) represents the total mass of the distribution, where \( \int P(k) \, dk = 1 \).
    \item \( q = 1 \): The first moment \( \langle k \rangle \) is the mean of the distribution, which corresponds to the average degree.
    \item \( q = 2 \): The second moment \( \langle k^2 \rangle \) relates to the variance of the distribution, \( \sigma^2 = \langle k^2 \rangle - \langle k \rangle^2 \), which measures the spread in the number of connections.
    \item \( q = 3 \): The third moment \( \langle k^3 \rangle \) indicates the skewness of the distribution, revealing the symmetry of \( P(k) \) around the mean. A symmetric distribution has zero skewness.
\end{itemize}

\subsection{Shortest paths, diameter and betweenness}

\begin{marginfigure}%
  \includegraphics[width=\linewidth]{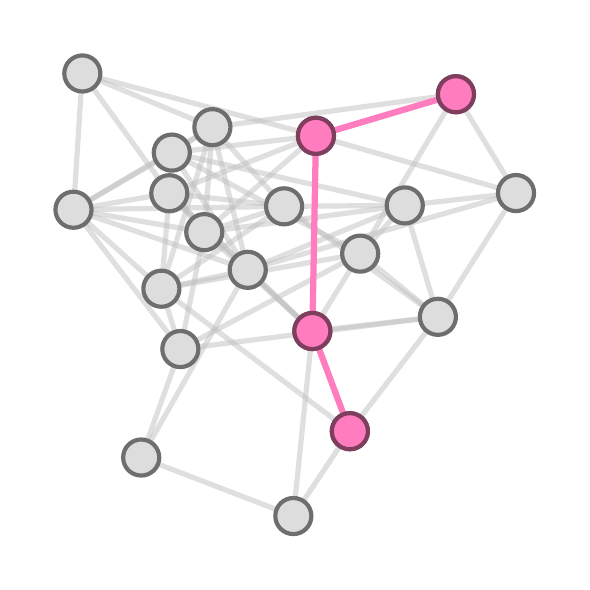}
  \caption[Shortest path]{A visualization of the network highlighting the shortest path between node 0 and node 10. The nodes and edges along this path are emphasized using a distinct color.}
  \label{fig:shortestpath}
\end{marginfigure}

Navigating and communicating across networks relies heavily on knowing the shortest paths. The distance between two nodes is measured by the number of links one must traverse to get from node \( i \) to node \( j \); this is known as the geodesic distance (we will see another definition of geodesic distance in section \ref{sec:hyperbolic}). It is important to note that there can be multiple shortest paths between two nodes.
The Figure \ref{fig:shortestpath} shows a shortest path between two nodes of the network of Figure \ref{fig:network}.

Watts\cite{watts1999networks} defined the typical separation between two nodes in a network as the average shortest path length, calculated as the mean geodesic distance \( \langle d_{ij} \rangle \) between all pairs of nodes,

\[
\langle d_{ij} \rangle = \frac{1}{\frac{N(N-1)}{2}} \sum_{i \neq j} d_{ij}.
\]

Consequently, the diameter of a network is defined as the longest geodesic distance between any two nodes. However, this definition encounters issues as the diameter \( d \) diverges if there are disconnected nodes or components. To address this, Latora and Marchiori\cite{latora2001efficient} proposed an alternative by defining the average efficiency, or harmonic mean distance, as,

\[
E = \frac{1}{\frac{N(N-1)}{2}} \sum_{i \neq j} \frac{1}{d_{ij}}.
\]

Even if \( d_{ij} \) diverges, the efficiency \( E \) will still be well-defined. The average distance provides insight into the size of a network but does not indicate which paths are crucial. Communication between two non-neighboring nodes \( i \) and \( j \) depends on other nodes along the paths that connect them. In this context, the importance of a node can be measured by counting the number of shortest paths that pass through it. The betweenness \( b_i \) of node \( i \) is defined as,

\[
b_i = \sum_{u \neq v \neq i} \frac{\sigma_{uv}(i)}{\sigma_{uv}},
\]

where \( \sigma_{uv} \) refers to the total number of shortest paths that connect nodes \( u \) and \( v \), while \( \sigma_{uv}(i) \) represents the number of these paths that pass through node \( i \). Analogous to node betweenness, the betweenness of an edge can similarly be defined as the number of shortest paths that traverse a specific edge \citep{costa2007}. Alongside degree, betweenness serves as a measure of a node's (or an edge's) centrality within a network. This concept is crucial for understanding the vulnerability of networks, as it quantifies the impact of structural failures on network performance \citep{holme2002, newman2003}. There are numerous ways to define centrality, which in turn gives rise to a multitude of centrality measures. These include closeness centrality \citep{sabidussi1966}, straightness centrality \citep{vragovic2005}, Katz centrality \citep{katz1953}, eigenvector centrality \citep{newman2008}, and information centrality \citep{porta2006}.

\begin{marginfigure}%
  \includegraphics[width=\linewidth]{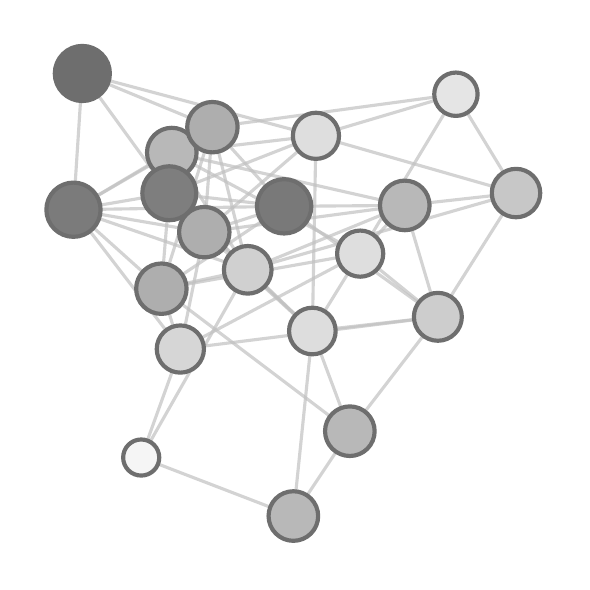}
  \caption[Clustering coefficient]{A visualization of the network where each node's size and color intensity represent its clustering coefficient. Nodes with higher clustering coefficients appear larger and are depicted with lighter colors, highlighting areas where nodes form tightly connected neighborhoods.}
  \label{fig:clustering}
\end{marginfigure}

\subsection{Clustering Coefficient}

In social networks, it is common for a friend's friend to also become your friend. This phenomenon is known as clustering or transitivity \citep{wasserman1994}. In network topology, clustering is characterized by a higher occurrence of triangles—subsets of three interconnected nodes. Transitivity \( T \) is defined by comparing the number of closed triangles to the number of open triplets (i.e., two nodes connected to a third one) within the network \citep{barrat2000},

\begin{align}
T = \frac{\text{number of triangles in the network}}{\text{number of connected triplets of nodes}}
\label{eq:clustering_frac}
\end{align}

A connected triplet refers to a single node linked to two others, forming an open triangle. The factor of three accounts for each triangle containing three triplets, ensuring \( T \) stays within \( 0 \leq T \leq 1 \). Essentially, transitivity measures the proportion of connected triplets that are closed by an additional edge, forming a triangle.

Watts and Strogatz\cite{watts1998collective} introduced an alternative definition called local clustering, which focuses on individual nodes. For a node \( i \), the clustering coefficient \( C_i \) is,

\begin{align}
    C_i = \frac{\text{number of triangles connected to node } i}{\text{number of triplets centered on node } i}.
    \label{eq:clustering_avg}
\end{align}

This definition is widely used in sociological literature as network density \citep{scott2011}. Figure \ref{fig:clustering} illustrates the clustering coefficient in the network depicted in Figure \ref{fig:network}. The global clustering coefficient is then the average of the local values,

\begin{align}
C = \frac{1}{n} \sum_{i=1}^{n} C_i.
\end{align}

By definition, \( 0 \leq C_i \leq 1 \) and \( 0 \leq C \leq 1 \). The difference between transitivity and clustering lies in their calculations: Equation \ref{eq:clustering_frac} computes a fraction of averages, while Equation \ref{eq:clustering_avg} averages fractions. Regardless of the measure used, social networks generally show higher clustering or transitivity than random networks, along with degree correlations—key features that distinguish social networks \citep{newman2003}.

\subsection{Random Network Models} 

Network models are developed to highlight various phenomena observed in real systems or serve as reference points or null models for comparing effects across different networks. This section introduces and explains the most fundamental network models and their main properties.

\begin{marginfigure}%
  \includegraphics[width=\linewidth]{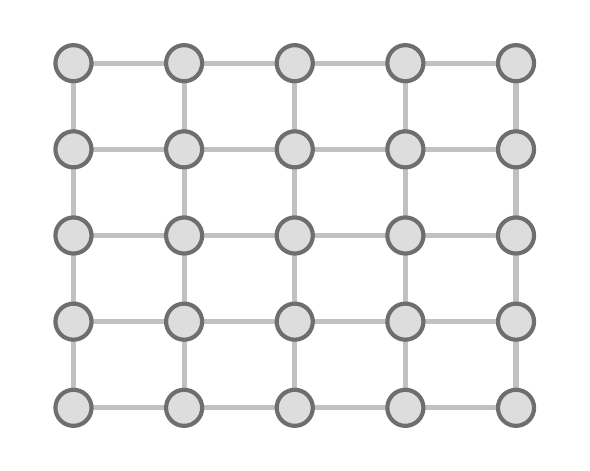}
  \caption[Lattice network]{A visualization of an lattice network random graph with 25 nodes.}
  \label{fig:marginfig}
\end{marginfigure}

\newthought{Lattice Networks} are simplest form of a network. While lattices are extremely useful for modeling physical systems, they do not resemble real-world networks because real-world networks are typically irregular, with both ordered and disordered structures coexisting.
Although some street networks exhibit characteristics reminiscent of lattice networks \citep{crucitti2006centrality}.

\begin{marginfigure}%
  \includegraphics[width=\linewidth]{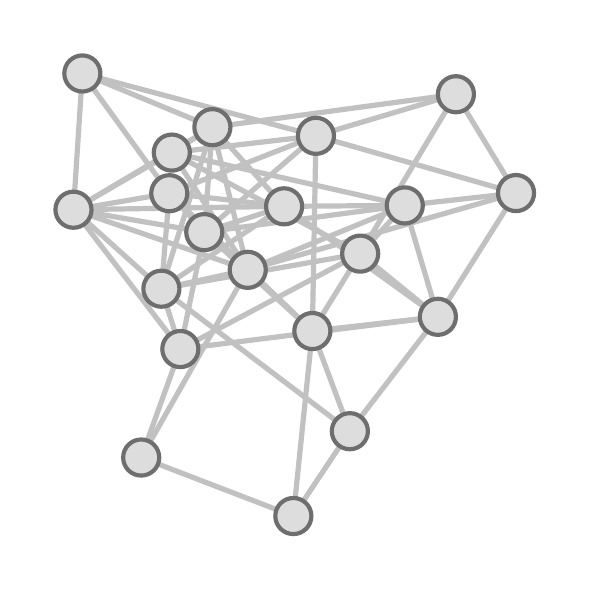}
  \caption[Erdős-Rényi random graph]{A visualization of an Erdős-Rényi random graph with 20 nodes and a connection probability of 0.3. Nodes are colored in a specified palette with darker borders, and edges are drawn in light grey. The spring layout algorithm positions the nodes to reflect the network's structure.}
  \label{fig:marginfig}
\end{marginfigure}

\newthought{Erdős–Rényi Random Networks}, the simplest stochastic network models is the Erdős–Rényi (ER) model, where in a network of \(N\) nodes, each pair of nodes is connected with a probability \(p_{ER}\). Despite its simplicity, the ER model is widely used as a reference model to analyze structural features in other network models and empirically observed static networks. In the ER model, the clustering coefficient is equal to the connection probability, \(C_{ER} = p_{ER}\), and the mean degree is \(\langle k \rangle = p_{ER} \times (N - 1)\).

The degree distribution in ER networks follows a binomial distribution, making node degrees relatively homogeneous. The average shortest path length scales as \(\langle s \rangle \propto \log N\), meaning nodes are quickly reachable within the network. However, the lack of clustering makes the ER model less suitable for modeling social networks.

\begin{marginfigure}%
  \includegraphics[width=\linewidth]{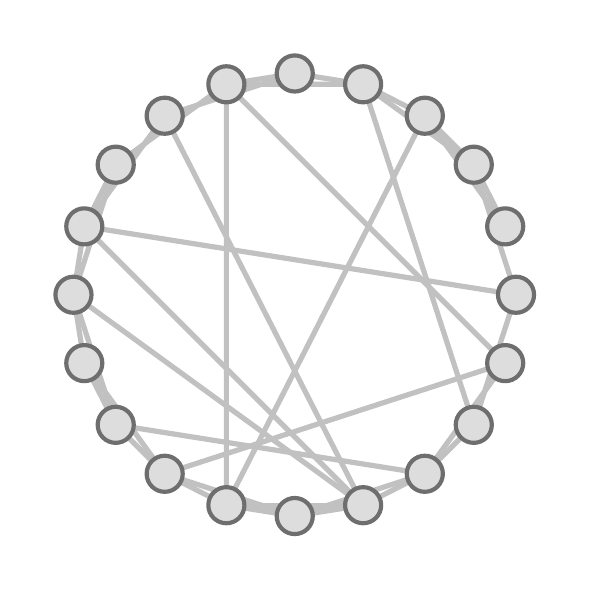}
  \caption[Small-world network]{A visualization of a small-world network created using the Watts-Strogatz model. The network consists of 20 nodes arranged in a circular layout, where each node is connected to 4 nearest neighbors with a rewiring probability of 0.3.}
  \label{fig:marginfig}
\end{marginfigure}

\newthought{Small-World Networks}, social networks exhibit two critical properties: they are highly clustered, and they display the small-world effect, i.e the average shortest path length is low.
In 1999, Watts and Strogatz introduced a model to capture these properties by starting with a \(k\)-regular nearest neighbor lattice and introducing a rewiring probability \(p_r\) \citep{watts1998collective}. This model provide networks with high clustering and with an average shortest path length that decrease as \(p_r\) increases. 
To address how people can efficiently identify short chains for communication without full knowledge of the network, Kleinberg introduced a variation of the small-world model that includes a notion of distance \citep{kleinberg2000small}. In this model, nodes are placed on a two-dimensional lattice, and long-range connections are added with a probability that decays algebraically with distance, optimizing the conditions under which short paths are most likely to be discovered.

\newthought{Modular Hierarchical Networks}, networks often display modular and hierarchical structures. Although the models introduced above do not reflect that. Watts, Dodds, and Newman proposed a model where nodes are grouped locally within a hierarchical tree \citep{watts2002identity}. Each pair of nodes has a hierarchical distance, and connections are formed based on this distance. Within this model, each node belongs to a local group of size \( g \), organized in a hierarchical tree. The hierarchical distance between any pair of nodes \( (i, j) \), denoted as \( l_{ij} \), is defined as the number of steps required to locate their lowest common ancestor in the tree. Nodes within the same group have a hierarchical distance of \( l = 1 \), while nodes in different groups but within the same module have a distance \( l = 2 \), and so forth. The probability that two nodes \( (i, j) \) are connected is given by:

\[
p_l = c \exp(-\alpha(l - 1))
\]

This probability function results in a nested-block structure in the average adjacency matrix.

\newthought{Barabási-Albert (BA) Networks:} Real-world networks often display a heavy-tailed degree distribution, where most nodes have a small degree, while a few nodes, termed hubs, possess a large degree \citep{barabasi1999emergence}. The Barabási-Albert model formalizes the process of generating such scale-free networks through two mechanisms: growth and preferential attachment.

\begin{marginfigure}%
  \includegraphics[width=\linewidth]{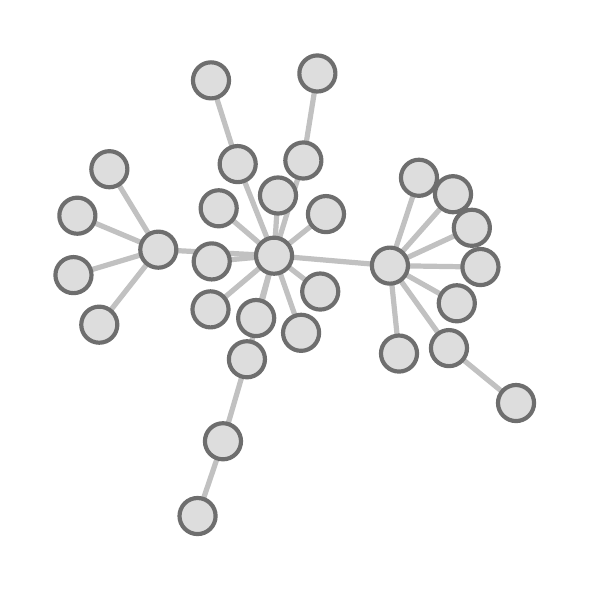}
  \caption[Barabási-Albert (BA) network]{A visualization of a Barabási-Albert (BA) network generated with 30 nodes, where each new node connects to 1 existing nodes. The network exhibits a scale-free topology characterized by the presence of hubs—nodes with significantly higher degrees.
}
  \label{fig:marginfig}
\end{marginfigure}

The model starts with a network of \( m_0 \) nodes. At each time step, a new node is added, and it forms \( m \leq m_0 \) links to existing nodes. The key to the model is preferential attachment, where the probability \( \Pi(k_i) \) that a new node connects to an existing node \( i \) with degree \( k_i \) is given by:

\[
\Pi(k_i) = \frac{k_i}{\sum_j k_j}
\]

This preferential attachment ensures that nodes with higher degrees (more connections) are more likely to gain new links, creating a rich-get-richer effect. As a result, the network evolves into a scale-free structure, characterized by a power-law degree distribution:

\[
P(k) \propto k^{-\gamma}
\]

where \( \gamma = 3 \) is the typical exponent for the BA model. This power-law distribution implies that hubs—nodes with very high degrees—are much more likely to occur compared to what would be expected in a random network (e.g., an Erdős-Rényi network).

The scale-free nature of BA networks has significant implications, particularly in network robustness. While random node failures typically have a limited effect on the network's overall connectivity, the failure of hubs can lead to catastrophic fragmentation. Moreover, the dynamics of spreading processes such as epidemics or information dissemination are heavily influenced by the presence of hubs, as they can act as super-spreaders \citep{barabasi1999emergence}.

The Barabási-Albert model provides a framework for understanding the emergence of power-law distributions in networks, particularly the degree distribution in scale-free networks. However, recognizing such distributions in real-world data is not always straightforward. While the BA model predicts a power-law degree distribution with a characteristic exponent, real-world networks often deviate from this ideal. Thus, to accurately validate and characterize power-law behavior in empirical data, it is essential to employ robust statistical techniques \citep{clauset_power-law_2009}.

\section{Heavy-Tailed Distributions}
\label{sec:tail}

\begin{marginfigure}
  \includegraphics[width=\linewidth]{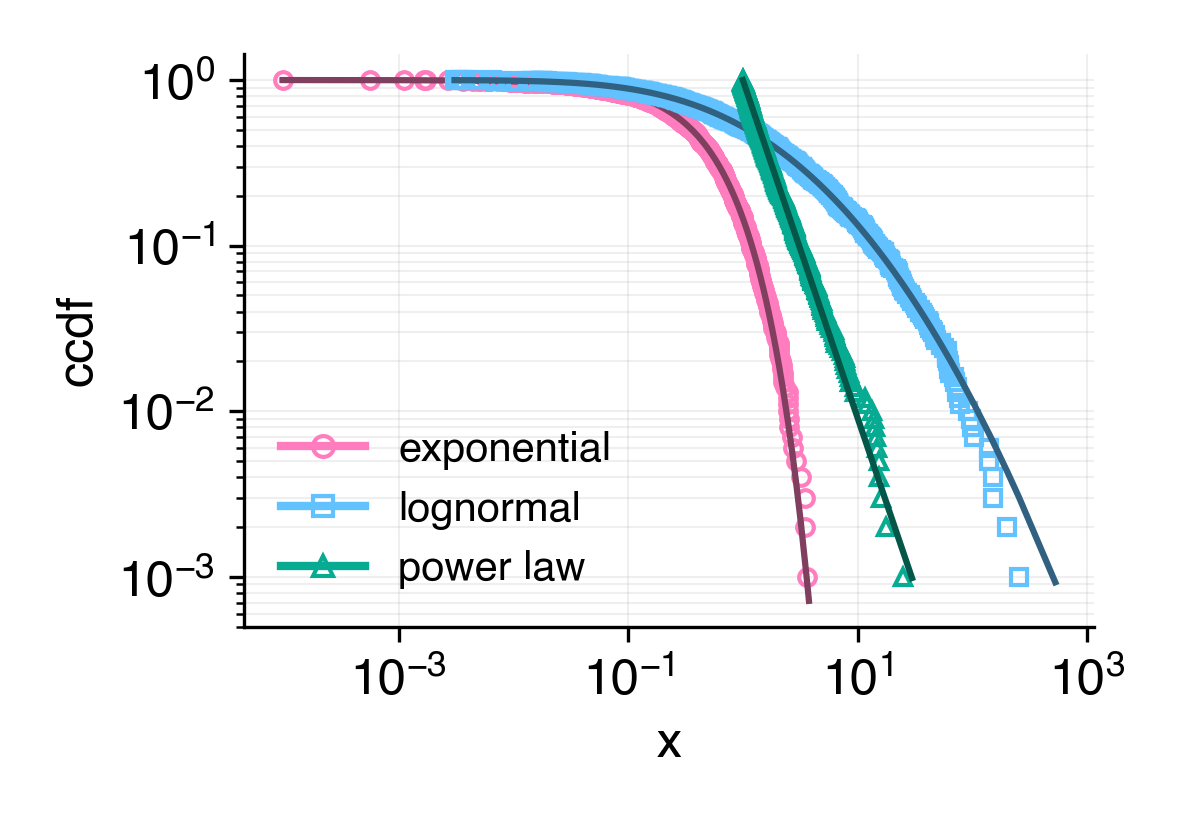}
  \caption[Heavy-tailed distributions]{Complementary cumulative distribution functions (CCDFs) of exponential, lognormal, and power-law distributions on a log-log scale. The empirical data for each distribution is represented by open circles, and the fitted lines are overlaid in matching but darker colors.}
  \label{fig:heavy_tailed}
\end{marginfigure}

Heavy-tailed distributions, particularly power law distributions, have emerged as a fundamental concept in complexity science, appearing across diverse fields such as network science, statistical physics, and human mobility studies. These distributions are characterized by their "fat tails," which decay more slowly than exponential distributions, indicating a higher probability of extreme events (see Figure \ref{fig:heavy_tailed}) \citep{clauset2009power}.

In network science, power law distributions are ubiquitous in the degree distribution of many real-world networks, including the World Wide Web, social networks, and biological networks \citep{barabasi1999emergence}. The presence of power laws in these systems often indicates a scale-free property, where the network structure remains similar at different scales of observation.Statistical physics has also embraced power law distributions as a key concept in understanding critical phenomena and phase transitions \citep{stanley1999scaling}. The universality of power laws in these contexts suggests that similar underlying mechanisms may be at play across seemingly disparate systems.

Human mobility patterns have been shown to exhibit power law distributions in various aspects, such as the distribution of travel distances and the frequency of visits to locations \citep{gonzalez2008understanding}. These findings have significant implications for urban planning, epidemiology, and the design of transportation systems. Some studies suggest human mobility follows power law distributions, while others indicate different patterns, such as exponential or stretched exponential distributions \citep{brockmann2006scaling}. The concept of Lévy flights, characterized by power law distributed step lengths, has been both supported and challenged in human mobility research. This paradox in findings has sparked debates about the true nature of human movement patterns.\footnote{Recent research has resolved this apparent paradox by showing that day-to-day human mobility does indeed contain meaningful scales, corresponding to spatial containers that restrict mobility behavior \citep{alessandretti2020scales}. The scale-free results arise from aggregating displacements across these containers.}

The prevalence of power law distributions in complex systems has led to discussions about universality in nature \citep{bak1987self}. Some researchers argue that the ubiquity of power laws points to common organizing principles across diverse phenomena, potentially leading to a unified theory of complex systems \citep{west2017scale}.

However, it is important to note that the identification and interpretation of power law distributions in empirical data can be challenging \citep{stumpf2012critical}. Rigorous statistical methods are necessary to distinguish true power law behavior from other heavy-tailed distributions and to avoid false positives in power law detection \citep{clauset2009power}.

\subsection{Estimating Power Law Distributions}

Estimating power-law distributions from empirical data is a crucial task in various fields, from physics to social sciences  \citep{virkar2014power}. The process of identifying and characterizing power laws in real-world data is non-trivial and requires careful consideration of various statistical techniques. While several methods exist, each with its strengths and limitations, the choice of method can significantly impact the conclusions drawn from the data \citep{clauset2009power}.

\newthought{Discrete vs Continuous} when estimating power-law distributions, it is important to distinguish between data drawn from continuous and discrete sample spaces. In practice, many empirical datasets, such as those representing wealth distributions or city populations, originate from discrete spaces, while others, like physical measurements or distances, are continuous. The distinction between continuous and discrete sample spaces affects the normalization of power-law distributions and, consequently, the method of parameter estimation.

For data from continuous sample spaces, \( O_c = [x_{\min}, x_{\max}] \), the normalization constant \(Z_{\gamma}\) for a power-law distribution \(p(x) = Z^{-1} x^{-\gamma}\) is given by:

\[
Z_{\gamma}(O_c) = \int_{x_{\min}}^{x_{\max}} x^{-\gamma} \, dx = \frac{x_{\max}^{1-\gamma} - x_{\min}^{1-\gamma}}{1-\gamma} \quad \text{for} \quad \gamma \neq 1.
\]

For data from discrete sample spaces, \( O_d = \{ z_1, z_2, \dots, z_W \} \), the normalization constant becomes a sum over the possible values \( z_i \):

\[
Z_{\gamma}(O_d) = \sum_{i=1}^{W} z_i^{-\gamma}.
\]

The choice between treating the data as continuous or discrete depends on the nature of the data. For example, even if the data is inherently continuous, it may be represented discretely in practice due to measurement precision or other factors. In such cases, it is often appropriate to treat the data as discrete for normalization and estimation purposes.

A practical rule of thumb is that data with many unique values (such as physical measurements) can be treated as continuous, while data with fewer unique values (such as population counts) can be treated as discrete.

\newthought{Least-squares on log data}, a dangerous approach to estimating power-law exponents is a least-squares linear regression on the log-transformed data. Indeed, while this method is straightforward, it is fundamentally flawed for power-law distributions due to biases that arise from the log-log transformation and the inappropriate weighting of data points.

\begin{figure}[h!]
\checkoddpage \ifoddpage \forcerectofloat \else \forceversofloat \fi
    \centering
    \includegraphics[width=\linewidth]{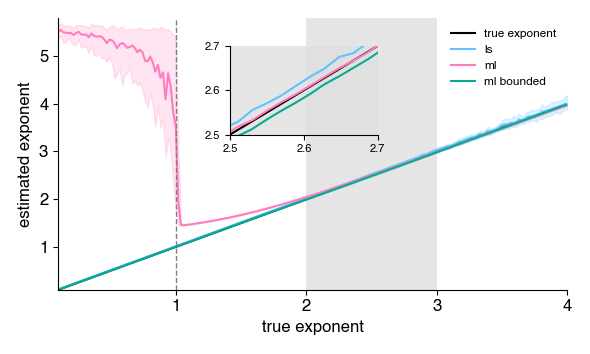}
    \caption[Estimating power-law]{Comparison of least-squares (LS), maximum-likelihood (ML) \citep{clauset2009power}, and bounded maximum-likelihood (ML bounded) \citep{hanel2017fitting} estimators for power-law exponents. The solid lines represent the estimated exponents across a range of true exponents (0.1 to 4). The shaded regions show the 10th and 90th percentiles for each estimator, illustrating the uncertainty in the estimates. The LS estimator is shown in blue, ML in pink, and ML bounded in teal. The black line represents the true exponent values. The inset zooms in on the range of exponents between 2.5 and 2.7, highlighting the performance of the estimators in this critical region. Most power laws have an exponent within the grey area, between 2 and 3 \citep{clauset2009power}.}
    \label{fig:power-law}
\end{figure}

The least-squares linear regression method involves taking the logarithm of both sides of the assumed power-law relationship \( y = Cx^{-\gamma} \), which yields:

\[
\log y = -\gamma \log x + \log C
\]

This linear equation suggests that by plotting \(\log y\) against \(\log x\), one can estimate \(\gamma\) as the slope of the resulting line using linear regression. Although simple, this method is problematic because the transformation amplifies errors for smaller values of \(x\). As a result, this approach tends to biased estimation of the exponent, especially when dealing with data spanning several orders of magnitude \citep{goldstein2004problems}. Figure \ref{fig:power-law} shows how the least-squares estimate the exponent with a bias (see the insert the blue line of the ls regression is always above the black line of the true exponent).

\subsection*{Maximum Likelihood Estimator}

To address the shortcomings of the least-squares approach, Clauset, Shalizi, and Newman \citep{clauset2009power} proposed a method based on Maximum Likelihood Estimation (MLE), which provides a more robust and unbiased estimate of the power-law exponent \(\gamma\). For a set of observed data points \(x_1, x_2, \dots, x_n\) that follow a power-law distribution \(p(x) = Cx^{-\gamma}\) for \(x \geq x_{\min}\), the MLE for \(\gamma\) is given by:

\[
\hat{\gamma} = 1 + n \left( \sum_{i=1}^{n} \ln \frac{x_i}{x_{\min}} \right)^{-1}
\]

where \(n\) is the number of data points with \(x \geq x_{\min}\). This method estimate the exponent without a bias\footnote{It's important to note that after \(x_{\min}\) has be estimated, the Clauset et al.'s method ignores the distribution of data below \(x_{\min}\), which can lead to biases if the data below this threshold carries important information. Recent work by Maier \citep{maier2023maximum} addresses this limitation by proposing a method to fit piece-wise Pareto distributions with a finite and non-zero core, effectively incorporating the information from the data below \(x_{\min}\). This method is used in the last chapter (\ref{sec:chapter3}).}.
The Figure \ref{fig:power-law} shows how the maximum-likelihood estimate the exponent for the range $[2,3]$ where lies most of the power law exponent of empirical observed \citep{clauset_power-law_2009}. Although the maximum likelihood method starts to be bias when the exponent is smaller than 2, i.e. when its the variance become infinite.

\subsection*{Bounded Power Law}

\begin{marginfigure}%
  \includegraphics[width=\linewidth]{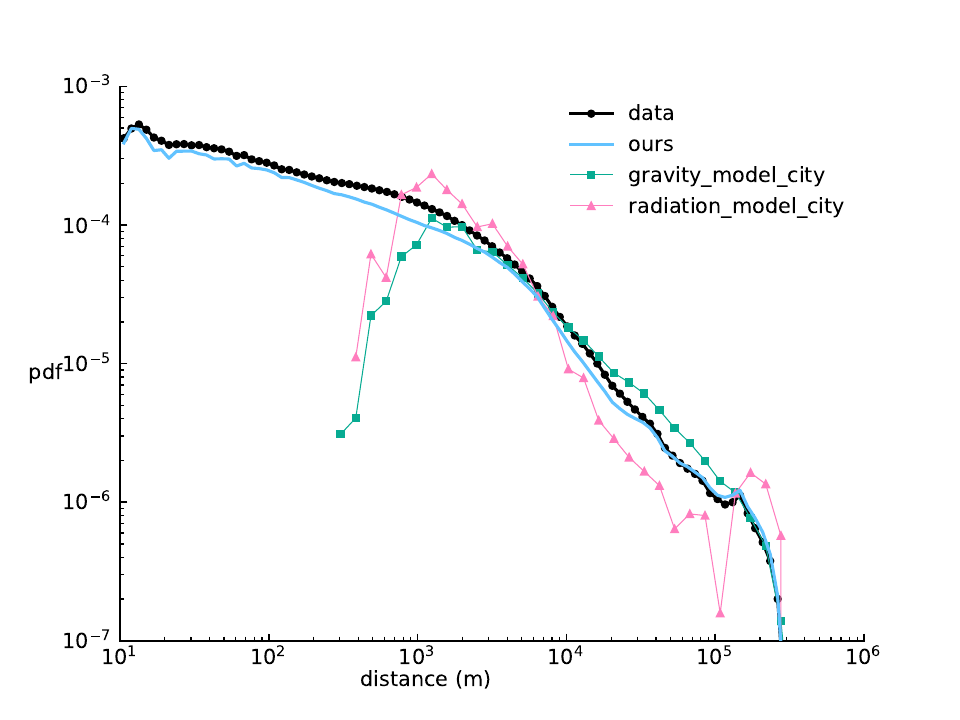}
  \caption[Gravity and Radiation model]{Residential Mobility in Denmark: Comparison between the empirical distance distribution, the gravity model, the radiation model and our framework (see section\ref{sec:decomposing}).}
  \label{fig:dk_grav}
\end{marginfigure}

In many practical scenarios, power-law distributions are not unbounded; instead, they are bounded due to physical, biological, or social constraints. For instance, the distance distribution of residential moves in Denmark is bounded by the maximal distance between two addresses in Denmark (465km, see Figure \ref{fig:dk_grav}). In such cases, the correct normalization of the distribution becomes crucial, particularly when the exponent \(\gamma\) is less than 2, which corresponds to distributions with infinite variance in the unbounded case.

For a bounded power-law distribution, the probability density function  can be written as:

\[
p(x) = \frac{(\gamma-1)x_{\min}^{\gamma-1}}{1 - \left(\frac{x_{\max}}{x_{\min}}\right)^{1-\gamma}} x^{-\gamma} \quad \text{for } x_{\min} \leq x \leq x_{\max}
\]

where \(x_{\max}\) is the upper bound of the distribution. The normalization constant ensures that the total probability integrates to 1 over the bounded interval \([x_{\min}, x_{\max}]\).

When \(\gamma < 2\), the distribution has a finite mean but an infinite variance, making the estimation of \(\gamma\) particularly sensitive to the choice of \(x_{\max}\). In such cases, the MLE method can still be applied, but it requires careful consideration of the bounded nature of the data \citep{hanel2017fitting}.
Figure \ref{fig:power-law} shows how the bounded maximum-likelihood estimate the exponent for the range $[0,2]$. However, it should be noted that the bounded maximum-likelihood approach also exhibits a bias, which is a consequence of the assumption that the support of the power law is bounded.

\section{Community detection}

Community detection is a fundamental problem in network science, where the goal is to identify the modular structure inherent in complex networks. In real-world networks, nodes are often organized into tightly knit groups, or communities, with dense intra-group connections and sparser inter-group connections. These communities can represent functional units in various systems, such as social circles in social networks, biological modules in metabolic networks, or clusters of related web pages in the internet graph. Detecting these structures is crucial for understanding the organization, dynamics, and function of networks. Several algorithms have been developed for community detection, including divisive methods like Girvan-Newman, modularity-based approaches such as Louvain, and inference methods like stochastic block models (SBMs). Each method has its advantages, depending on the nature of the network and the specific definition of "community" \citep{fortunato2010community,peixoto2014hierarchical}.

\begin{marginfigure}
  \includegraphics[width=\linewidth]{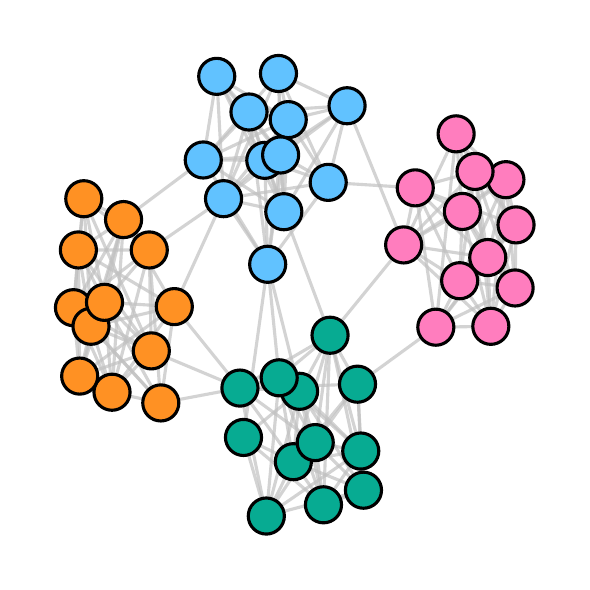}
  \caption[Modularity]{Stochastic block model model, illustrating high modularity with four distinct communities. Each community is represented by a different color, showcasing the strong intra-community connections and sparse inter-community links. This structure emphasizes the modular nature of the network, where nodes within the same community are more densely connected compared to those in different communities}
  \label{fig:marginfig}
\end{marginfigure}

\subsection{Modularity}

Modularity is one of the most widely used metric for community detection, introduced by Newman and Girvan \citep{girvan2002}. It provides a measure of the strength of the division of a network into communities, comparing the density of edges within communities to that is expected ramdomly. The modularity \( Q \) is defined as:

\begin{equation}
    Q = \frac{1}{2m} \sum_{ij} \left( A_{ij} - \frac{k_i k_j}{2m} \right) \delta(c_i, c_j)
\end{equation}

In this equation, \( A_{ij} \) is the adjacency matrix that represents the presence or absence of an edge between nodes \( i \) and \( j \). The variables \( k_i \) and \( k_j \) represent the degrees of nodes \( i \) and \( j \), respectively, where the degree \( k_i \) is the total number of edges connected to node \( i \). The term \( m \) denotes the total number of edges in the network. The Kronecker delta \( \delta(c_i, c_j) \) is equal to 1 if nodes \( i \) and \( j \) belong to the same community and 0 otherwise.

This equation measures how well the network is divided into communities by comparing the observed number of edges inside communities with the number of edges that would be expected under a random null model, where edges are placed between nodes according to their degree distribution.

\newthought{Modularity Optimization}, maximizing \( Q \) directly is computationally challenging due to the vast number of possible partitions of nodes into communities. Heuristic approaches are often used to find approximate solutions. For example, a greedy algorithm that iteratively merges communities to achieve a local maximum of \( Q \)\cite{clauset2004} . Another widely adopted method is the Louvain algorithm\cite{blondel2008}, which optimizes modularity in two steps. First, each node is placed in the community of a neighbor that maximizes the local modularity gain. After all nodes are assigned to communities, the algorithm aggregates nodes within communities into a super-node and repeats the process until no further improvement is possible.

One of the important concepts in modularity optimization is the change in modularity \( \Delta Q \) when moving a node \( i \) from one community to another. This change can be expressed as:

\begin{equation}
    \Delta Q = \left[ \frac{\sum_{in} + k_i^{in}}{2m} - \left( \frac{\sum_{tot} + k_i}{2m} \right)^2 \right] - \left[ \frac{\sum_{in}}{2m} - \left( \frac{\sum_{tot}}{2m} \right)^2 - \left( \frac{k_i}{2m} \right)^2 \right]
\label{eq:louvain_step}
\end{equation}

\newthought{Limitations of Modularity}, modularity maximization has many problems, one of them is the resolution limit. It means that methods that optimize modularity fail to detect small communities within large networks \citep{fortunato2007}. This occurs because modularity inherently favors large community structures, which can lead to the merging of small but meaningful communities. The resolution limit can be understood by considering how the modularity function compares the density of connections inside communities to that of the null model. In large networks, small communities may have a high internal density but may still be combined with larger communities due to the scale at which modularity operates. The modularity function with a resolution parameter is given by,

\[
Q = \frac{1}{2m} \sum_{ij} \left[ A_{ij} - \gamma \frac{k_i k_j}{2m} \right] \delta(c_i, c_j),
\]

where the new parameter \( \gamma \) is the resolution parameter, controlling the size of the detected communities (lower values favor larger communities, higher values favor smaller communities).

Another limitation of modularity is its tendency to detect spurious communities even in random networks. This phenomenon, often referred to as overfitting, arises because modularity maximization can find high values of \( Q \) even when applied to random graphs that lack true community structure\cite{peixoto2023descriptive}. This indicates that modularity alone is not a reliable measure of community structure, especially in networks where no clear community structure exists. 

Efforts to overcome these limitations include multi-scale approaches and resolution parameter adjustments, but these methods often introduce their own challenges. In particular, adjusting the resolution parameter allows detection of smaller communities but can lead to artificial partitioning of larger communities. Therefore, while modularity remains a popular tool for community detection, it should be used with caution. To ensure the robustness of the methods, it should be compared with other community detection methods such as InfoMap.

\subsection{Infomap}

InfoMap is an information-theoretic approach to community detection introduced by Rosvall and Bergstrom\cite{rosvall2008}. Unlike modularity-based methods, InfoMap relies on the dynamics of random walks to reveal community structure. The main idea is that a random walker tends to remain within communities for long periods, making it possible to identify communities by encoding the walker’s trajectory efficiently. By minimizing the description length of the random walk, InfoMap uncovers a network’s hierarchical and overlapping community structure. The main equation of InfoMap is the map equation that separates the total description length into two terms. The map equation is defined as:

\begin{equation}
    L(M) = q_{\curvearrowleft} H(Q) + \sum_{i=1}^{m} p_i H(P^i)
\end{equation}

Here, \( L(M) \) represents the total description length of the random walk. The term \( q_{\curvearrowleft} \) is the probability of exiting a community, and \( H(Q) \) is the entropy associated with inter-community movements:

\[
H(Q) = - \sum_{i} q_i \log q_i
\]

Meanwhile, \( p_i \) corresponds to the stationary distribution of the random walk within community \( i \), and \( H(P^i) \) is the entropy of movements within the community, defined as:

\[
H(P^i) = - \sum_{j} p_{ij} \log p_{ij}
\]

The map equation works by balancing these two costs—encoding the movement between communities and within them—to find the optimal division of the network.

\newthought{Random walks} are central to InfoMap for detecting communities. A random walker tends to stay within densely connected groups of nodes, known as communities, for extended periods before transitioning to another community. The stationary distribution \( p_i \), which is determined by the flow of probability through the network, reflects how important each node is in keeping the random walker within its community. The probability of exiting a community, \( q_{\curvearrowleft} \), represents the total likelihood of the random walker crossing community boundaries. This is minimized when the network is partitioned optimally into communities.

One of InfoMap's advantages is its ability to handle hierarchical structures. It naturally captures community hierarchies, where larger communities are subdivided into smaller ones, allowing for multiple levels of organization. This feature is particularly effective in real-world networks, where hierarchical and overlapping community structures are common (see section \ref{paper1}).

InfoMap has been extended in various ways to adapt to different network properties. InfoMap can be applied to directed and weighted networks, where the direction of edges and edge weights influence the random walk dynamics\cite{rosvall2010}. For example, in a directed network, the movement probabilities are determined by the edge directionality, while in weighted networks, the edge weights affect transition probabilities. Furthermore, InfoMap has been extended to incorporate memory effects, allowing for more accurate community detection in cases where the transition probabilities depend on the walker’s previous steps. This memory-infomap extension captures richer dynamics, such as the tendency of certain paths to recur, which is important in networks like transportation systems or biological pathways\footnote{Memory-infomap allows for trajectories with memory, improving community detection in systems with non-Markovian dynamics, where the transition probabilities depend on past states \citep{rosvall2014}.
The transition probabilities of a sparse Markov chain with memory can be represented as,

\begin{scriptsize}
\begin{align*}
P(X_{t+1} =& x \mid X_t = x_t, \dots, X_{t-n} = x_{t-n}) = \\
&\sum_{j=1}^{n} p_j P(X_{t+1} = x \mid X_{t-j} = x_{t-j})
\end{align*}
\end{scriptsize}

where \( P \) is the transition probability, \( X_t \) represents the state at time \( t \), and \( p_j \) defines the weight of past states on the current transition. This approach captures systems where current transitions are influenced by a limited memory of previous states, offering a more complex but realistic representation of network dynamics.
}.

Multi-level InfoMap extends the original framework to allow for the identification of communities at multiple hierarchical levels. This approach facilitates a multi-scale analysis, enabling the simultaneous discovery of both smaller, tightly connected sub-communities and larger, overarching structures within the network. Such capabilities are especially beneficial for large datasets like citation networks \citep{rosvall2011}.

\subsection{Stochastic Block Model}
\label{sec:dcSBM}
The Stochastic Block Model (SBM) is a generative model for networks that assumes that nodes belong to distinct communities, and the probability of an edge between two nodes depends on their respective communities. SBM is widely used for community detection because it provides a probabilistic framework that explicitly models the community structure. The edge probabilities between nodes are determined by the community memberships of the nodes.

The likelihood function for the SBM is given by:

\begin{equation}
    P(A_{ij} = 1 | z_i, z_j, \theta) = \theta_{z_i z_j}
\end{equation}

where \( z_i \) and \( z_j \) are the community labels of nodes \( i \) and \( j \), and \( \theta_{z_i z_j} \) is the probability of an edge between communities \( z_i \) and \( z_j \). SBM assumes that edges are independent given the community memberships of the nodes, and the matrix \( \theta \) captures the connectivity patterns between different communities.

\newthought{Degree-Corrected Stochastic Block Model}, one limitation of the basic SBM is that it does not account for variations in the degrees of nodes within the same community. In many real networks, nodes within the same community may have vastly different degrees. To address this issue, the Degree-Corrected Stochastic Block Model (DC-SBM) was introduced by Karrer and Newman\cite{karrer2011}, which incorporates node-specific degree parameters. The likelihood function for DC-SBM is:

\begin{equation}
    P(A_{ij} = 1 | z_i, z_j, \theta, \kappa_i, \kappa_j) = \frac{\kappa_i \kappa_j \theta_{z_i z_j}}{1 + \kappa_i \kappa_j \theta_{z_i z_j}}
\end{equation}

where \( \kappa_i \) and \( \kappa_j \) are the degree parameters for nodes \( i \) and \( j \). This modification allows DC-SBM to better fit networks with highly skewed degree distributions, which are common in social, biological, and technological networks.

\newthought{Bayesian Estimation} is often used in conjunction with SBM to estimate the posterior distribution of the community structure given the observed network. The goal is to compute the posterior distribution of the community assignments \( z \) given the adjacency matrix \( A \) of the network. Using Bayes’ theorem, the posterior is given by:

\begin{equation}
    P(z | A) = \frac{P(A | z) P(z)}{P(A)}
\end{equation}

Here, \( P(A | z) \) is the likelihood of observing the network given the community assignments, \( P(z) \) is the prior probability of the community assignments, and \( P(A) \) is the marginal likelihood, which ensures the posterior is properly normalized. This Bayesian framework allows us to incorporate prior information about the network and quantify uncertainty in the inferred community structure.

The inference procedure often involves maximizing the posterior distribution, or sampling from it using techniques like Markov Chain Monte Carlo (MCMC). When combined with the degree-corrected SBM, the Bayesian framework becomes even more powerful because it can naturally incorporate degree heterogeneity and provide more robust community detection results. The Bayesian formulation of SBM also supports model selection by allowing for the estimation of the most appropriate number of communities. By integrating over all possible configurations of the parameters and community assignments, one can compute the evidence (marginal likelihood) for different models, selecting the model with the highest evidence. This avoids the need for manually specifying the number of communities, which can be problematic in real-world networks where this information is not known a priori.

\begin{equation}
    P(A | \theta) = \sum_z P(A | z, \theta) P(z)
\end{equation}

In this equation, \( P(A | \theta) \) is the marginal likelihood, and \( P(z) \) is the prior over the community assignments. The Bayesian approach provides a principled way to balance model complexity and fit by integrating over the possible configurations of community assignments.

While SBM and its extensions provide a flexible and powerful framework for community detection, they are not without limitations. One challenge is that the basic SBM assumes that edges are independent, which may not always be true in real-world networks where there may be higher-order dependencies or non-random mixing patterns. Another challenge is computational: Bayesian inference for SBM typically requires sampling or optimization over a large space of possible community assignments, which can be computationally expensive for large networks.

Despite these challenges, SBM and Bayesian estimation remain one of the most robust approaches for community detection, especially when combined with degree correction and Bayesian model selection methods\cite{peixoto2023,peixoto2023descriptive}


\subsection{Link Clustering}

\begin{marginfigure}
\centering
\checkoddpage \ifoddpage \forcerectofloat \else \forceversofloat \fi
  \includegraphics[width=0.7\linewidth]{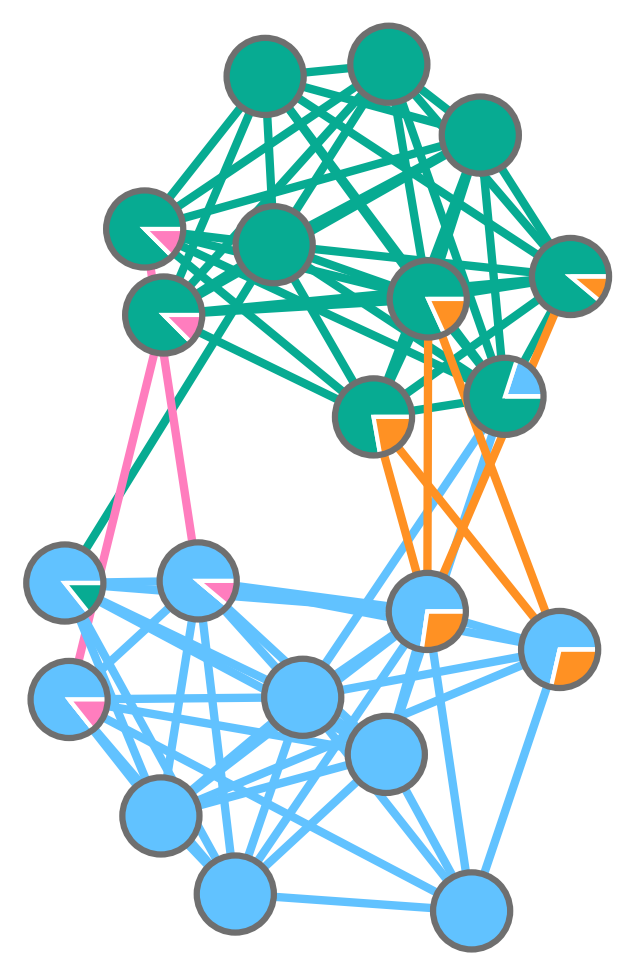}
  \caption[Link clustering example]{Network with edges colored according to the link communities identified using Link Clustering \citep{ahn2010link}. The nodes are represented as pie charts, showing the proportion of their edges belonging to each community.}
  \label{fig:link-clustering}
\end{marginfigure}

Link clustering is a method designed to handle overlapping community structures by clustering links rather than nodes. This approach is useful in networks where nodes can belong to multiple communities, which is common in social and biological networks. Ahn et al.\cite{ahn2010link} introduced this framework, which redefines the concept of community by focusing on edges. Instead of assuming that a community is a set of nodes with many links between them, link clustering treats communities as sets of highly interconnected links. This allows the detection of overlapping communities, as nodes naturally belong to multiple groups based on the relationships their links represent. The Figure \ref{fig:link-clustering} shows a 2-block stochastic block model network with link communities. 

In the link clustering, the similarity between two links sharing a common node is defined as the number of shared neighbors between the two connected nodes, normalized by the total number of neighbors. The similarity between links \( e_{ik} \) and \( e_{jk} \), which share a common node \( k \), is given by:

\begin{equation}
    S(e_{ik}, e_{jk}) = \frac{|\mathcal{N}(i) \cap \mathcal{N}(j)|}{|\mathcal{N}(i) \cup \mathcal{N}(j)|}
\end{equation}

where \( \mathcal{N}(i) \) and \( \mathcal{N}(j) \) denote the set of neighbors of nodes \( i \) and \( j \), respectively. This similarity measure reflects the extent to which two links are part of the same local structure, with higher similarity indicating that the links are likely to belong to the same community. By applying hierarchical clustering to the similarity matrix of links, a dendrogram of links is built. Cutting this dendrogram at an appropriate threshold yields the desired link communities.

To evaluate the quality of these communities, Ahn et al. introduced the concept of partition density \( D \), which measures how densely interconnected the links within a community are, relative to the total number of possible connections between nodes in the community. The partition density for a single community is given by:

\begin{equation}
    D_c = \frac{m_{in} - (n_c - 1)}{n_c (n_c - 1) / 2 - (n_c - 1)}
\end{equation}

where \( m_{in} \) is the number of internal links within the community, and \( n_c \) is the number of nodes in the community. The partition density \( D \) for the entire network is the average of the partition densities of all individual communities, weighted by the number of links in each community:

\begin{equation}
    D = \frac{2}{M} \sum_{c} m_c D_c
\end{equation}

where \( M \) is the total number of links in the network, \( m_c \) is the number of links in community \( c \), and \( D_c \) is the partition density of community \( c \).

One significant advantage of link clustering is that it naturally handles networks with overlapping communities. In highly overlapping networks, where each node may belong to several groups, link clustering captures these multiple affiliations by assigning a single membership to each link rather than each node. This method also reveals the hierarchical organization of communities, allowing for the detection of structures at different levels of resolution without suffering from the resolution limit that affects modularity-based approaches\footnote{The partition density avoids the resolution limit issue by evaluating the internal density of each community independently, thus allowing it to detect small, dense communities that are often merged in traditional methods \citep{ahn2010link}.}.

\newpage
\section{Adaptive Cut}
\label{paper1}
\begin{figure*}[h!]
\checkoddpage \ifoddpage \forcerectofloat \else \forceversofloat \fi
    \centering
    \includegraphics[width=\textwidth]{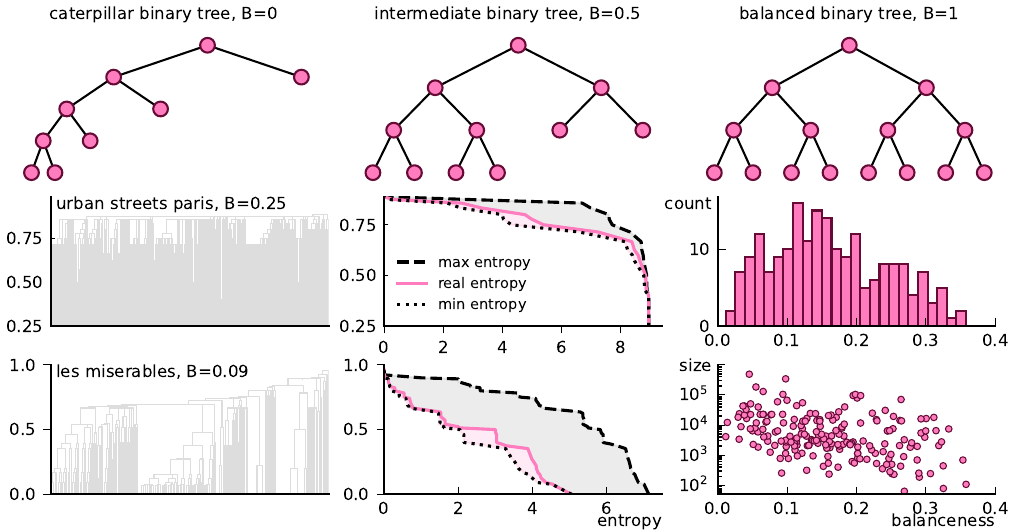}
    \caption[Balanceness measure]{\textbf{Explanation of the Balanceness Measure.}
\textbf{(a, b, c)} Illustrations of different tree structures: (a) an unbalanced caterpillar tree, (b) an intermediate tree, and (c) a balanced tree.
\textbf{(d)} Dendrogram representing the "Les Miserables" character network based on link similarities \citep{ahn2010link}. The dendrogram is unbalanced, as shown in (e).
\textbf{(e)} The progression of the real, maximal and minimal entropies (x axis) across different similarity levels (y axis). The three entropies are used to compute the balanceness metric (Eq. \ref{eq:balanceness}).
\textbf{(f)} The distribution of balanceness scores for 200 real networks \ref{SI:list_network}.
\textbf{(g)} A balanced dendrogram for the urban street network of Brasilia.
\textbf{(h)} The progression of the real, maximal and minimal entropies across different levels.
\textbf{(i)} A plot of the balanceness metric against network size (number of nodes), demonstrating that the balanceness score is independent of network size.}
    \label{fig:balanceness}
\end{figure*}

\addcontentsline{toc}{subsection}{\textit{Summary of Paper I}}
\section*{Summary of Paper I}

Hierarchical clustering and community detection are essential in various fields such as machine learning and network analysis. Traditional clustering methods often rely on cutting dendrograms at a single level, which may not capture the complex and hierarchical nature of the data, especially when dealing with unbalanced dendrograms. The Adaptive Cut method aims to overcome these limitations by performing multi-level cuts along the dendrogram, thus allowing for more accurate and refined partitions of the data.

The key innovation of the Adaptive Cut is to propose a multilevel cut by optimizing an objective function using a Markov Chain Monte Carlo (MCMC) with a simulated annealing scheme. Another contribution of the Adaptive Cut is the introduction of the Balanceness score, an information-theoretic metric that quantifies the balance of a dendrogram. The Balanceness score is calculated based on the entropy of the dendrogram structure at each level. The entropy at each level \( l \) is defined as:

\[
H(\pi_l) = - \sum_{i=1}^{|\pi_l|} p(B_i) \log_2(p(B_i)) \,,
\]

where \( p(B_i) = \frac{|B_i|}{n} \) is the probability of a branch \( B_i \) having an ancestor at that level. The Balanceness score \( B \) is then computed by comparing the entropies across all levels, using the following equation:

\begin{align}
B = \frac{1}{L}\sum_{i=1}^{L} \frac{H_{}(\pi_l)-H_{min}(\pi_l)}{H_{max}(\pi_l)-H_{min}(\pi_l)}\,.
\label{eq:balanceness}
\end{align}

This score, ranging from 0 (completely unbalanced) to 1 (perfectly balanced), allows the method to indicate the potential of the multi-level cut to improve the clustering compared to the single level cut alternative.

In Figure \ref{fig:balanceness} the balanceness score is examined for two real networks: the character network from \textit{Les Miserables} (unbalanced) and the street network of Paris (balanced), as shown in Figure \ref{fig:balanceness}d,g. To compute the balanceness, the maximum, minimum, and actual entropy values at each dendrogram level are calculated (Equation \ref{eq:balanceness}). The balanceness score represents the area between the minimum and maximum entropy curves, with a score of 1 indicating perfect balance and 0 indicating complete imbalance. Most real-world networks tend to have unbalanced dendrograms (Figure \ref{fig:balanceness}f), and the balanceness score is independent of network size (Figure \ref{fig:balanceness}i).

\begin{figure*}[h!]
\checkoddpage \ifoddpage \forcerectofloat \else \forceversofloat \fi
    \centering
    \includegraphics[width=\textwidth]{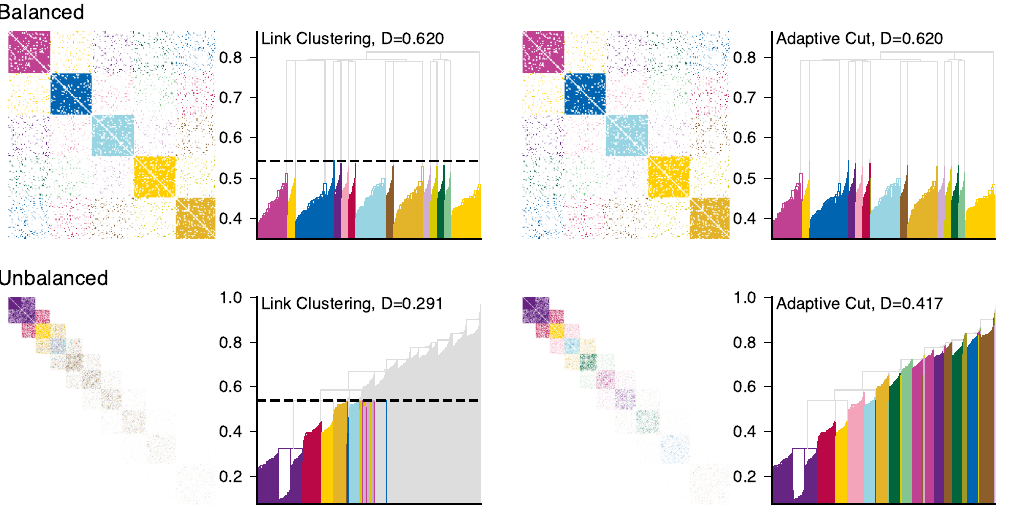}
    \caption[Adaptive cut]{\textbf{(a)} The adjacency matrix of a stochastic block model network, with nodes colored according to edge communities identified by the link clustering method \citep{ahn2010link}. \textbf{(b)} The corresponding dendrogram for the same network, with partitions or communities defined by the single-level link clustering cut \citep{ahn2010link}. The similarity level at which the cut is made is indicated by the dashed line. \textbf{(c, d)} Similar network to (a) and (b), but using an adaptive cut method instead of link clustering. \textbf{(e, f)} The same type of analysis applied to a stochastic block model with decreasing density}
    \label{fig:adaptive-cut}
\end{figure*}

The Markov Chain defined in the Adaptive Cut method allows for movement up or down the dendrogram, where clusters either merge (up) or split (down). Each move is accepted or rejected based on the objective function difference $\Delta f$. The method ensures ergodicity, meaning every partition in the dendrogram can be accessed, and detailed balance is maintained. The transition probabilities for moving up or down are adjusted accordingly so that the Markov chain follows detailed balance, via the \textit{Metropolis-Hastings} algorithm \citep{metropolis1953equation,hastings1970monte}.

To sample the Markov chain, the Metropolis-Hastings algorithm is applied, with the acceptance probability calculated based on the change in the objective function. If the objective function increases, the move is always accepted; otherwise, it is accepted with a probability proportional to $\exp(-|\Delta f|/T)$, where $T$ is a temperature parameter used to escape local optima. The cooling schedule follows a fast annealing to gradually reduce the temperature and avoid local maxima \citep{szu1987fast} .

The Markov chain starts at the partition obtained from a single-level cut, ensuring faster convergence by leveraging the dendrogram structure \citep{neal1993probabilistic}. By restricting the state space to the dendrogram, the Adaptive Cut converges more efficiently than general MCMC approaches in larger networks. Figure \ref{fig:adaptive-cut} illustrates the difference between traditional Link Clustering and the Adaptive Cut for a stochastic block model network.

To illustrate the limitations of single-level cuts, we focus on two toy networks with known community structures, a stochastic block model \citep{holland1983stochastic,karrer2011stochastic} and a varying density stochastic block model. In the stochastic block model (SBM), the probability of an edge between two nodes depends on whether the nodes belong to the same or different communities (see adjacency matrix in Fig. \ref{fig:adaptive-cut}a). The model is described as:
\begin{align}
P(A_{ij} = 1 \mid z_i = z_j) = \theta_{\text{intra}}, \quad P(A_{ij} = 1 \mid z_i \neq z_j) = \theta_{\text{inter}} \,,
\end{align}
where \( A_{ij} \) is the adjacency matrix entry, \( z_i \) and \( z_j \) are the community assignments, and \( \theta_{\text{intra}} \) and \( \theta_{\text{inter}} \) represent intra- and inter-community edge probabilities. In the SBM, both \( \theta_{\text{intra}} \) and \( \theta_{\text{inter}} \) are constant across all communities.

In the Varying Density Stochastic Block Model, we extend this by varying \( \theta_{\text{intra}} \) and \( \theta_{\text{inter}} \) across communities, allowing for internal structural differences and variable connectivity between communities (Fig. \ref{fig:adaptive-cut}e).

The stochastic block model (Fig. \ref{fig:adaptive-cut}a) demonstrates clear, identical-size communities that the single-level cut can distinguish effectively (Fig. \ref{fig:adaptive-cut}b), and the adaptive cut does not offer a significant improvement (Fig. \ref{fig:adaptive-cut}d). Both methods result in the same partition density \( D = 0.620 \).

However, in the Varying Density Stochastic Block Model (Fig. \ref{fig:adaptive-cut}e), the single-level cut struggles to partition communities with varying densities. The Adaptive Cut provides an improvement with a partition density of \( D = 0.417 \) compared to the single-level cut's \( D = 0.291 \) (Fig. \ref{fig:adaptive-cut}g,h). The dendrogram reflects this improved partitioning, where the communities align better with the adjacency matrix (Fig. \ref{fig:adaptive-cut}g). The balanceness scores for the stochastic block model (\( B = 0.6 \)) and the varying density model (\( B = 0.4 \)) further highlight the differences in dendrogram structure.

\begin{marginfigure}
\checkoddpage \ifoddpage \forcerectofloat \else \forceversofloat \fi
  \includegraphics[width=\linewidth]{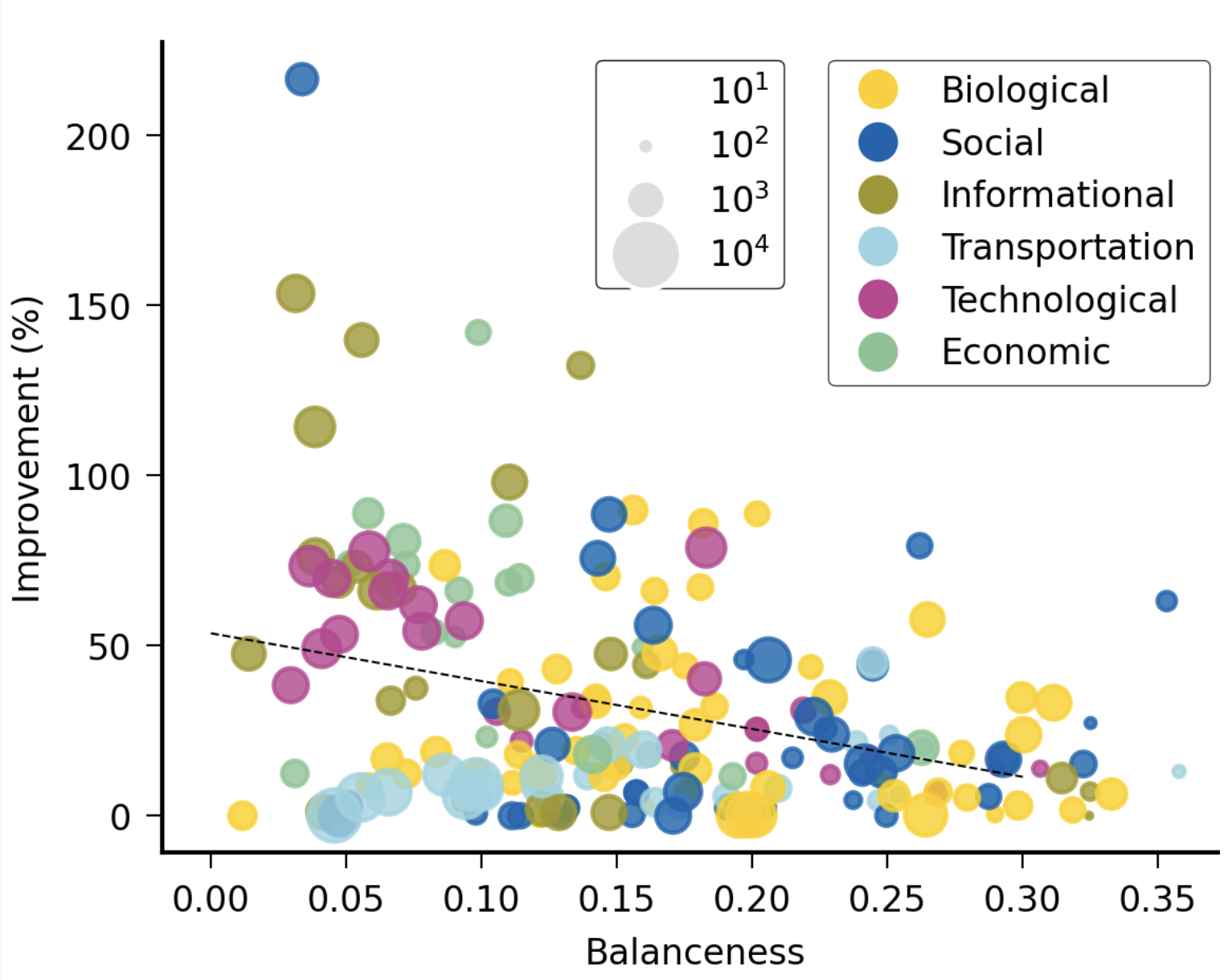}
  \caption[Adaptive Cut on 200 real networks]{Improvement (in \%) of the partition density between the single-level cut (Link Clustering) and the multi-level Adaptive Cut as a function of the dendrogram balanceness. The symbol colors indicate the domain of the network, and the size the number of nodes in the network, as show in legends. }
 \label{fig:200linkclustering}
\end{marginfigure}

We compared Link Clustering and Adaptive Cut across 200 real-world networks to evaluate the effect of balanceness on community detection. Results show that networks with lower balanceness scores gain more from the Adaptive Cut, improving partition density across different types of networks, including economic, transportation, and social networks (see Figure \ref{fig:200linkclustering}).

The Louvain method \citep{blondel2008fast}, commonly used for optimizing modularity, was also modified to include the Adaptive Cut. The standard Louvain method often stops merging communities when modularity stop increasing, but we construct a full dendrogram, allowing further merges even if modularity decreases. Our experiments show that the Adaptive Cut improves modularity, particularly in unbalanced networks.

\begin{figure}[h!]%
\checkoddpage \ifoddpage \forcerectofloat \else \forceversofloat \fi
  \includegraphics[width=\linewidth]{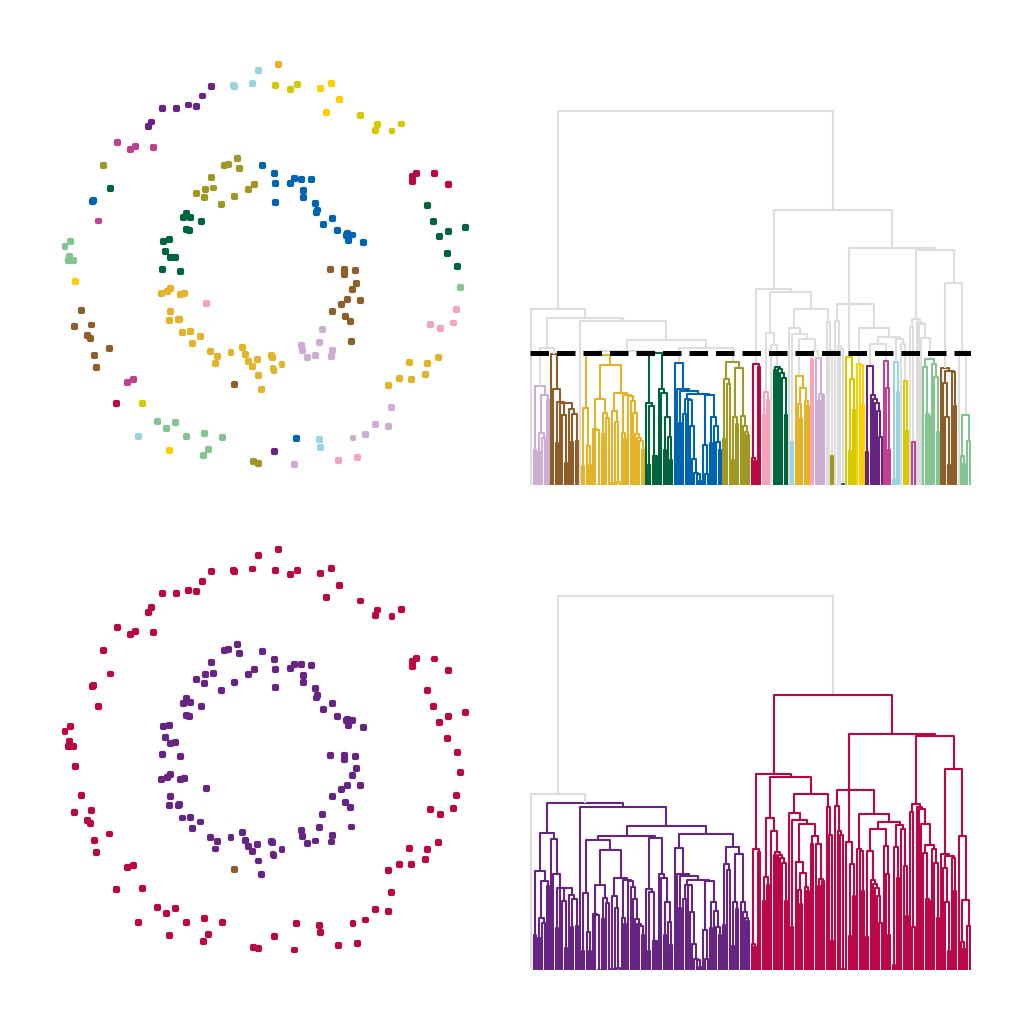}
  \caption[Circle cluster]{Comparison of clustering results on the two-circle dataset using a single-level cut and the Adaptive Cut. The Adaptive Cut effectively identifies the two concentric clusters, which the single-level cut fails to separate.}
 \label{fig:clustering_circle}
\end{figure}

The Adaptive Cut framework is versatile and can be applied to any clustering task that generates a dendrogram. This requires two components: a method to construct the dendrogram (e.g., single linkage \citep{johnson1967hierarchical}, average linkage \citep{sokal1958statistical}, or Ward’s method \citep{ward1963hierarchical}) and an objective function to optimize (e.g., the Davis-Bouldin index \citep{davies1979cluster}, silhouette score \citep{rousseeuw1987silhouettes}, or within-cluster sum of squares). 

To demonstrate its generality, we applied the Adaptive Cut to the two-circle dataset \citep{pedregosa2011scikit}, where traditional single-level cuts struggle to separate the clusters. The Adaptive Cut effectively identifies the two concentric circles thanks to its multi-level cuts (Figure \ref{fig:clustering_circle}).

\chapter{The Geometry of Complex Network}
\label{sec:chapter2}
As we saw in the first chapter, networks are ubiquitous in representing complex systems across various domains, including social sciences, biology, and technology. Understanding the underlying geometry of networks is crucial for tasks such as visualization, clustering, and link prediction. Network geometry refers to the embedding of network nodes into a geometric space such that the geometric relationships reflect the structural properties of the network. A key advancement in this area is graph representation learning, which leverages machine learning techniques to embed nodes into low-dimensional spaces while preserving important topological features. These embeddings facilitate downstream tasks like node classification, link prediction, and community detection, by capturing the complex relationships between nodes in a continuous space the network\cite{hamilton2017representation}\cite{ bronstein2017geometric}.

Traditional approaches to network embeddings often rely on Euclidean spaces, utilizing methods like matrix factorization \citep{belkin2003laplacian}, latent distance models \citep{hoff2002latent}, and random walk-based algorithms like node2vec \citep{grover2016node2vec}. These techniques aim to capture the similarity or proximity between nodes by representing them as points in a lower-dimensional space.

However, embeddings in an Euclidean geometry face limitations when representing hierarchical or tree-like data, where the underlying structure exhibits exponential growth. Hyperbolic space, characterized by constant negative curvature, naturally accommodates such hierarchies due to its exponential volume growth properties \citep{krioukov2010hyperbolic}. Embedding networks into hyperbolic space has shown promise in effectively capturing the hierarchical organization of data \citep{nickel2017poincare}.

In this context, hyperbolic embedding methods, such as Poincaré embeddings \citep{nickel2017poincare}, have been proposed to learn representations that reflect the latent hierarchical structure of networks. These methods enable more efficient representations of complex networks, particularly those with a tree-like structure.

\section{Network Embedding}

Building on the concepts of network embeddings and their evaluation criteria, we now explore methods for representing networks in lower-dimensional spaces.  This section discusses basic approaches such as matrix factorisation methods and random walk-based algorithms.

Matrix factorization methods, like the Laplacian Eigenmaps\cite{belkin2003laplacian}, involve decomposing the graph Laplacian to obtain embeddings that preserve local neighborhood information. These methods are effective in capturing the global structure of the network while reducing dimensionality.

Latent distance models\cite{hoff2002latent,nakisphd} assume that the probability of an edge between two nodes depends on the distance between them in an unobserved latent space. By modeling these distances probabilistically, latent distance models can capture complex relational patterns inherent in the network data.

Node2vec\cite{grover2016node2vec} leverages biased random walks to capture diverse network neighborhoods and applies the Skip-Gram model from natural language processing to learn continuous feature representations for nodes. This approach balances the exploration of homophily and structural equivalence in networks.

\subsection{Matrix Factorization Methods}

\begin{marginfigure}%
  \includegraphics[width=\linewidth]{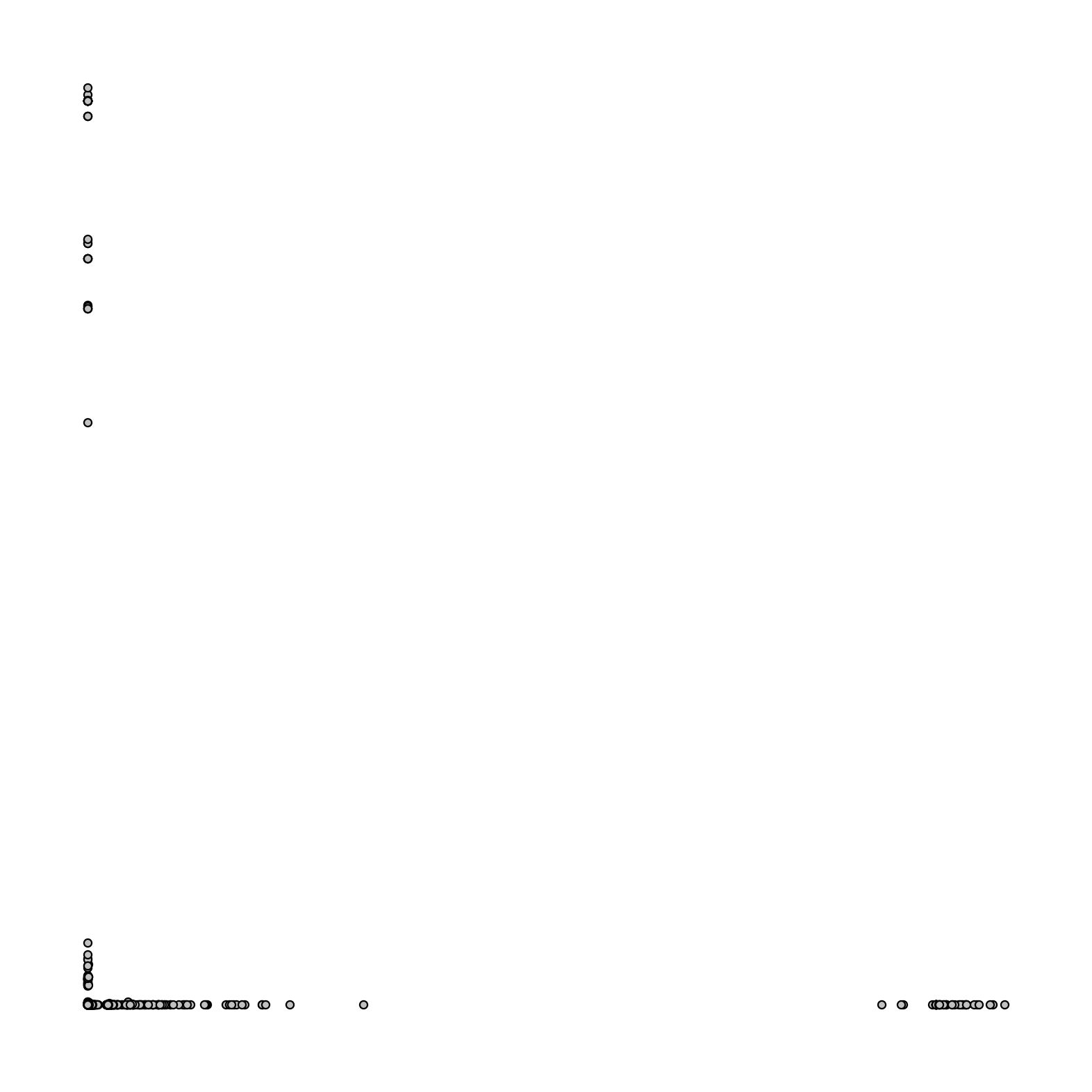}
  \caption[Matrix Factorization embedding]{Embedding of the GrQc collaboration network with Matrix Factorization \citep{leskovec2007graph}}
  \label{fig:Matrix Factorization}
\end{marginfigure}

Matrix factorization methods serve as foundational techniques in the field of network embeddings. These methods aim to represent the high-dimensional adjacency matrix of a graph in a lower-dimensional space while preserving the structural properties of the network.

\newthought{Adjacency Matrix Factorization} methods involve factorizing the adjacency matrix \( A \) of a graph into two lower-dimensional matrices \( U \) and \( V \) such that:

\[
A \approx UV^T
\]

In this formulation, \( A \) is an \( N \times N \) matrix representing the network with \( N \) nodes, where \( A_{ij} = 1 \) if there is an edge between node \( i \) and node \( j \), and \( 0 \) otherwise. The matrices \( U \) and \( V \) are \( N \times d \) matrices, with \( d \) being the target embedding dimension, typically much smaller than \( N \). The rows of \( U \) and \( V \) represent the embedded vectors for each node in the network\footnote{The choice of \( d \) is important and often determined empirically, balancing between dimensionality reduction and preservation of network properties.}.

This factorization captures the latent features of the network, where the dot product \( U_i^T V_j \) approximates the adjacency between nodes \( i \) and \( j \). However, this method is not without its limitations. For large networks, the factorization becomes computationally expensive, especially when \( N \) is large, leading to scalability issues. Moreover, the linear nature of this factorization might fail to capture complex, non-linear structures that are often present in real-world networks.

A notable extension of this method is Singular Value Decomposition (SVD), where the adjacency matrix \( A \) is decomposed into three matrices:

\[
A = U \Sigma V^T
\]

In this decomposition, \( U \) and \( V \) are orthogonal matrices, and \( \Sigma \) is a diagonal matrix containing the singular values. By truncating the singular values, we can obtain a lower-dimensional approximation of \( A \). This technique is closely related to Principal Component Analysis (PCA) and has been widely used in network science \citep{newman_networks_2010} (see Figure \ref{fig:Matrix Factorization}

\subsection{Laplacian Eigenmaps}

\begin{marginfigure}%
  \includegraphics[width=\linewidth]{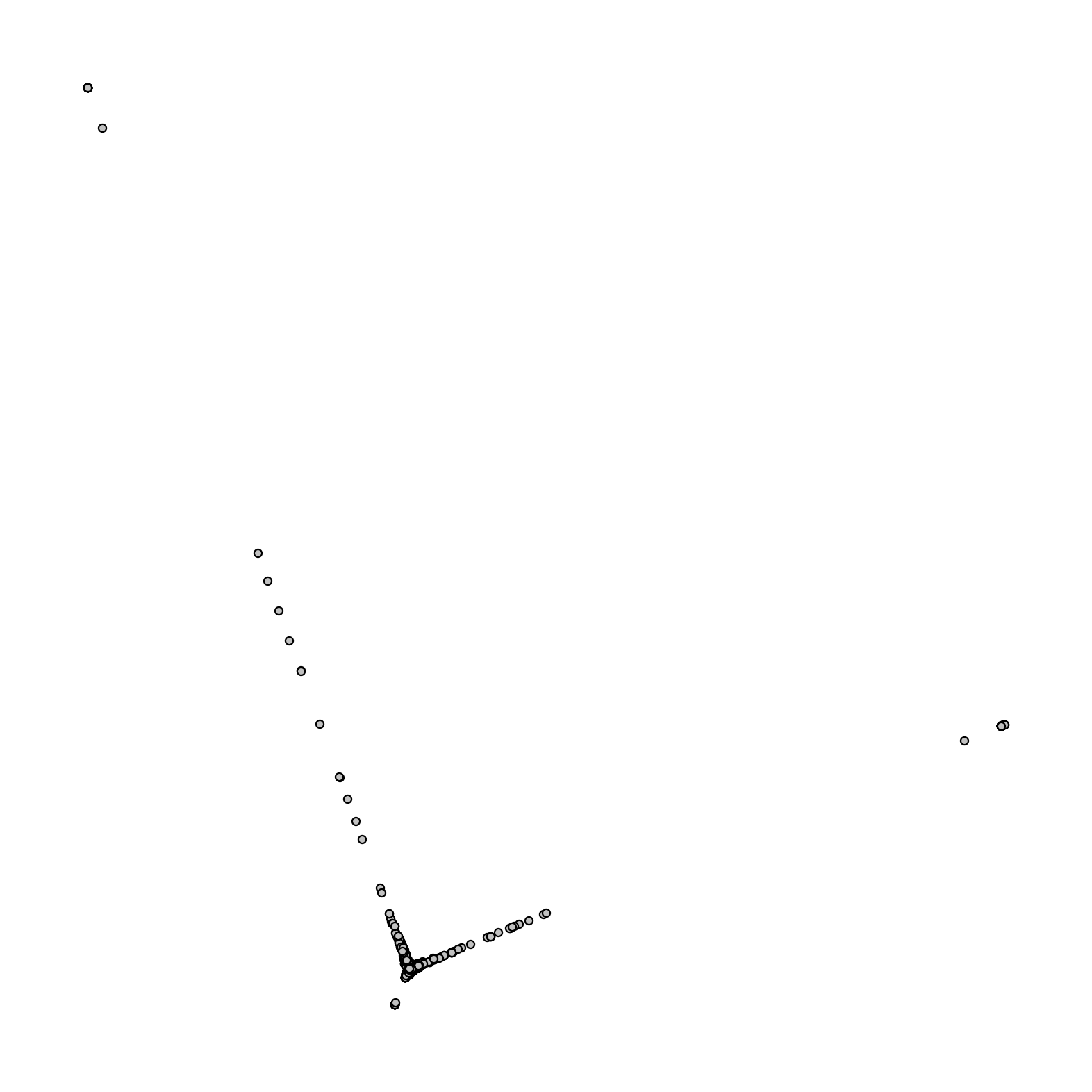}
  \caption[Laplacian eigenmap embedding]{Embedding of the GrQc collaboration network with Laplacian Eigenmaps}
  \label{fig:lapalcian}
\end{marginfigure}

Laplacian Eigenmaps, proposed by Belkin and Niyogi \citep{belkin_laplacian_2002}, is a matrix factorization technique based on spectral graph theory. The method aims to preserve the local neighborhood structure of the graph in the embedded space, effectively capturing the manifold's geometry in the data (see Figure \ref{fig:lapalcian}).

For a given graph \( G = (V, E) \), where \( V \) is the set of nodes and \( E \) is the set of edges, the graph Laplacian matrix \( L \) is defined as:

\[
L = D - A
\]

Here, \( A \) is the adjacency matrix, and \( D \) is the degree matrix, a diagonal matrix where \( D_{ii} = \sum_j A_{ij} \). The objective of Laplacian Eigenmaps is to find an embedding \( Y \) that minimizes the following cost function,

\[
\min_{Y} \sum_{i,j} (Y_i - Y_j)^2 A_{ij}
\]

This can be rewritten in matrix form as,

\[
\min_{Y} \,\, \text{Tr}(Y^T L Y)
\]

subject to the constraint \( Y^T D Y = I \), where \( I \) is the identity matrix. The solution to this optimization problem is given by computing the eigenvectors of the Laplacian matrix \( L \). The smallest non-zero eigenvalues correspond to the most informative dimensions for the embedding\footnote{The number of eigenvectors used in the final embedding is a hyperparameter that can be tuned based on the specific application and desired dimensionality.}.

The resulting eigenvectors provide the coordinates of the nodes in the lower-dimensional space.

\subsection*{Advantages and Limitations of Matrix Factorization Methods}

Matrix factorization techniques, such as Adjacency Matrix Factorization and Laplacian Eigenmaps, come with several notable advantages. Their simplicity and ease of implementation make them accessible for a wide range of applications. Moreover, these methods are highly interpretable, which is particularly beneficial in fields like social network analysis and bioinformatics, where understanding the structure of data embeddings is crucial\footnote{The interpretability of these methods enhances their applicability in domains where understanding the process is just as important as the results.}.

Despite these advantages, matrix factorization methods also have some significant limitations. One major drawback is their computational inefficiency when dealing with large-scale networks, due to the intensive matrix operations and eigenvalue decomposition required. This can create challenges in scaling up to larger datasets. Additionally, these techniques inherently assume that the network structure can be represented through linear transformations. However, this assumption often fails in complex real-world networks that exhibit non-linear relationships, limiting the methods' effectiveness\footnote{Recent developments, such as deep learning-based graph embedding techniques, address these limitations by capturing non-linearities more effectively (see Section \ref{sec:gnn})}.

\subsection{Node2Vec Algorithm}

To understand the \texttt{node2vec} algorithm, it is helpful to first consider the principles behind Word2Vec, a model originally developed for natural language processing. Word2Vec, introduced by Mikolov et al.\cite{mikolov2013distributed}, is a neural network-based model that learns vector representations of words from large text corpora. The primary goal is to capture the semantic relationships between words by placing similar words closer together in the embedding space.

A key technique within Word2Vec is the skip-gram model, which predicts context words surrounding a target word within a given window in the text. Formally, the skip-gram model seeks to maximize the likelihood of predicting the context of a word given its embedding. This approach effectively captures both syntactic and semantic patterns in language data. The skip-gram objective is mathematically defined as:

\[
\max_{\theta} \sum_{u \in V} \sum_{v \in N(u)} \log P(v \mid u; \theta)
\]

where \( N(u) \) represents the set of context words for the word \( u \), and \( P(v \mid u; \theta) \) denotes the probability of word \( v \) appearing in the context of word \( u \), given their embeddings parameterized by \( \theta \). This probability is typically modeled using a softmax function:

\[
P(v \mid u; \theta) = \frac{\exp(\mathbf{z}_v \cdot \mathbf{z}_u)}{\sum_{v' \in V} \exp(\mathbf{z}_{v'} \cdot \mathbf{z}_u)}
\]

Here, \( \mathbf{z}_u \) and \( \mathbf{z}_v \) are the vector representations of the words \( u \) and \( v \), respectively. The Word2Vec model, particularly the skip-gram method, laid the groundwork for embedding techniques that could be adapted to various domains, including network analysis. We will come back to Word2Vec model for studying human mobility in Chapter 3.

\begin{marginfigure}%
  \includegraphics[width=\linewidth]{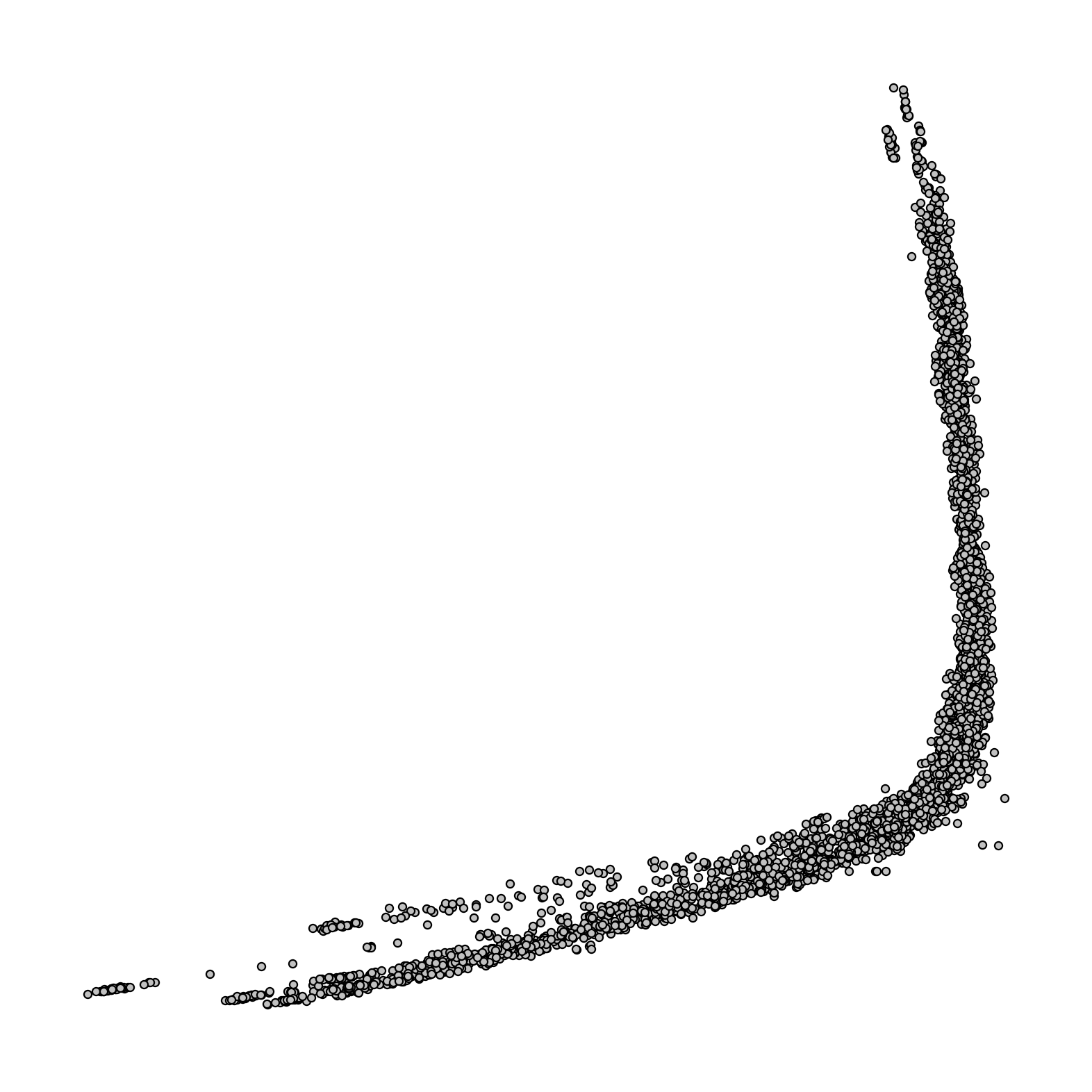}
  \caption[Node2vec embedding]{Embedding of the GrQc collaboration network with node2vec \citep{leskovec2007graph}}
  \label{fig:n2v}
\end{marginfigure}

Building on the principles of Word2Vec, the \texttt{node2vec} algorithm, introduced by Grover and Leskovec\cite{grover2016node2vec}, represents a significant advancement in random walk-based embedding techniques for graphs. \texttt{node2vec} extends the basic random walk approach by introducing a biased random walk strategy that generates node sequences through a mixture of breadth-first search (BFS) and depth-first search (DFS) behaviors. This innovative approach allows \texttt{node2vec} to capture both local and global structural features of the graph effectively (see Figure \ref{fig:n2v}).

The key innovation in \texttt{node2vec} lies in the introduction of two parameters: the return parameter \( p \) and the in-out parameter \( q \). These parameters offer fine-grained control over the behavior of the random walks. The return parameter \( p \) governs the likelihood of revisiting a node during the walk. A high value of \( p \) discourages the walk from returning to the previous node, encouraging exploration of new nodes, akin to depth-first search, which captures more global structural information in the graph. Conversely, the in-out parameter \( q \) determines whether the walk prefers exploring nodes closer to or farther from the current node. A high value of \( q \) biases the walk towards exploring further nodes, behaving more like breadth-first search and capturing local structural equivalences and community structures. Unlike \texttt{DeepWalk}, which uses uniform random walks, \texttt{node2vec} introduces this flexibility in walk dynamics, enabling it to capture a richer set of structural features within the network\footnote{\texttt{DeepWalk} employs uniform random walks and primarily focuses on capturing local neighborhoods through a Skip-Gram model \citep{perozzi2014deepwalk}.}.

The transition probability from node \( v \) to node \( u \), given the previous node in the walk \( t \), is defined in \texttt{node2vec} as:

\[
P(v_{i+1} = u \mid v_i = v, v_{i-1} = t) = \begin{cases}
\frac{1}{p} & \text{if } d_{tu} = 0, \\
1 & \text{if } d_{tu} = 1, \\
\frac{1}{q} & \text{otherwise},
\end{cases}
\]

where \( d_{tu} \) represents the shortest path distance between nodes \( t \) and \( u \). By adjusting the parameters \( p \) and \( q \), \texttt{node2vec} can interpolate between BFS-like and DFS-like behaviors, providing a flexible mechanism to generate walks that capture the desired structural features.

\begin{marginfigure}%
  \includegraphics[width=\linewidth]{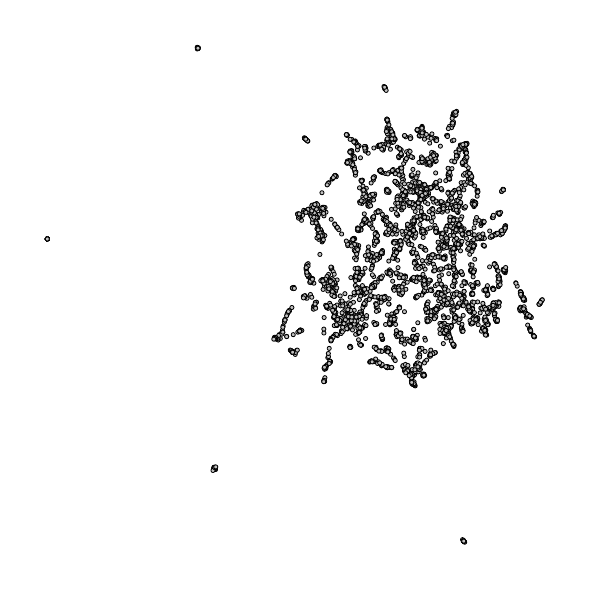}
  \caption[Matrix Factorization embedding]{Embedding of the GrQc collaboration network with node2vec in higher dimension and then reduce with UMAP \citep{leskovec2007graph}.}
  \label{fig:marginfig}
\end{marginfigure}

Once the sequences of nodes are generated, \texttt{node2vec} employs the skip-gram model to learn the embeddings. The objective of this step is to maximize the likelihood of predicting a node's neighbors given its embedding. Formally, the skip-gram objective is defined as,

\[
\max_{\theta} \sum_{u \in V} \sum_{v \in N(u)} \log P(v \mid u; \theta)
\]

Here, \( N(u) \) denotes the neighborhood of node \( u \), and \( P(v \mid u; \theta) \) represents the probability of node \( v \) being a neighbor of \( u \) given their embeddings, parameterized by \( \theta \)\footnote{The optimization of \( \theta \) is typically performed using stochastic gradient descent or its variants. This approach allows for efficient training on large-scale networks, as it updates the model parameters incrementally based on small batches of data, rather than processing the entire graph at once.}. This probability is typically modeled using a softmax function,
\begin{align}
P(v \mid u; \theta) = \frac{\exp(\mathbf{z}_v \cdot \mathbf{z}_u)}{\sum_{v' \in V} \exp(\mathbf{z}_{v'} \cdot \mathbf{z}_u)}
\end{align}

where \( \mathbf{z}_u \) and \( \mathbf{z}_v \) are
 the embeddings of nodes \( u \) and \( v \). The adaptability of \texttt{node2vec}, enabled by its tunable parameters \( p \) and \( q \), often leads to superior performance in tasks like node classification, link prediction, and community detection\footnote{The flexibility provided by \( p \) and \( q \) allows the algorithm to be tailored to the characteristics of a specific network. For example, in social networks where local community structures are significant, setting a lower \( q \) value can be beneficial. On the other hand, in citation networks where capturing long-range dependencies is important, a higher \( p \) value might be more suitable. This adaptability makes \texttt{node2vec} versatile across different types of networks and analytical tasks.}. The ability to balance local and global exploration in \texttt{node2vec} results in a more refined approach to learning graph representations, making it a powerful tool in network analysis and representation learning.

Many graph embeddings methods, like \texttt{node2vec}, are biased because of degree heterogeneity, as random walks favors high-degree nodes. This bias leads to the over-representation of such nodes in the embedding space. The \texttt{residual2vec}\cite{kojaku2021residual2vec} model remove the bias by modeling a baseline transition probability, \(P_0(j|i)\), from a null model, removing structural biases while preserving node properties like degree. The debiased transition probability in \texttt{residual2vec} is given as:

\[
P_{\text{r2v}}(v \mid u) = P_0(v \mid u) \cdot \frac{\exp(\mathbf{z}_v \cdot \mathbf{z}_u)}{Z_u'}
\]

where \( P_0(v \mid u) \) is derived from a degree-corrected stochastic block model (dcSBM) (see section \ref{sec:dcSBM}) or other random graph models, and \(Z_u\) is a normalization term. This approach ensures that the embedding focuses on residual information—i.e., structural properties not captured by the null model—leading to improved embeddings in tasks like link prediction and community detection, effectively debiasing the \texttt{node2vec} framework.

\subsection{Latent Distance Models}

Latent Space Models (LSMs) for the representation of graphs have been well established over the past years \citep{hoff2002latent,newman2003structure,}. LSMs utilize the generalized linear model framework to obtain informative latent node embeddings while preserving network characteristics. The choice of latent effects in modeling the link probabilities between the nodes leads to different expressive capabilities for characterizing the network structure. Popular choices include the Latent Distance Model \citep{hoff2002latent}, which defines a probability of an edge based on the Euclidean distance of the latent embeddings, and the Latent Eigen-Model \citep{hoff2005bilinear}, which generalizes stochastic blockmodels as well as distance models. Various non-Euclidean geometries of LSMs have also been studied \citep{nickel2017poincare}, with the hyperbolic case being of particular interest \citep{nickel2018learning}.

Here, we will focus on the Latent Distance Model (LDM) \citep{hoff2002latent,nakis2022hm}, where network nodes are positioned close in the latent space if they are connected or share additional similarities, such as high-order dependency or proximity. LDMs embed network nodes into a latent space where the proximity of nodes reflects the likelihood of edges between them. The central idea is that similar or connected nodes are positioned closer together, naturally capturing key properties like homophily\footnote{Homophily refers to the tendency of nodes with similar attributes to connect, which is a common phenomenon in social networks.} and transitivity\footnote{Transitivity indicates the likelihood that two nodes connected to a common third node are also connected. This can be explained via the triangle inequality in Euclidean space.}. Each node is represented by a latent variable \( z_i \in \mathbb{R}^D \) in a \( D \)-dimensional latent space, where the latent variables are sufficient to describe and explain the relationships between nodes. The probability of an edge between two nodes is modeled based on the Euclidean distance between their latent positions.

Formally, the probability of the network \( Y \), given the latent positions \( Z \) and additional parameters \( \theta \), is:

\[
P(Y | Z, \theta) = \prod_{i<j} p(y_{ij} | z_i, z_j, \theta),
\]

where \( p(y_{ij} | z_i, z_j, \theta) \) is the probability of an edge between nodes \( i \) and \( j \), parameterized by the latent positions and other potential covariates \( \theta \)\footnote{The parameter \( \theta \) can include other network-level factors such as node-specific covariates or regressors.}.

For binary networks, a logistic model is often used to parameterize the edge probabilities, but in this study, we focus on a Poisson LDM\footnote{The Poisson LDM is useful for modeling networks with weighted or count data, as it generalizes binary networks to allow for integer-weighted edges \citep{nakis2022hm}.}. The Poisson LDM models the rate of an edge as a function of the Euclidean distance between the latent positions,

\[
\lambda_{ij} = \exp(\gamma_i + \gamma_j - d(z_i, z_j)),
\]

where \( \gamma_i \in \mathbb{R} \) represents a node-specific random effect, and \( d(z_i, z_j) = \|z_i - z_j\|_2^2 \) is the Euclidean distance between nodes \( i \) and \( j \). This ensures that nodes closer in the latent space have higher probabilities of being connected, effectively capturing homophily.

The total log-likelihood of the Poisson LDM is then given by,

\[
\log P(Y | \Lambda) = \sum_{i < j} \left[ y_{ij} \log \lambda_{ij} - \lambda_{ij} - \log(y_{ij}!) \right],
\]

where \( \Lambda = \{\lambda_{ij}\} \) is the matrix of Poisson rates for all node pairs. The first term \( y_{ij} \log \lambda_{ij} \) forces similar nodes to be placed closer together, as it is maximized when \( d(z_i, z_j) \) is minimized. The second term, \( \lambda_{ij} \), acts as a repelling force, pushing dissimilar nodes further apart, and is maximized when \( d(z_i, z_j) \to \infty \) for unconnected nodes.

A critical property of LDMs is their ability to model transitivity through the triangle inequality in Euclidean space. Specifically, for nodes \( i \), \( j \), and \( k \), the distances between them satisfy,

\[
d(z_i, z_j) \leq d(z_i, z_k) + d(z_k, z_j),
\]

which ensures that if two nodes share a similar relationship with a third node, they will be positioned closely in the latent space, reinforcing the network’s transitivity structure\footnote{This property is crucial in real-world networks, such as social or collaboration networks, where indirect connections often suggest a higher likelihood of direct connections.}.

The Latent Distance Model is not scalable. The model's computational complexity scales quadratically, \( O(N^2) \), with the number of nodes \( N \), making it difficult to apply to large networks. Furthermore, LDMs do not explicitly account for hierarchical structures within the network, limiting their ability to capture multiscale relationships.

To overcome these limitations, the Hierarchical Block Distance Model (HBDM) introduces a hierarchical structure to the LDM \cite{nakis2023hierarchical}. This model groups nodes into clusters using divisive clustering based on their latent positions, allowing for the identification of hierarchical structures at different scales. The HBDM reduces the computational complexity to \( O(N \log N) \), making it suitable for large-scale network analysis. The log-likelihood of the HBDM is defined as,

\begin{align}
\log P(Y | Z, \gamma) = & \sum_{l=1}^{L} \sum_{k=1}^{K_l} \sum_{i \in C_k^{(l)}} \sum_{j \in C_{k'}^{(l)}} \exp(\gamma_i + \gamma_j - \|z_i-z_j\|^2) \\
& - \sum_{k=1}^{K_l} \sum_{k' > k} \exp(-\|\mu_k^{(l)} - \mu_{k'}^{(l)}\|^2),
\end{align}

where \( \mu_k^{(l)} \) represents the centroid of cluster \( k \) at level \( l \), and \( \gamma \) denotes random effects. This hierarchical extension enables the model to uncover both local and global patterns in the network while maintaining efficiency\footnote{The HBDM introduces the concept of "cluster-homophily" and "cluster-transitivity," where closely related nodes are grouped into clusters, and these clusters are positioned according to their relationships in the latent space.}.

While the classical LDM provides an effective framework for modeling network homophily and transitivity, its hierarchical extensions like the HBDM offer a scalable solution for large and complex networks, preserving key properties while uncovering deeper hierarchical structures.

\newpage
\subsection*{Summary of Paper \citep{nakis2023characterizing}}

\begin{marginfigure}%
    \centering
  \includegraphics[width=0.8\linewidth]{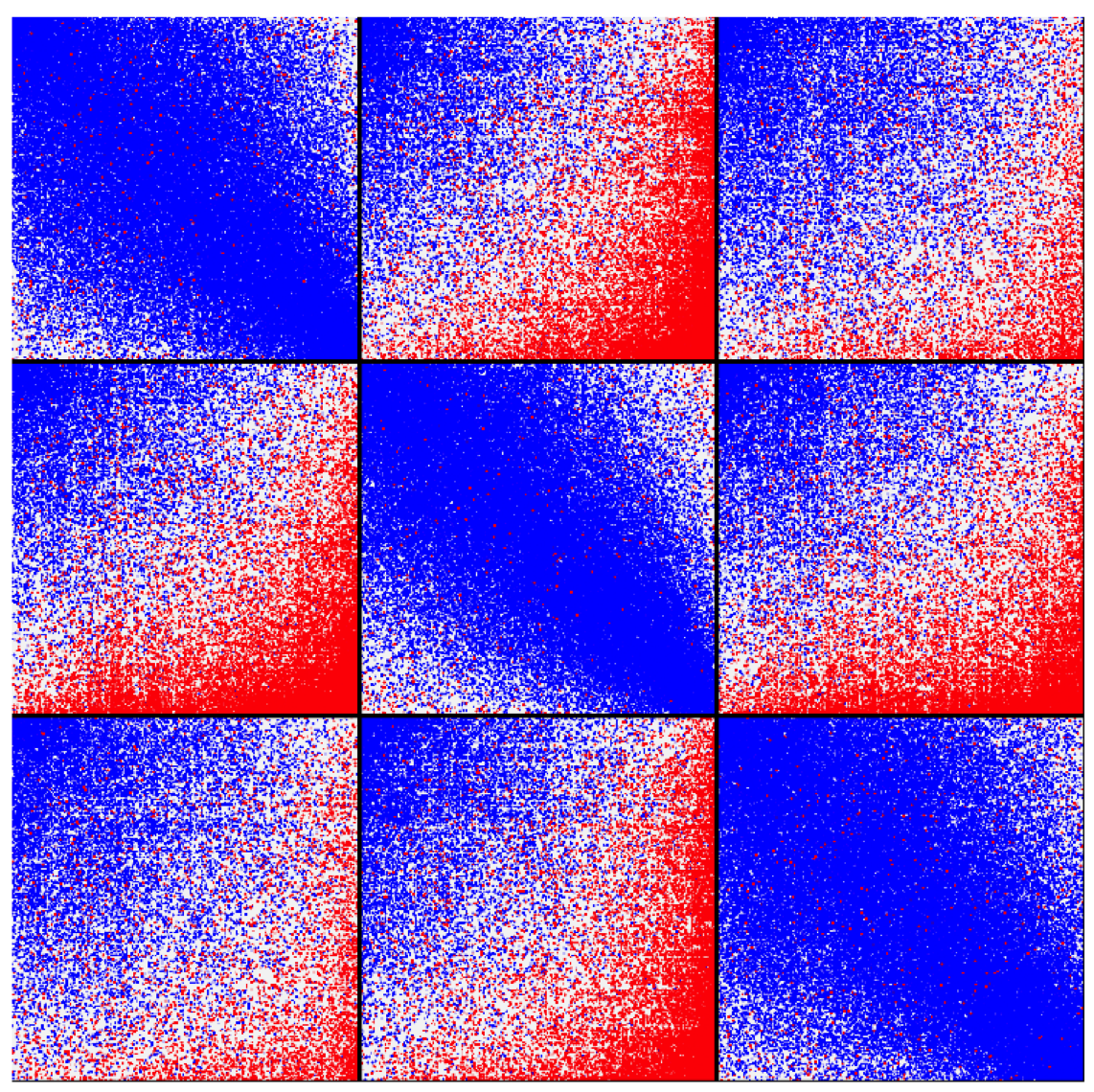}
  \caption[Adjacency matrix of an artificially generated polarized network]{Adjacency matrix of an artificially generated networks with a levels of polarization ($z_i \sim Dir(1))$. Th network is of size $N = 5000$ nodes and $K = 3$ archetypes. The network adjacency matrices is ordered based on $z_i$, in terms of maximum archetype membership and internally according to the magnitude of the corresponding archetype most used for their reconstruction. \citep{nakis2023characterizing}}
  \label{fig:polorized_network}
\end{marginfigure}

A major current concern in social networks is the emergence of polarization and filter bubbles promoting a mindset of "us-versus-them" that may be defined by extreme positions believed to ultimately lead to political violence and the erosion of democracy. Such polarized networks are typically characterized in terms of signed links reflecting likes and dislikes. In this paper, we propose the latent Signed relational Latent dIstance Model (SLIM) utilizing for the first time the Skellam distribution as a likelihood function for signed networks and extend the modeling to the characterization of distinct extreme positions by constraining the embedding space to polytopes. On four real social signed networks of polarization, we demonstrate that the model extracts low-dimensional characterizations that well predict friendships and animosity while providing interpretable visualizations defined by extreme positions when endowing the model with an embedding space restricted to polytopes \citep{nakis2023characterizing}.

\subsection*{Summary of Paper \citep{nakis2024time}}

\begin{marginfigure}
  \includegraphics[width=\linewidth]{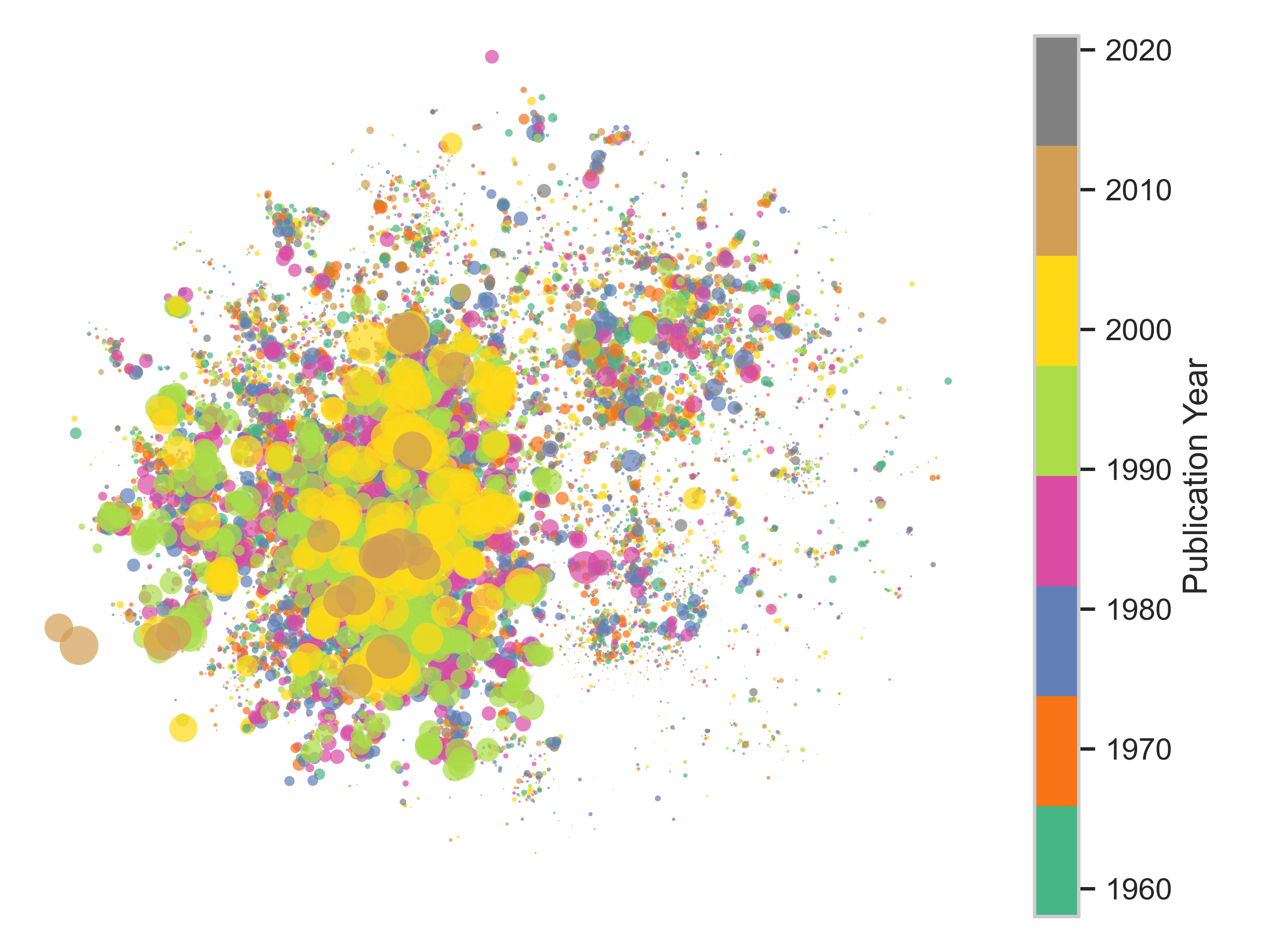}
  \caption[Citation Network]{Machine Learning: DISEE 2-dimensional embedding space Log-Normal yearly evolution. Node sizes are based on each paper’s mass, $fi(t)exp(\alpha i)$. Nodes are color-coded based on their publication year. \citep{nakis2024time}}
  \label{fig:disee}
\end{marginfigure}

Understanding the structure and dynamics of citation network has become an important area of research in order to address imminent questions including how scholars interact to advance science, how disciplines are related and evolve, and how research impact can be quantified and predicted. In this paper, we study two  modeling methodologies to assess the citation impact dynamics of papers, using parametric distributions, and embedding the citation networks in a latent space optimal for characterizing the static relations between papers in terms of their citations. Citation networks are a prominent example of single-event dynamic networks, i.e., networks for which each dyad only has a single event (i.e., the point in time of citation). We presently propose a novel likelihood function for the characterization of such single-event networks. Using this likelihood, we propose the Dynamic Impact Single-Event Embedding model (DISEE). The DISEE model characterizes the scientific interactions in terms of a latent distance model in which random effects account for citation heterogeneity while the time-varying impact is characterized using existing parametric representations for assessment of dynamic impact. We highlight the proposed approach on several real citation networks finding that the DISEE well reconciles static latent distance network embedding approaches with classical dynamic impact assessments \citep{nakis2024time}.

\subsection{Advanced Machine Learning Techniques}
\subsection{Advanced Machine Learning Techniques}
\label{sec:gnn}
As networks grow in complexity and scale, traditional embedding approaches often fall short in capturing their intricate patterns and structures. To address these challenges, advanced machine learning techniques, particularly those based on deep learning, have been developed. This section focuses on two leading approaches: Graph Neural Networks (GNNs) and Variational Graph Autoencoders (VGAEs), emphasizing their advantages over shallow methods. Traditional approaches like matrix factorization, DeepWalk, and node2vec rely on linear or shallow non-linear transformations, often limited by fixed features or structural properties\footnote{Shallow methods may struggle to effectively capture deep hierarchical patterns within complex networks \citep{perozzi2014deepwalk}.}. In contrast, GNNs and VGAEs utilize deep architectures, enabling the capture of complex hierarchical representations through multi-layer transformations and non-linearities.

\newthought{Graph Neural Networks} extend traditional embedding approaches by operating directly on graph structures, facilitating message passing between neighboring nodes\footnote{For a comprehensive overview of GNN architectures and their applications, refer to \citep{wu2020comprehensive} and \citep{hamilton2020graph}.}. This message-passing mechanism allows GNNs to incorporate both local and global graph structures by iteratively aggregating information from a node’s neighborhood. The general update rule for a GNN is formulated as:

\[
\mathbf{h}_v^{(k+1)} = \sigma \left( \mathbf{W}^{(k)} \cdot \text{AGGREGATE} \left( \left\{ \mathbf{h}_u^{(k)} : u \in \mathcal{N}(v) \right\} \right) \right),
\]

where \( \mathbf{h}_v^{(k)} \) denotes the feature vector of node \( v \) at layer \( k \), \( \mathcal{N}(v) \) is the set of its neighbors, \( \mathbf{W}^{(k)} \) is a learnable weight matrix, and \( \sigma \) is a non-linear activation function.

\newthought{Variational Graph Autoencoders} (VGAEs) introduce a probabilistic framework to the embedding process, combining the strengths of autoencoders with the ability to model uncertainty within graph data \citep{kipf2016variational}. A VGAE comprises an encoder, which maps nodes into a latent space, providing distributions over potential embeddings, and a decoder, which reconstructs the graph by predicting edge probabilities based on these embeddings. The loss function for a VGAE is defined as:

\begin{align}
    \mathcal{L} = \mathbb{E}_{q(\mathbf{Z}|\mathbf{X},\mathbf{A})} \left[ \log p(\mathbf{A}|\mathbf{Z}) \right] - \text{KL}\left( q(\mathbf{Z}|\mathbf{X},\mathbf{A}) \| p(\mathbf{Z}) \right),
\end{align}

where \( q(\mathbf{Z}|\mathbf{X},\mathbf{A}) \) denotes the approximate posterior, \( p(\mathbf{A}|\mathbf{Z}) \) represents the likelihood of the data given the latent variables, and \( \text{KL} \) is the Kullback-Leibler divergence\footnote{The Kullback-Leibler (KL) divergence between two distributions \( q \) and \( p \) is given by:

\[
\text{KL}(q(\mathbf{Z}) \parallel p(\mathbf{Z})) = \int q(\mathbf{Z}) \log \frac{q(\mathbf{Z})}{p(\mathbf{Z})} \, d\mathbf{Z}
\]

It quantifies the divergence between the approximate posterior \( q(\mathbf{Z}|\mathbf{X},\mathbf{A}) \) and the true posterior \( p(\mathbf{Z}|\mathbf{A}) \).}.
These advanced techniques offer notable improvements over shallow methods, including deep hierarchical representation learning, capturing complex non-linear interactions, and end-to-end training tailored to specific tasks. Additionally, they efficiently process dynamic data without retraining and integrate diverse node and edge features, making them well-suited for analyzing complex, multimodal networks.

In summary, GNNs and VGAEs represent a substantial step forward in network embedding, providing versatile tools for representing and analyzing complex network structures due to their ability to capture non-linear relationships and adapt to evolving data.

\section{Example: Co-habitation Network}

We now turn our attention to the co-habitation network in Denmark\footnote{This section is still a work in progress.}. This dataset covers every residential moves in Denmark since 1986, approximately 40 millions residential moves. It can be represented as a temporal bipartite network between individuals and the house they reside in. For our analysis, however, we treat it as a static network of addresses, the nodes are addresses and the edges weighted by the number of moves between two houses. The network is composed of 3.3 millions nodes/addresses.

\section{Non-Euclidean Space Hyperbolic Embedding}
\label{sec:hyperbolic}
\begin{marginfigure}%
  \includegraphics[width=\linewidth]{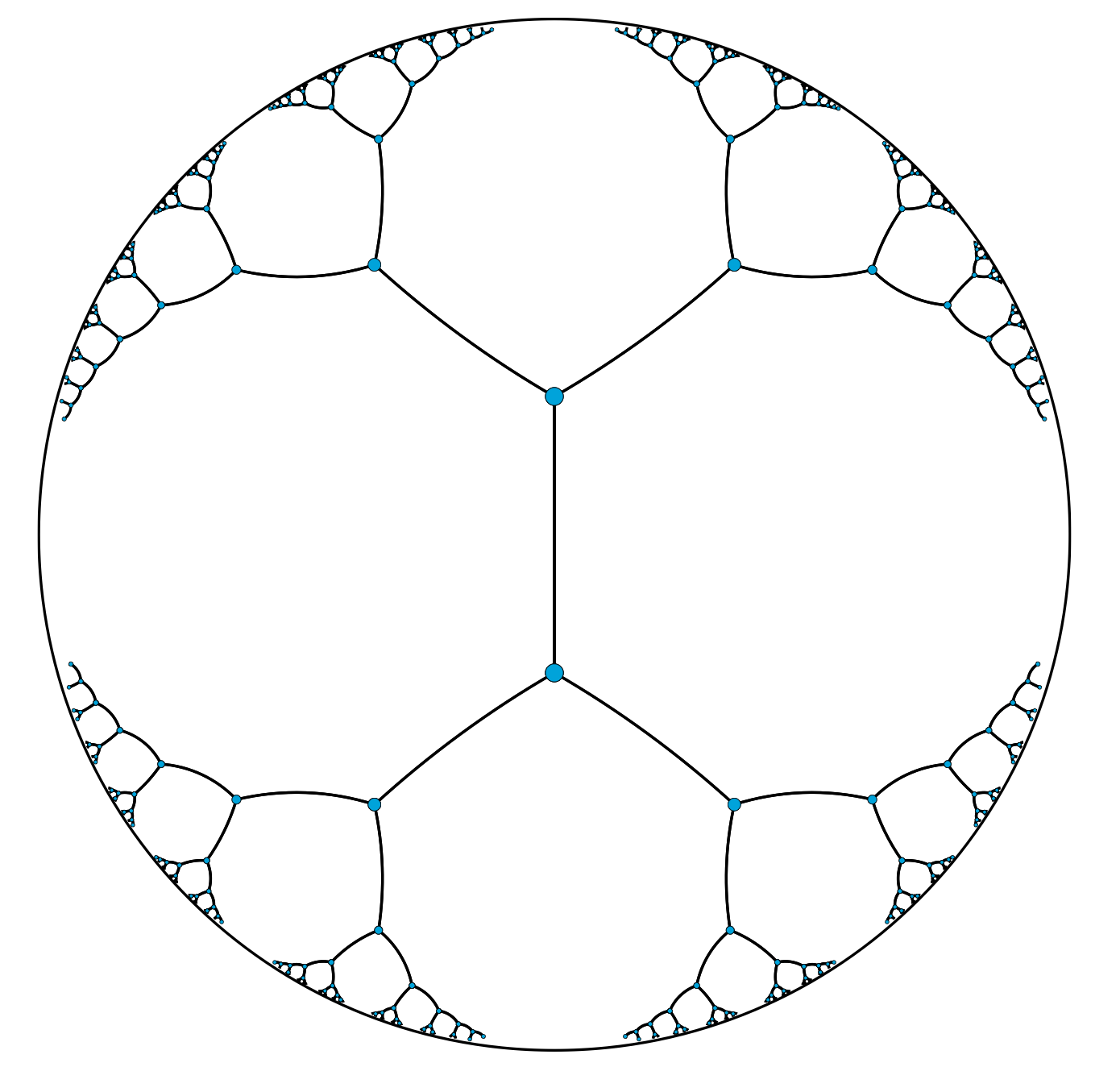}
  \caption[Poincaré disk with a tree]{Embedding of a regular tree in \( \mathbb{B}^2 \) such that all connected nodes are spaced equally far apart (i.e., all black line segments have identical hyperbolic length) \citep{nickel2017poincare}.}
  \label{fig:poincare_disk}
\end{marginfigure}

Until this point, all discussed embedding techniques have operated within Euclidean space. While Euclidean embeddings can efficiently capture certain structures, they often face challenges when representing hierarchical systems, particularly those with exponential growth patterns, like trees. As an example, drawing a tree on paper often results in overlapping branches after a few levels, illustrating the limitations of Euclidean space for such tasks. In contrast, hyperbolic space is well-suited for capturing hierarchies due to its exponential volume growth. In Euclidean space, the volume of a ball grows polynomially with its radius, but in hyperbolic space, the volume grows exponentially, allowing for the efficient representation of hierarchical structures with minimal distortion, as shown in Figure \ref{fig:poincare_disk}.

\begin{marginfigure}%
  \includegraphics[width=\linewidth]{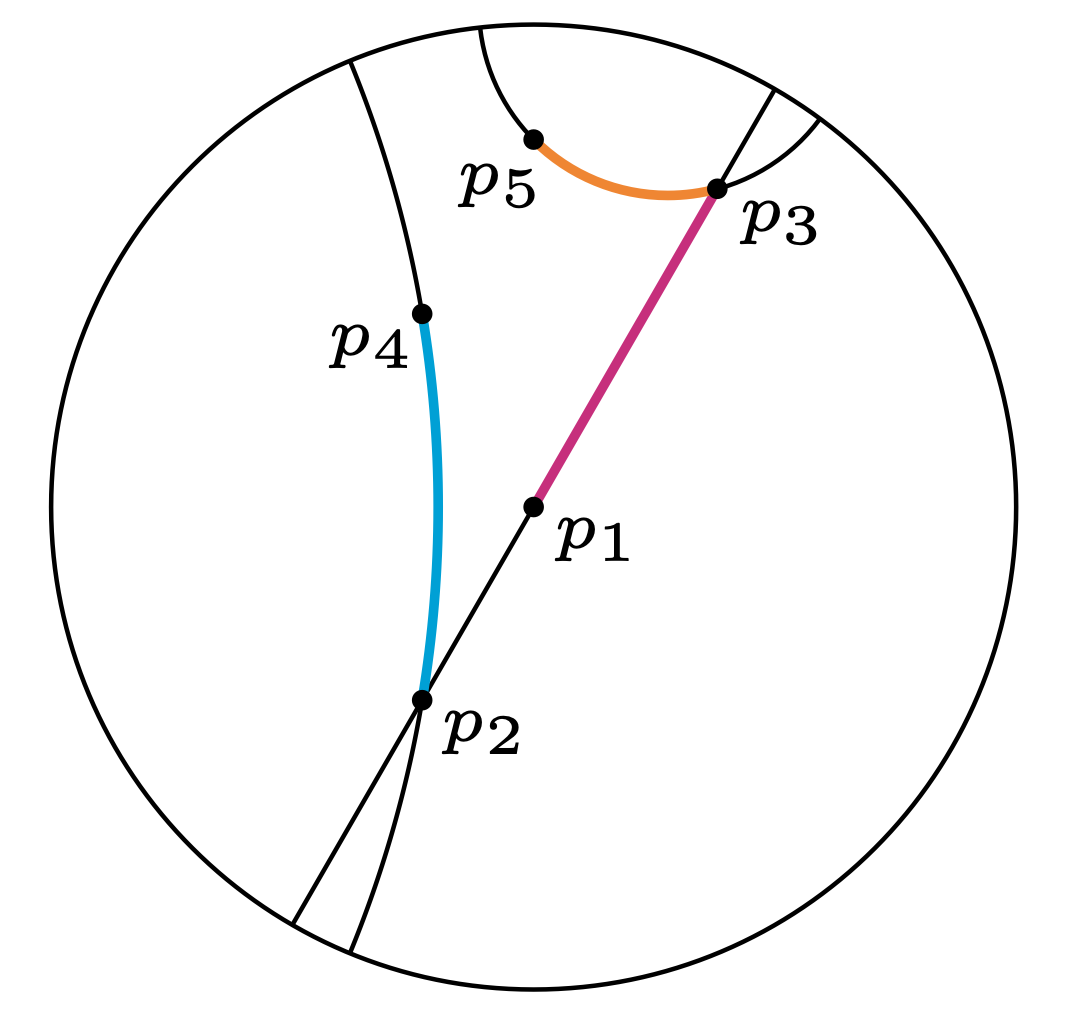}
  \caption[Geodesics in the Poincaré disk]{Due to the negative curvature of hyperbolic space, the distance between points increases exponentially (relative to their Euclidean distance) as they approach the boundary \citep{nickel2017poincare}.}
  \label{fig:poincare_distance}
\end{marginfigure}

The most popular model for hyperbolic embeddings is the Poincaré ball, denoted \( \mathbb{B}^d \), which is defined as:

\[
\mathbb{B}^d = \{ \mathbf{x} \in \mathbb{R}^d : \|\mathbf{x}\| < 1 \}
\]

where \( \|\cdot\| \) is the Euclidean norm. The distance between two points \( \mathbf{x}, \mathbf{y} \in \mathbb{B}^d \) is calculated using the hyperbolic distance function:

\[
d_{\mathbb{B}}(\mathbf{x}, \mathbf{y}) = \text{arcosh}\left(1 + 2 \frac{\|\mathbf{x} - \mathbf{y}\|^2}{(1 - \|\mathbf{x}\|^2)(1 - \|\mathbf{y}\|^2)}\right)
\]

This distance grows exponentially as points move closer to the boundary, making the Poincaré ball ideal for embedding hierarchical data. Hyperbolic embeddings preserve both local and global relationships, minimizing distortion even for deep hierarchies, as demonstrated in Figure \ref{fig:poincare_disk}.

Given a hierarchical network, Poincaré embeddings use Riemannian optimization to map nodes into hyperbolic space, ensuring that connected nodes remain close in terms of hyperbolic distance. The loss function minimized during this process is,

\[
L(\Theta) = \sum_{(u, v) \in D} \log \frac{e^{-d_{\mathbb{B}}(u,v)}}{\sum_{v' \in N(u)} e^{-d_{\mathbb{B}}(u,v')}}
\]

where \( D \) is the set of observed edges, and \( N(u) \) represents the set of negative samples for node \( u \). The objective is to learn embeddings \( \Theta \) such that the hyperbolic distance between nodes reflects their true relationships in the original graph.

The Poincaré ball model is well-suited for modeling tree-like structures, such as taxonomies or hierarchical link prediction, due to its capacity to represent exponential growth patterns with low distortion. Unlike Euclidean embeddings, which need high dimensionality for accurate representation of deep hierarchies, hyperbolic embeddings achieve high fidelity in fewer dimensions. This makes them effective for biological hierarchies, taxonomy embedding, and large-scale network analysis \citep{nickel2017poincare, chami2019hyperbolic}. In summary, hyperbolic embeddings, particularly the Poincaré ball model, provide a compact and effective framework for representing hierarchical data, outperforming Euclidean methods in capturing complex hierarchical relationships with fewer dimensions.

\subsection{Geodesic interpolation}

\begin{figure*}[h!]
\checkoddpage \ifoddpage \forcerectofloat \else \forceversofloat \fi
    \centering
    \includegraphics[width=\textwidth]{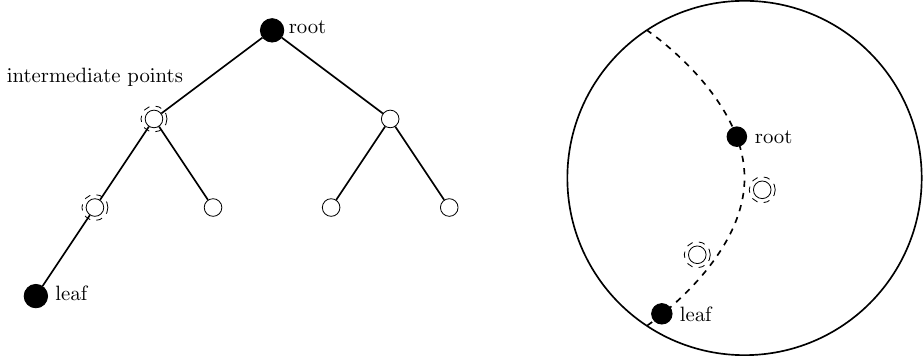}
    \caption[Geodesic interpolation]{An example of geodesic interpolation \citep{bhasker2024contrastive}.}
    \label{fig:geodesic_interpolation}
\end{figure*}

In order to investigate the intrinsic structure of hyperbolic embeddings, it can be beneficial to interpolate between points along geodesics, which constitute the shortest paths in this space. In the hyperbolic geometry of the Poincaré ball model, geodesics are arcs of circles that intersect the boundary of the ball orthogonally (see the black lines of Figure \ref{fig:poincare_distance}). For any two points in the Poincaré ball, there exists a unique geodesic connecting them. Geodesic interpolation consist of selecting a pairs of root and leaf nodes, computing the unique geodesic path that connects these two points and computing the distance of intermediary points (see Figure \ref{fig:geodesic_interpolation}.

This interpolation allows to examine the relationship between data points and the underlying structure of the embedded data. By considering the root and leaf nodes, we are effectively exploring the extremes of the hierarchical structure encoded by our hyperbolic embeddings. The intermediate points along the geodesic offer valuable insights into the transition from the root to the leaf, and potentially reflect the tree's structure (see Figure \ref{fig:geodesic_interpolation}).

For each pair of root and leaf nodes, we perform the following steps:

\begin{enumerate}
    \item Compute the coordinates of the root and leaf nodes in the hyperbolic embedding space.
    \item Compute the unique geodesic $\gamma(t)$ that connects the root to the leaf.
    \item Select a set of $t$ values to sample intermediate points along the geodesic.
    \item For each intermediate point, compute its hyperbolic distance to the geodesic.
\end{enumerate}

We expect that if the hyperbolic embeddings have captured the tree structure well, then the distances from the intermediate points to the geodesic should be relatively small, reflecting a high consistency of the tree structure with the geodesic path in the embedding space. On the other hand, if these distances are large, it would suggest that the embedding may not fully capture the hierarchical relationships in the original data.

The computation of geodesics is as follows: Given two points $x, y \in \mathbb{P}^d$, the geodesic $\gamma: [0,1] \rightarrow \mathbb{P}^d$ from $x$ to $y$ is formulated by,
\begin{equation}
\gamma(t) = {\rm Exp}_x(t \cdot {\rm Log}_x(y)) \; ,
\end{equation}
where ${\rm Exp}_x(v)$ and ${\rm Log}_x(y)$ refer to the exponential and logarithm maps at $x$, respectively. These mappings can be expressed in terms of hyperbolic cosine and sine functions and the Minkowski product. In particular,
\begin{equation}
{\rm Exp}_x(v) = \cosh(\| v \|_M) x + \sinh(\| v \|_M) \frac{v}{\| v \|_M} \; ,
\end{equation}
and
\begin{equation}
{\rm Log}_x(y) = \frac{d_{\mathbb{P}} (x,y)}{\sinh(d_{\mathbb{P}} (x,y))} (y - \cosh(d_{\mathbb{P}} (x,y)) x) \; .
\end{equation}

These equations allow for the calculation of a point along the geodesic from $x$ to $y$ at any parameter $t \in [0,1]$. For $t=0$, $\gamma(t)$ returns $x$; for $t=1$, it returns $y$. For values of $t$ in between, it yields a point along the geodesic from $x$ to $y$. Geodesic interpolation offers a useful way to navigate the learned representation space, helping to understand and visualize the structure and relationships within the embedded data in hyperbolic space.

\subsection*{Poincaré Half-Plane Model and Geodesic Calculation}

We can project the Poincaré ball model to the Poincaré half-plane model using the Cayley transform. The Cayley transform is an isometry, it preserves the hyperbolic distances and angles between points, ensuring the integrity of the underlying geometrical structures of the embeddings during the transformation.

The Cayley transform is given by,

\begin{equation}
C: \mathbb{B}^d \rightarrow \mathbb{H}^d, \quad C(x) = \frac{x + i}{ix + 1},
\end{equation}

where $x \in \mathbb{B}^d$ is a point in the Poincaré ball model, $i$ is the imaginary unit, and $\mathbb{H}$ is the Poincaré half-plane model. In the Poincaré half-plane model, geodesics are represented as either vertical lines (when they pass through the point at infinity) or semi-circles that are orthogonal to the boundary line (the x-axis). The geodesic $\gamma: [0,1] \rightarrow \mathbb{H}^d$ connecting two points $x, y \in \mathbb{H}^d$ is given by:

\begin{equation}
\gamma(t) = a \cosh(t) + b \sinh(t),
\end{equation}

where $a$ and $b$ are computed to satisfy the boundary conditions $\gamma(0) = x$ and $\gamma(1) = y$. By utilizing these equations, we can compute the geodesic in the half-plane model that connects any two given points.

\subsection{Contrastive Poincaré Maps for Single-Cell Data Analysis}
\section*{Summary of Paper \citep{bhasker2024contrastive}}

\begin{figure*}[h]
\checkoddpage \ifoddpage \forcerectofloat \else \forceversofloat \fi
    \centering
    \includegraphics[width=\textwidth]{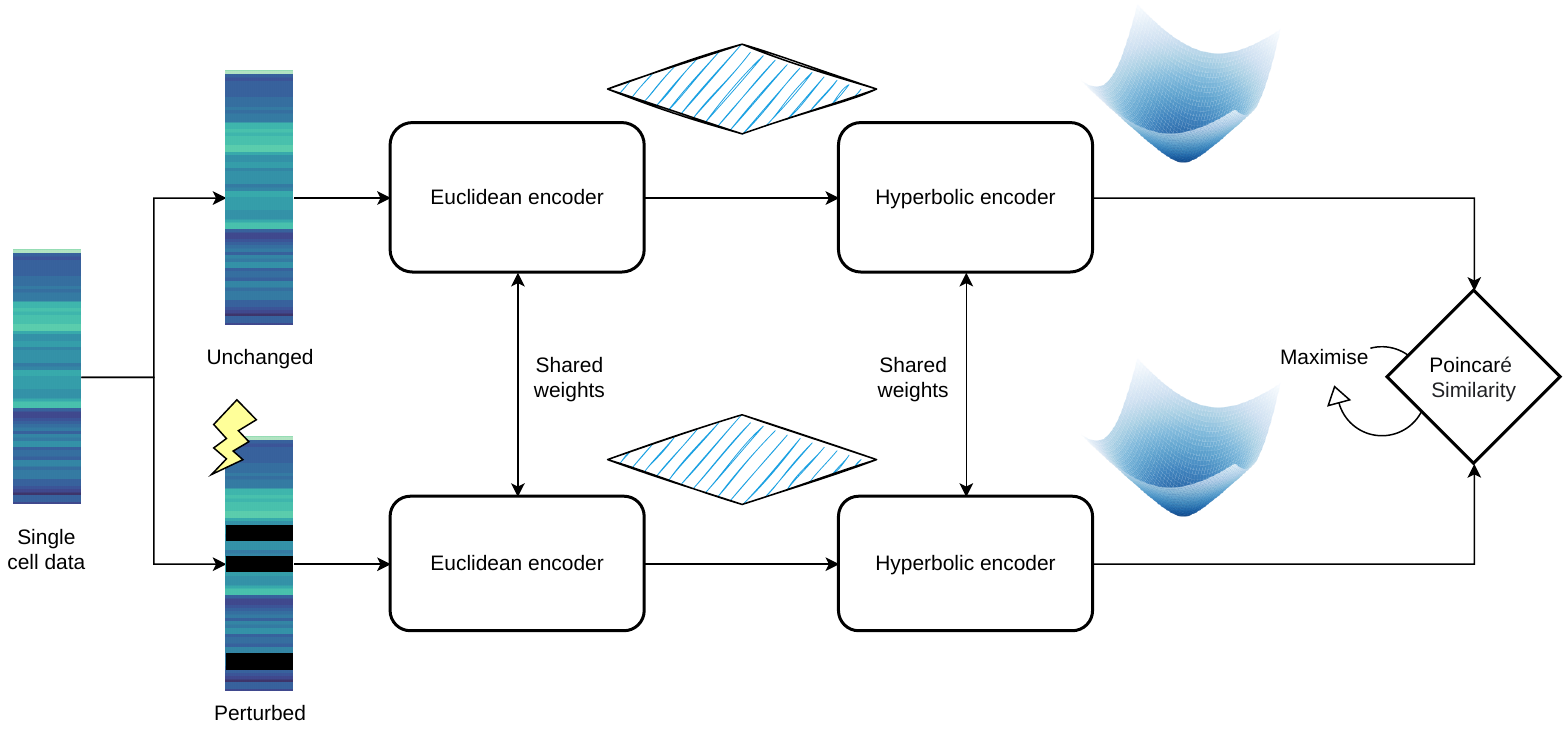}
    \caption[Contrastive Poincaré map]{Archictecture of the contrastive Poincaré map model. The model is composed of an Euclidean encoder and a hyperbolic encoder (see Figure \ref{model}). The Euclidean encoder consists of linear layers with ReLU activation. The representations learnt by the Euclidean encoder are then mapped to the Poincar\'e ball. The hyperbolic encoder is composed of hyperbolic linear layers with ReLU activation \citep{bhasker2024contrastive}.}
    \label{fig:cpm_model}
\end{figure*}

In this work, we introduce \emph{Contrastive Poincaré Maps} (CPM), a novel contrastive learning-based approach for embedding tabular data into hyperbolic space, specifically designed for single-cell RNA sequencing (scRNA-seq) data. Single-cell data, often characterized by complex branching patterns and hierarchies, poses significant challenges when represented in Euclidean space. Previous methods like Poincaré Maps (PM) have demonstrated the potential of hyperbolic spaces for capturing hierarchical structures, but they suffer from various limitations such as performance degradation on deep trees and computational inefficiency \citep{klimovskaia2020poincare}.

The CPM model is motivated by the need to improve representation accuracy for large and deep hierarchical structures while addressing the shortcomings of prior methods like PM. However, PM tends to lose representation accuracy for deep trees, which limits its application to complex single-cell datasets \citep{klimovskaia2020poincare}. Moreover, PM requires substantial feature engineering and is computationally intensive, which further constrains its scalability.

\newthought{Model Architecture:} CPM leverages a contrastive learning framework \citep{bahri2022scarf}, where positive and negative pairs of data points are sampled to optimize the learned embeddings. The architecture consists of two main components: a Euclidean projection block and a hyperbolic block (see Figure \ref{fig:cpm_model}). The Euclidean projection block transforms the input data into an intermediate latent space, which is subsequently projected into hyperbolic space using the Poincaré ball model via the exponential map introduced in the last section.

The hyperbolic block further processes these embeddings by applying hyperbolic neural layers, allowing the model to learn non-linear hierarchical relationships in the data. The encoder learns the representation through the InfoNCE loss, which helps in maximizing the similarity of positive pairs while minimizing that of negative pairs.

The following equations describe a forward pass in the CPM model:

\[
H = \sigma(W_e X^e + b_1)
\]
\[
Z_h = \sigma^{\otimes}(W_h \otimes \text{Exp}_o(H) \oplus \text{Exp}_o(b_2)),
\]

where \(H\) is the intermediate Euclidean embedding, \(Z_h\) is the final hyperbolic embedding, and \(\text{Exp}_o\) is the exponential map that projects points from Euclidean space to the Poincaré ball \citep{chami2019hyperbolic}.

\newthought{Efficiency and Scalability} CPM significantly improves memory efficiency and scalability compared to PM. As demonstrated through experiments on synthetic and real single-cell datasets, CPM requires fewer computational resources and scales better with increasing sample sizes. This makes it more suitable for large-scale single-cell data analysis, where the number of cells can grow exponentially.

\newthought{Results} In experiments on both synthetic trees and real single-cell data (e.g., chicken heart development data), CPM outperforms PM in terms of both local and global representation accuracy. CPM maintains a high global representation accuracy across deep trees, as indicated by Q-scores (Qlocal and Qglobal), MAP scores, and distortion metrics. For example, in synthetic tree experiments, CPM consistently shows better performance as the depth of the tree increases, maintaining both local and global accuracy, while PM's performance degrades significantly \citep{klimovskaia2020poincare}.

\begin{figure*}[t]
\checkoddpage \ifoddpage \forcerectofloat \else \forceversofloat \fi
    \centering
    \includegraphics[width=\textwidth]{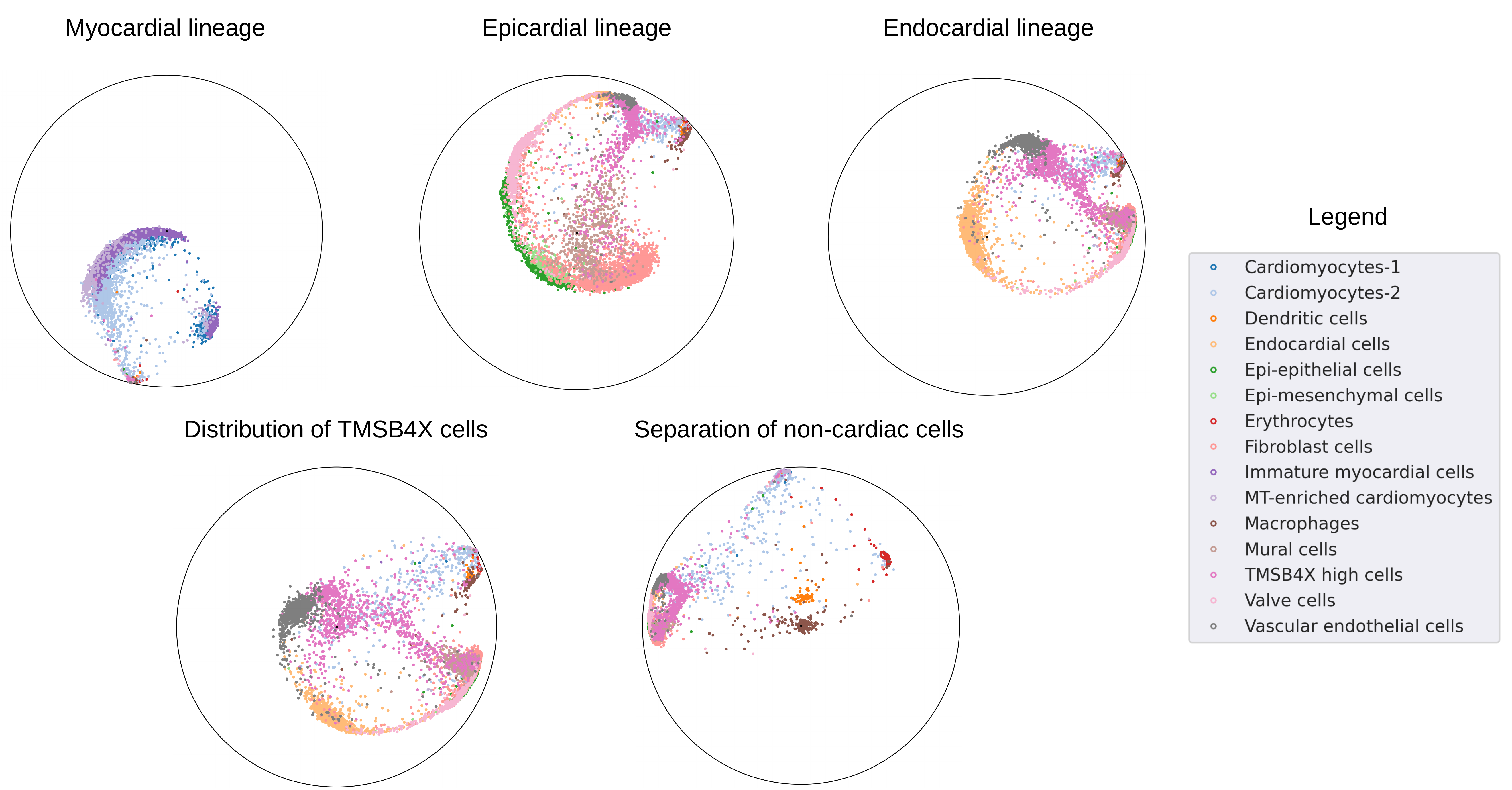}
    \caption[Chicken heart development]{ Illustration of the contrastive Poincaré map resolving multiple cardiac cell lineages in the chicken heart development dataset \citep{mantri2021spatiotemporal}.}
    \label{fig:chicken}
\end{figure*}

\newthought{Application to Single-Cell Data} In single-cell RNA sequencing analysis, CPM accurately captures the hierarchical branching of cellular differentiation, preserving both intra-lineage and inter-lineage relationships. For instance, in the chicken heart dataset, CPM effectively resolves distinct lineages (e.g., myocardial, epicardial, and endocardial lineages) and maintains the relative hierarchy among these lineages without the need for additional feature processing, such as PCA.

By introducing contrastive learning into the hyperbolic embedding framework, CPM addresses the limitations of prior methods, offering a robust, scalable, and accurate method for embedding complex hierarchical data, particularly in the context of single-cell biology. CPM provides an efficient tool for analyzing large-scale single-cell datasets, ensuring accurate representation of both local and global structures.

\chapter{Human Mobility and Geography}
\label{sec:chapter3}

The origins of human mobility studies can be traced back to the late 18th century with Gaspard Monge’s \textit{Théorie des remblais et des déblais}, where was laid the foundations for understanding the optimal transportation of materials\cite{monge1781theorie}. Although this work was focused on earthworks and logistics, it introduced critical concepts of movement, distance, and resource allocation that would later influence the study of human mobility and optimal transport. Monge's ideas marked the beginning of theoretical approaches to spatial flows, laying the groundwork for future explorations of how geography shapes movement.

In the late 19th century, the pioneering work of Ernst Ravenstein on migration laws further developed these ideas. The \textit{Laws of Migration} (1885) analyzed human movements, highlighting the push-and-pull factors that influenced why people move and how distance plays a central role in migration patterns\cite{ravenstein1885laws}. Following this, Samuel Stouffer’s (1940) theory of intervening opportunities\cite{stouffer1940intervening} and George Zipf’s (1946) inverse distance law\cite{zipf1946p} offered further insight into the statistical regularities governing migration and mobility. These theories suggested that human mobility is not just a function of distance but also influenced by intervening factors, such as the availability of jobs or services.

In the meantime, W.J. Reilly’s \textit{Law of Retail Gravitation} (1931) applied principles of gravitational attraction to commercial activity, showing that the draw of cities and markets followed patterns similar to physical forces\cite{reilly1931law}. Reilly's work introduced a model where the interaction between two urban centers could be predicted based on their population sizes and the distance separating them, foreshadowing the later development of gravity models of human mobility.

In recent decades, the study of human mobility has been revolutionized by the availability of vast geolocated datasets, allowing researchers to quantitatively explore the spatiotemporal patterns in human trajectories at an unprecedented scale\cite{brockmann2006scaling}\cite{gonzalez2008understanding}. This has led to breakthroughs in understanding the regularities and complexities of human movement, with applications ranging from migratory flow estimation and traffic forecasting to urban planning and epidemic modeling\cite{brockmann2013hidden}.

\section{Data-Driven Human Mobility}

Recent advancements in technology and data collection, particularly through mobile devices, have revolutionized the field of human mobility studies. These new data-driven approaches allow for unprecedented insights into the regularities and scaling properties of human movement. Several influential studies have shaped the current understanding of human mobility at both individual and population levels.

In 2006, Brockmann et al. \citep{brockmann2006scalinglaws} introduced the concept of scaling laws in human travel by analyzing banknote tracking data. They revealed that human movement follows a power-law distribution, suggesting that most people tend to travel short distances frequently, while a few travel long distances. The probability \( P(d) \) that a person travels a distance \( d \) can be modeled as,

\[
P(d) \propto d^{-\alpha}
\]

where \( \alpha \) is the scaling exponent, typically found to be between 1.5 and 2, indicating a heavy-tailed distribution \citep{brockmann2006scalinglaws} (see Section \ref{sec:tail}). This discovery provided a solid foundation for later studies on mobility patterns across various scales. Gonzalez et al. \citep{gonzalez2008understanding} followed with a study of individual human mobility, showing that people’s movements are remarkably regular and predictable, with most individuals exhibiting a high degree of regularity in their daily trajectories.

Building on these findings, Song et al. \citep{song2010limits} explored the limits of predictability in human mobility, highlighting that even though human movement appears highly regular, about 20\% of human mobilty remains unpredictable due to various socio-environmental factors\footnote{Song et al.'s work emphasized that even though humans have habits, the 20\% unpredictability factor suggests that some surprises are inevitable, due to spontaneous decisions and environmental disruptions.}. In a subsequent study, Song et al. \citep{song2010scaling} modeled the scaling properties of human mobility, further confirming that human movement is not random but follows distinct scaling laws that reflect individual travel patterns.

Simini et al. \citep{simini2012universal} introduced the radiation model for human mobility, which provides an alternative to traditional gravity models by focusing on intervening opportunities rather than distance alone. The model predicts the probability of human movement between locations based on population and the number of opportunities between them.

In 2015, Pappalardo et al.\cite{pappalardo2015returners} explored the dichotomy between returners and explorers in human mobility. They showed that while many individuals frequently return to a few specific locations, others explore a wide range of places. This behavior introduces complexity to human mobility models, as it highlights the distinction between habitual and exploratory movement.

Most recently, Alessandretti et al.\cite{alessandretti2020scales} addressed the paradox between the power-law distribution of human mobility and the discrete scales imposed by the built environment. By analyzing large-scale mobile phone data, they resolve this apparent paradox by showing that day-to-day human mobility does indeed contain meaningful scales, corresponding to spatial ‘containers’ that restrict mobility behaviour. The scale-free results arise from aggregating displacements across containers.

Schläpfer et al.\cite{schlapfer2021universal} introduced the universal visitation law, which captures both the spatial and temporal dimensions of human mobility using large-scale data from cities worldwide. Unlike previous models such as the gravity or radiation models, which primarily focus on spatial distance, the visitation law accounts for the frequency of recurrent visits to the same locations. The law states that the number of visitors to any location decreases as the inverse square of the product of their visiting frequency and travel distance.

\section{Human Mobility: Population-Level Models}

\subsection{Gravity Models}

As we saw in the last section, gravity models are one of the earliest and most widely used approaches to modeling population-level mobility patterns \citep{zipf1946p,carey1858principles}. These models are based on an analogy with Newton's law of gravitation, where the interaction between two locations is proportional to their population sizes and inversely proportional to the distance between them. Formally, the flow $T_{ij}$ between two locations $i$ and $j$ is modeled as,

\[
T_{ij} = G \frac{P_i P_j}{D_{ij}^\beta}
\]

where $P_i$ and $P_j$ are the populations of locations $i$ and $j$, $D_{ij}$ is the distance between them, $G$ is a proportionality constant, and $\beta$ is an empirically determined exponent that captures the effect of distance decay on mobility flows.\footnote{The gravity model is often considered simplistic, but its ease of application and its foundational role in understanding mobility patterns make it indispensable for large-scale population studies \citep{wilson1971geography}.}

As demonstrated by Murray et al.\cite{murray2023unsupervised}, the gravity model can be related to the \texttt{word2vec} model introduced in Chapter 2. Indeed, the \texttt{word2vec} algorithm mirrors this structure. By treating locations as words and migration trajectories as sentences, \texttt{word2vec} embeds each location into a vector space where distances reflect migration rates. The probability \( P(v \mid u) \) of migration from location \( u \) to \( v \) is given by:

\[
P(v \mid u) = \frac{e^{\mathbf{z}_u^\top \mathbf{z}_v}}{\sum_{v' \in V} e^{\mathbf{z}_u^\top \mathbf{z}_{v'}}}
\]

where \( \mathbf{z}_u \) and \( \mathbf{z}_v \) represent the embeddings of locations \( u \) and \( v \), and the inner product \( \mathbf{z}_u^\top \mathbf{z}_v \) reflects the attraction between locations, akin to the gravity model. This mathematical similarity suggests that \texttt{word2vec} can be viewed as a generalization of the gravity model, allowing it to incorporate not only spatial factors but also non-spatial elements like culture, language, and prestige into a cohesive framework.

\subsection{Intervening Opportunities Models}

The intervening opportunities model, introduced by Stouffer \citep{stouffer1940intervening}, offers an alternative perspective to the gravity model by focusing on the opportunities available between origin and destination locations.\footnote{This model posits that human mobility is more influenced by the socio-economic opportunities encountered along the way, rather than the physical distance between locations. This idea is particularly useful in urban planning, where distances are often short but opportunity density varies significantly.} Unlike gravity models, which emphasize the importance of distance, the intervening opportunities model suggests that the likelihood of an individual traveling to a particular destination is influenced more by the availability of opportunities along the way than by the distance to the destination.

Mathematically, the model can be expressed as,

\[
P(T_{ij}) = \frac{O_j}{\sum_{k=1}^{j} O_k}
\]

where $P(T_{ij})$ is the probability of traveling from location $i$ to location $j$, and $O_j$ represents the opportunities available at destination $j$.

\newthought{Radiation Model as an Extension of the Opportunities Model} \\
The Radiation model \citep{simini2012universal} can be considered an extension of the intervening opportunities framework. In this model, the decision to move from one location to another depends on the number of opportunities both at the destination and within the area between the origin and destination. The key difference is that, instead of relying on the concept of distance decay, the Radiation model evaluates the intervening opportunities by considering the population density in the area.

The Radiation Model can be expressed using the opportunities notations,

\[
T_{ij} = T_i \frac{P_i P_j}{(P_i + O_{ij})(P_i + P_j + O_{ij})}
\]

where $T_{ij}$ is the flow of individuals from origin \(i\) to destination \(j\), \(P_i\) and \(P_j\) are the populations of the origin and destination, and \(O_{ij}\) represents the number of opportunities between the two locations. This model assumes that people choose their destination based on the highest utility they can achieve, influenced by the number of intermediate opportunities, much like the original intervening opportunities model.

Unlike gravity models, the radiation model does not require a parameter to control the decay with distance, providing a parameter-free approach to modeling mobility patterns \citep{simini2012universal}. 

\section{Geography \& The Built Environment}

The study of the built environment has been a central focus in geography, as the human-made surroundings that provide the setting for human activity are crucial for understanding urban planning, economic development, and sustainability. The built environment encompasses the spatial and physical components of urban areas, including the layout of streets, the positioning of houses, and the distribution of public spaces \citep{batty2007cities}. 

From a complex systems perspective, the built environment is viewed as an emergent property of the interactions between various elements within a city, such as people, infrastructure, and economic activities\cite{batty2007cities}. This perspective highlights the non-linear, dynamic, and often unpredictable nature of urban growth and development, where small changes in one part of the system can lead to significant and widespread effects across the entire urban landscape. Modeling the built environment using complex systems approaches can provide valuable insights into these dynamics, offering a more holistic understanding of urbanization processes \cite{batty2005fractal}.

\subsection{Historical Perspective}
The study of the built environment in geography has evolved significantly over time. Early approaches primarily focused on static, deterministic models of urban form, which were based on idealized assumptions about the regularity and uniformity of urban spaces\footnote{These early models often assumed isotropy and homogeneity, simplifying the complexities of real-world environments to facilitate mathematical modeling \citep{christaller1933central}.}. However, as the complexity of urban systems became more apparent, geographers began to adopt more dynamic, complex systems-based models that better captured the irregular and multifaceted nature of cities. One of the most influential theories from the earlier period is Central Place Theory, introduced by the German geographer Walter Christaller in the 1930s. This theory sought to explain the distribution, size, and number of cities in a region based on economic principles, and it laid the groundwork for much of the subsequent research in urban geography\footnote{Christaller's Central Place Theory was groundbreaking in its attempt to systematically explain the spatial organization of settlements, though it has since been criticized for its oversimplification of urban dynamics \citep{christaller1933central, beavon1977central}.} \citep{christaller1933central}.

\subsection{Central Place Theory}

Central Place Theory was introduced as a framework to understand the spatial distribution of cities, towns, and villages. It aimed to explain the number, size, and location of human settlements in an urban system, suggesting that settlements function as 'central places' providing goods and services to surrounding areas. Christaller proposed that these central places are organized hierarchically, with larger cities offering more specialized and higher-order services that are not available in smaller towns or villages. 

\begin{marginfigure}%
  \includegraphics[width=\linewidth]{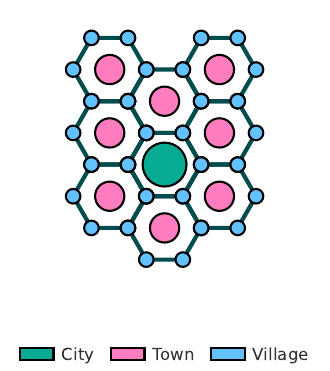}
  \caption[Central place theory]{A hexagonal grid illustrating the spatial arrangement of settlements according to Central Place Theory. In this model, larger cities are represented by widely spaced hexagons, while smaller towns and villages are clustered more closely together. The hexagonal pattern reflects the idealized spatial distribution proposed by Christaller, where settlements are evenly spaced to efficiently serve surrounding areas.}
  \label{fig:marginfig}
\end{marginfigure}

The Central Place Theory is based on the idea that human settlements serve as 'central places' where economic activities are concentrated. The theory predicts that settlements are evenly spaced, forming a hexagonal lattice pattern, with larger cities spaced further apart and smaller towns more closely clustered around them. While Central Place Theory has been widely applied in urban planning and regional development, particularly in the planning of retail locations and public services, it has limitations, particularly in its assumptions of uniformity and rational behavior. Modern cities often exhibit complex and irregular patterns that do not conform to the idealized hexagonal lattice predicted by the theory \citep{beavon1977central}.

\subsection{Fractal Cities}

In contrast to the regular patterns suggested by Central Place Theory, fractal theory has emerged as a powerful tool for modeling the complexity of urban forms. Fractals are mathematical constructs that exhibit self-similarity across different scales, making them particularly suited to capturing the irregular and intricate patterns found in the built environment. Researchers have demonstrated that many aspects of urban structure, such as the layout of streets and the distribution of building heights, can be modeled as fractals, providing a way to analyze and understand the inherent complexity of urban environments\footnote{Fractal geometry allows for the modeling of urban forms that are too complex and irregular to be captured by traditional Euclidean geometry, offering new insights into the spatial organization of cities \citep{batty1994fractal}.} \citep{batty1994fractal}.

\newthought{Fractal Dimension} is one of the key aspects of fractal modeling in urban settlements. It quantifies the complexity of urban patterns. Unlike traditional Euclidean dimensions, which are integers (e.g., a line has dimension 1, a plane has dimension 2), the fractal dimension can be a non-integer, reflecting the space-filling capacity of the fractal. The fractal dimension \( D \) is often calculated using the box-counting method, which is defined as:

\begin{marginfigure}
  \includegraphics[width=\linewidth]{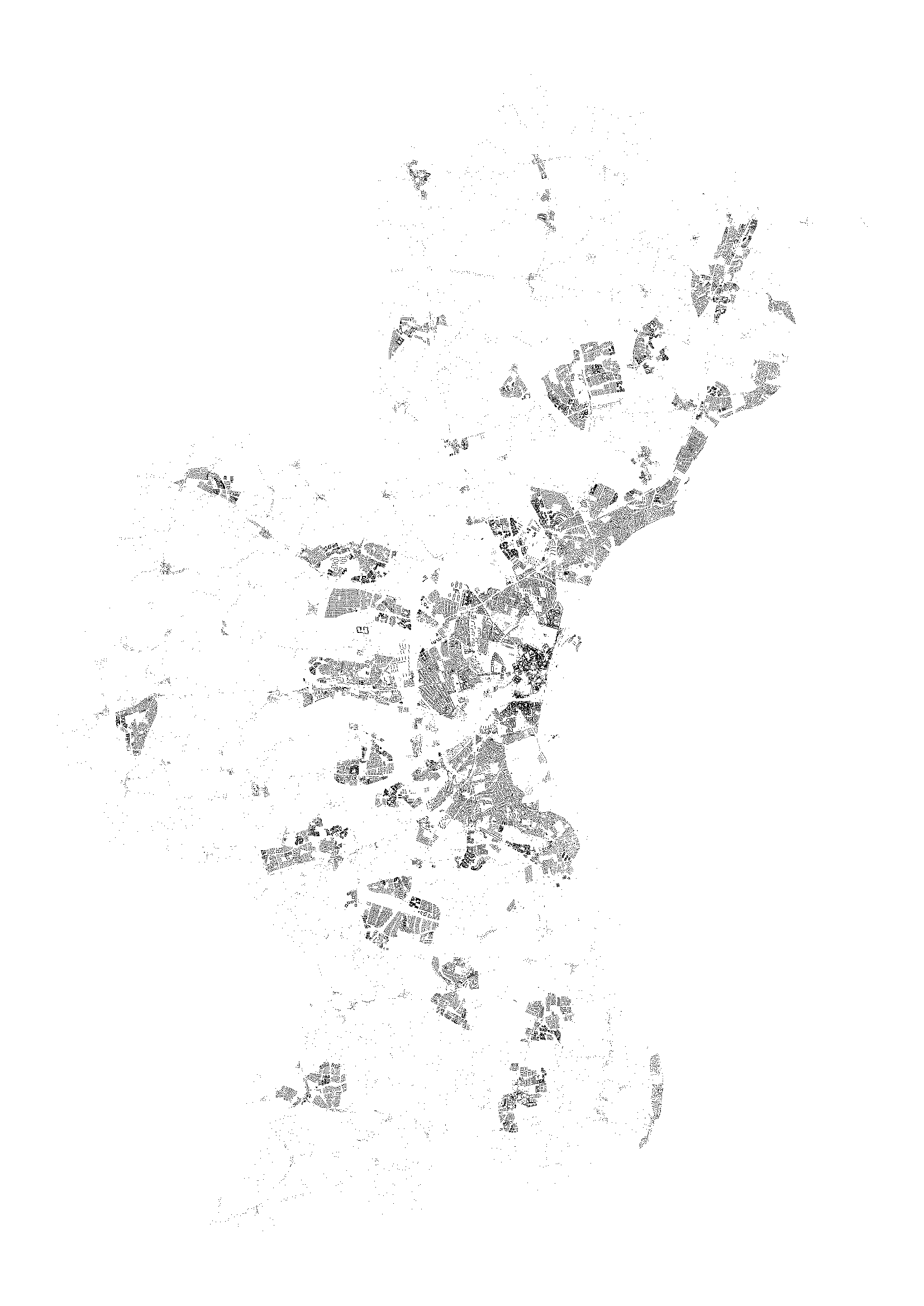}
  \caption[Addresses in Aarhus]{Every addresses in Aarhus to showcase of urban settlements can look like fractal, here the fractal dimension given by the box-counting method is 1.5.}
  \label{fig:fractal}
\end{marginfigure}

\[
D = \lim_{\epsilon \to 0} \frac{\log N(\epsilon)}{\log \left(\frac{1}{\epsilon}\right)}
\]

where \( N(\epsilon) \) is the number of boxes of size \( \epsilon \) required to cover the urban pattern. The relationship between the box size and the number of occupied boxes typically follows a power law, and the slope of the log-log plot of \( N(\epsilon) \) versus \( \frac{1}{\epsilon} \) gives the fractal dimension. For instance, a fractal dimension close to 1.7 has been observed in many cities, suggesting a balance between space-filling growth and the maintenance of open spaces \citep{batty2008size}. Figure \ref{fig:fractal} illustrates of the city of Aarhus is fractal.


\newthought{Multifractality} extends the concept of fractal dimension by recognizing that many urban systems exhibit multifractal behavior, meaning that different parts of the city may have different fractal dimensions. This reflects the heterogeneity of urban development, where areas like central business districts, residential neighborhoods, and industrial zones may each exhibit distinct scaling properties.

Multifractality is characterized by a spectrum of fractal dimensions, rather than a single value. The multifractal formalism involves computing the generalized dimension \( D_q \), which depends on a parameter \( q \) that emphasizes different parts of the measure. It is defined as:

\[
D_q = \frac{1}{q-1} \lim_{\epsilon \to 0} \frac{\log \sum_{i} \mu_i^q(\epsilon)}{\log \left(\frac{1}{\epsilon}\right)}
\]

where \( \mu_i(\epsilon) \) represents the measure (e.g., the population density or the concentration of buildings) in the \( i \)-th box of size \( \epsilon \). The parameter \( q \) controls the weighting of different measures, with \( q > 1 \) emphasizing regions with higher values of \( \mu_i(\epsilon) \) and \( q < 1 \) emphasizing regions with lower values. This spectrum of dimensions can provide a more detailed description of the spatial variability within a city \citep{chen2013multifractal}.

\newthought{Application in Urban Studies} The fractal dimension and multifractal analysis have profound implications for understanding urban growth patterns and planning. For instance, by examining historical maps and satellite imagery, researchers can track changes in the fractal dimension of cities over time, providing insights into the dynamics of urban expansion. Studies have shown that rapidly growing cities in developing countries often exhibit increasing fractal dimensions, reflecting more chaotic and space-filling growth patterns \citep{terzi2019urban}.

Furthermore, fractal models have been used to simulate urban growth scenarios. These models often incorporate principles of cellular automata or agent-based modeling, where simple rules governing local interactions can lead to the emergence of complex, fractal-like urban patterns. Such simulations can help planners explore different growth scenarios and assess their potential impacts on urban form and function \citep{portugali2000self}.

Complementary to fractal analysis, raster analysis has also been employed to understand urban form. Ratti and Richens \citep{ratti2004raster} have demonstrated how raster-based techniques can be used to analyze urban morphology, providing a different but related perspective to fractal approaches. Their work highlights the importance of considering multiple analytical methods in comprehending the complexities of urban structures.

The study of road networks has further contributed to our understanding of urban structure through complex systems theory. Marshall \citep{marshall2006structure} examined the structure and evolution of morphogenetic patterns in road networks, adding another layer to the analysis of urban form that complements fractal-based approaches. This research underscores the intricate relationships between different elements of urban systems and their collective impact on urban morphology.

\subsection{Urban Scaling Laws}

Zipf’s law states that the size \( S_r \) of a city ranked \( r \) is inversely proportional to its rank:

\begin{align}
S_r \propto \frac{1}{r}
\end{align}

This suggests that the largest city in a country is roughly twice as large as the second-largest, three times larger than the third-largest, and so on. Zipf’s law provides a foundational empirical framework for understanding the hierarchical organization of cities, consistent with patterns observed in broader urban scaling laws\cite{gabaix1999zipf}.

\subsection*{Superlinear Scaling: Innovation and Wealth Creation}

Urban scaling laws describe how various urban properties change with population size, typically following power-law relationships. The general form of these scaling laws is:

\[
Y = Y_0 P^\beta
\]

where \( Y \) represents an urban metric (e.g., GDP, infrastructure), \( P \) is the population, \( Y_0 \) is a normalization constant, and \( \beta \) is the scaling exponent that determines how the metric \( Y \) responds to changes in population size. The exponent \( \beta \) indicates whether the urban metric grows superlinearly or sublinearly with population. If \( \beta > 1 \), the metric grows faster than population (superlinear scaling), while \( \beta < 1 \) indicates sublinear scaling, meaning the metric grows slower than the population.

\newthought{Superlinear scaling} occurs when \( \beta > 1 \), meaning that as a city's population grows, its output increases disproportionately. Bettencourt and colleagues \cite{bettencourt2007growth} found that metrics such as economic productivity, innovation (measured by the number of patents), and social interactions scale superlinearly. For example, the relationship between GDP (\( Y_{\text{GDP}} \)) and population follows:

\[
Y_{\text{GDP}} = Y_0 P^{1.15}
\]

This scaling implies that a city with twice the population of another generates more than twice the economic output or innovation. The superlinear scaling reflects the increased interactions and efficiencies found in larger populations, driving productivity and creativity.

\newthought{Sublinear scaling} occurs when \( \beta < 1 \). Bettencourt and West found that physical infrastructure, such as roads, electrical cables, and water networks, scales sublinearly with population size. For instance, the length of a city's road network (\( Y_{\text{Road}} \)) follows:

\[
Y_{\text{Road}} = Y_0 P^{0.85}
\]

This relationship shows that larger cities can provide infrastructure more efficiently, requiring fewer resources per capita as they grow\cite{bettencourt2010unified}.

Understanding these scaling laws has important implications for urban planning and sustainability. Superlinear scaling highlights the role of large cities as centers of innovation and wealth, but it also suggests challenges like increased crime and pollution. In contrast, sublinear scaling demonstrates how larger cities benefit from economies of scale in infrastructure, using fewer resources per capita as they expand. These insights are crucial for policymakers and urban planners as they work to build more efficient and sustainable cities \citep{rybski2019urban,bettencourt2010unified,pumain2006evolutionary}.

\subsection{The Definition of a City}
\label{sec:HDBSCAN}

Defining a city’s boundaries is a significant challenge in urban studies, as different definitions can lead to varying interpretations of scaling laws\cite{arcaute2015constructing}\cite{bettencourt2013origins}. City boundaries can be based on administrative, geographic, or functional criteria, and each method influences the scaling relationships between population size and urban metrics such as GDP and infrastructure.

As seen in the previous section, scaling laws often take the form \( Y = Y_0 P^\beta \), where \( Y \) is the urban metric, \( P \) is the population, and \( \beta \) is the scaling exponent. Different city definitions can lead to different values of \( \beta \), causing discrepancies in the observed scaling behavior. For example, administrative boundaries may underestimate the true extent of a city, while functional definitions based on agglomerations may provide more accurate scaling relationships \citep{arcaute2015constructing}.

\newthought{Log-Normal vs. Power Law Distributions}, there is ongoing debate about whether city sizes follow a power law, as described by Zipf’s Law, or a log-normal distribution as suggested by Gibrat’s Law \citep{eeckhout2004gibrats}. Under Gibrat’s Law, city sizes \( S \) follow:

\[
\log S \sim \mathcal{N}(\mu, \sigma^2)
\]

indicating a log-normal distribution. In contrast, Zipf’s Law states that the size \( S_r \) of a city is inversely proportional to its rank \( r \):

\[
S_r \propto \frac{1}{r}
\]

Recent studies by Corral et al. (2019) have shown that, when cities are defined by clustering algorithms based on population density, the size of these clusters follows a log-normal distribution, supporting Gibrat’s Law \citep{corral2019lognormal} (see Figure \ref{fig:2} for the distributions of city size in Denmark). This suggests that scaling behavior is sensitive to how cities are defined and at what spatial scale the analysis is conducted.

\newthought{Hierarchical Urban Systems}, urban systems exhibit hierarchical structures, where smaller towns are nested within larger metropolitan areas. As noted by Pumain, this hierarchical organization influences how cities grow and interact \citep{pumain2006hierarchy}. The clustering of population centers can reveal different scaling behaviors depending on the scale of analysis, showing that urban growth is not a simple linear process. Instead, it follows complex, multi-scale patterns.

The method used to define city boundaries—whether administrative, functional, or through clustering—affects the scaling exponents \( \beta \) and the observed size distributions. Definitions that better reflect functional relationships or spatial agglomerations tend to produce more accurate scaling laws, supporting the idea that urban scaling is context-dependent. The challenge is to develop consistent methods that capture the complexity of cities across different regions. While administrative boundaries provide a well-established framework for defining cities, they often fail to capture the functional and spatial dynamics of urban areas. To overcome these limitations and better understand the complexity of city growth and interactions, alternative methods are needed.

In the case of Denmark, we first adopted the definition of a city provided by Danmarks Statistik \citep{dst_befolkning_valg}, which identifies 1,473 cities across Denmark. This official classification served as our initial benchmark. To ensure the robustness of our results, we compared the Danmarks Statistik definition with cities defined using density-based clustering techniques. These methods are designed to distinguish densely clustered urban or urbanized areas from sparsely populated or rural regions. The Density-Based Spatial Clustering of Applications (DBSCAN) is particularly effective for identifying clusters of varying shapes\cite{schubert2017dbscan}, which is crucial for capturing the fractal geometries of urban areas. DBSCAN categorizes data points into core points, boundary points, and noise based on local data density. However, the accuracy of this method hinges on the appropriate selection of parameters, such as $\epsilon$ (the neighborhood search radius) and the minimum number of points, which determine the granularity of the identified urban zones \citep{ester1996density}.

For datasets covering cities with varying population densities, HDBSCAN (Hierarchical Density-Based Spatial Clustering of Applications with Noise) provides a better approach. Building on DBSCAN's foundation, HDBSCAN removes the need for a fixed $\epsilon$ value by employing a hierarchical clustering strategy. This allows HDBSCAN to effectively differentiate between densely populated cities and smaller towns within the same dataset. The method's ability to adjust to varying densities within the data is analogous to hierarchical observed in urban systems \citep{campello2013density, mcinnes2017hdbscan}.

An interesting case study is Copenhagen, where the city's influence extends beyond its administrative boundaries, as evidenced by commuting patterns and economic activity. Consequently, neighboring towns have been merged into the Copenhagen metropolitan area, or Hovedstadsområdet, as defined by Danmarks Statistik. HDBSCAN clustering results also agregate Copenhagen and its neighboring towns.

To quantify the similarity between different city definitions, we employed Normalized Mutual Information (NMI) and Adjusted Mutual Information (AMI). These metrics measure the overlap between two partitions, indicating how much information is shared between the official city labels and the clusters identified by the algorithm. NMI is calculated as the mutual information between two clusterings divided by the geometric mean of their entropies \citep{strehl2002cluster}. However, a limitation of the NMI is that it ignores the random grouping of clusters, i.e., random cluster assignments can produce a non-zero NMI value. On the other hand, the AMI corrects for this limitation by adjusting the score to account for noise, ensuring that random cluster assignments result in an AMI score close to zero \citep{vinh2010information} .

We choose the parameters for HDBSCAN that optimize the NMI between the clustering and the real city labels. Figure \ref{fig:NMI_HDBSCAN} shows the values for different \textit{minimum samples} and \textit{minimum cluster size} settings; other parameters follow the default values of this implementation \citep{campello2013density}.

\begin{table}[ht]
\centering
\begin{tabular}{@{}lcc@{}}
\toprule
Algorithm & NMI & AMI \\
\midrule
DBSCAN    & 0.88 & 0.87 \\
HDBSCAN   & 0.91 & 0.89 \\
\bottomrule
\end{tabular}
\caption{NMI and AMI values between empirical city clustering (Copenhagen merged) and algorithmic clustering (DBSCAN and HDBSCAN).}
\label{tab:nmi_ami_values_HOV}
\end{table}
\begin{table}[ht]
\centering
\begin{tabular}{@{}lcc@{}}
\toprule
Algorithm & NMI & AMI \\
\midrule
DBSCAN    & 0.71 & 0.67 \\
HDBSCAN   & 0.67 & 0.65 \\
\bottomrule
\end{tabular}
\caption{NMI and AMI values between the empirical cities (Copenhagen not merged) clustering and the algorithmic clustering (DBSCAN and HDBSCAN)}
\label{tab:nmi_ami_values_CPH}
\end{table}

Figure \ref{fig:dk_cluster-cities} shows all addresses colored by the cluster they belong to. The background color indicates the official city border.

\newpage

\section{Decomposing Geographical and Universal Aspects of Human Mobility}
\label{sec:decomposing}
\addcontentsline{toc}{subsection}{\textit{Summary of Paper II}}
\section*{Summary of Paper II}

\begin{figure*}[h!]
\checkoddpage \ifoddpage \forcerectofloat \else \forceversofloat \fi
    \centering
    \includegraphics[width=\textwidth]{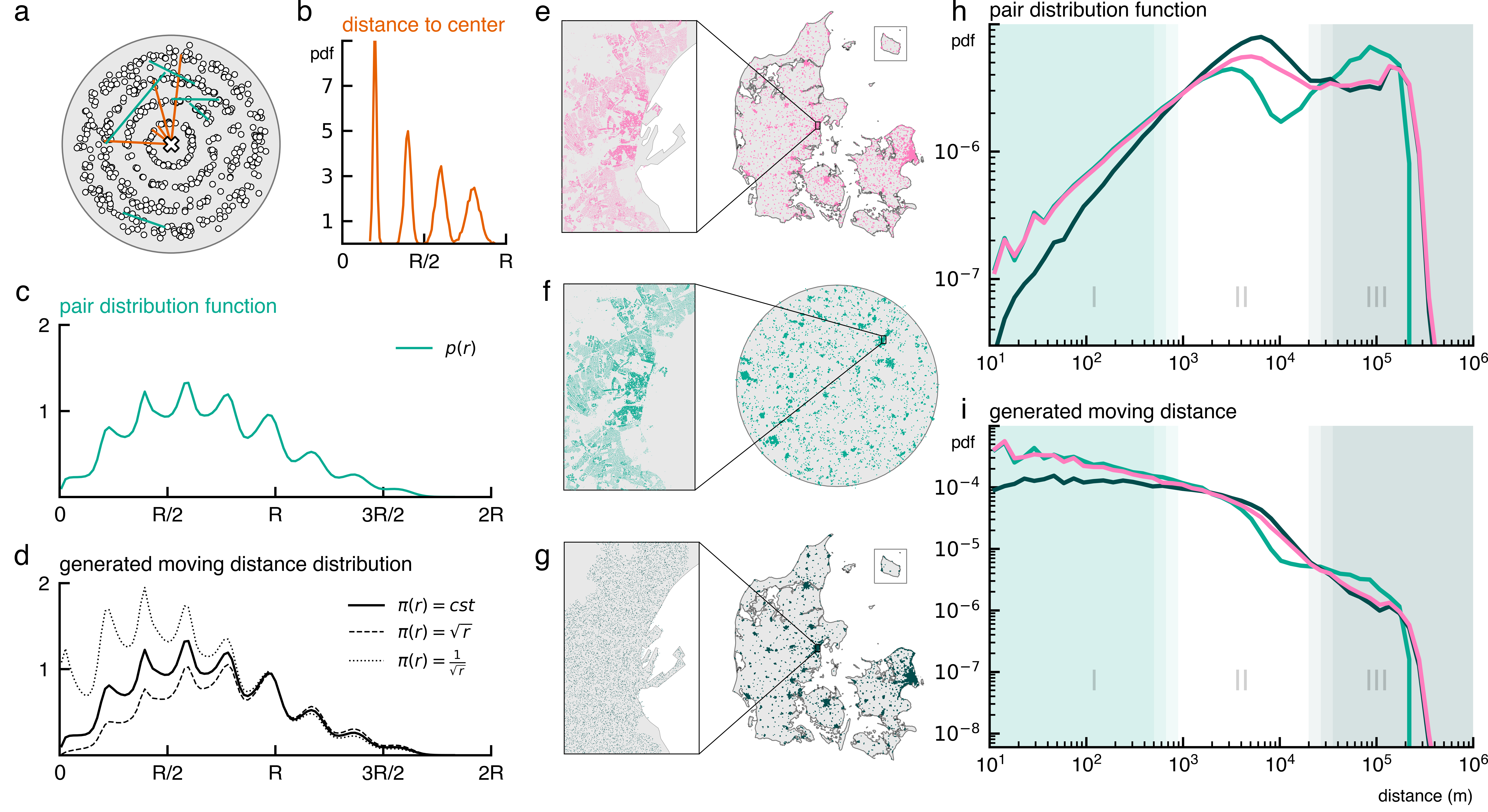}
    \caption[Introduction to the pair distribution function]{Illustration of how geography influences human mobility: comparing Real Denmark, Disk Denmark, and Uniform Denmark. The pair distribution function reveals scale-dependent differences in mobility patterns across geographic layouts.}
    \label{fig:1}
\end{figure*}

The last decade has seen a surge in human mobility research, propelled by the availability of large datasets detailing movement patterns. Previous studies have extensively characterized human mobility as scale-free, proposing models to explain the observed distributions of movement distances. However, these studies have largely overlooked the role of geography—how the physical world shapes and is shaped by human movement. This paper aims to bridge this gap by analyzing how geographical constraints influence human mobility patterns.

We aim to separate the impact of geography from other factors influencing human movement patterns by utilizing high-resolution datasets to reveal the spatial structures that shape mobility. We analyze 39 million residential migration records in Denmark over 40 years, recorded with 2-meter accuracy. We use the pair distribution function between locations, a statistical physics tool treating locations as particles to quantify spatial structures\footnote{A simple, yet deeply clever idea from Benjamin Maier, also used in \citep{wiedermann_spatial_2016,maier_generalization_2019,maier_modular_2019,maier_spreading_2020}}. The pair distribution function at distance $r$ can be expressed,

\[
p(r) = \int_\Omega d  x\ \int_\Omega d  y\ \varrho^d(  x) \varrho^d(  y) \delta(r - \lVert x-y\rVert),
\]

This method reveals mobility patterns across scales with unprecedented precision. We also include a dataset of day-to-day mobility between points of interest from cell phones.

\begin{figure*}[t!]
\checkoddpage \ifoddpage \forcerectofloat \else \forceversofloat \fi
    \centering
    \includegraphics[width=\textwidth]{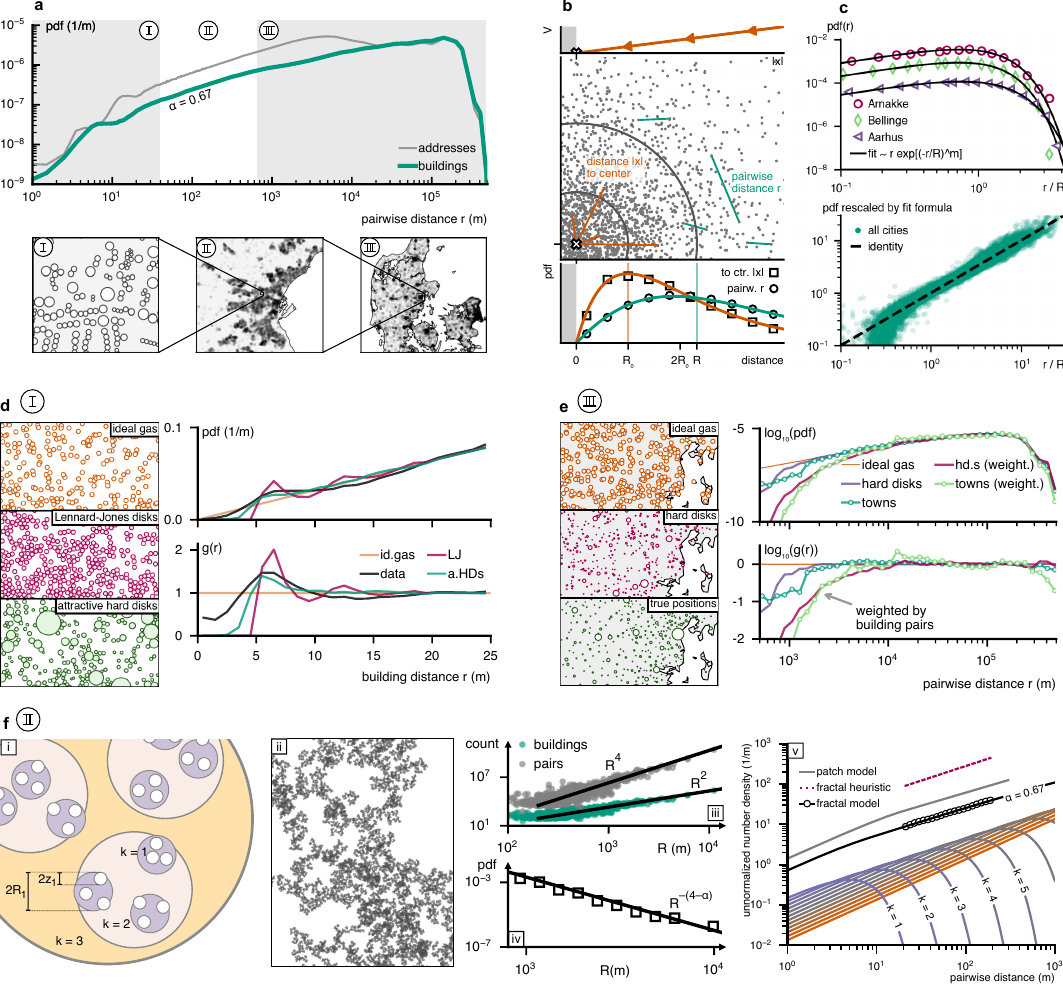}
    \caption[Geography and pair distribution function]{\textbf{(a)} Pair distribution of residential locations in Denmark. (I) Neighborhood-scale with linear onset and oscillations, (II) city-scale with power-law growth ($\alpha = 0.67$), and (III) country-scale with slower growth and fast decay.
    \textbf{(b)} Ideal gas of buildings forms a city-like structure with radial density following an Erlang distribution. Pair distribution approximates a generalized Gamma distribution.
    \textbf{(c)} Maximum-likelihood fits of generalized pair distribution against Danish cities.
    \textbf{(d)} Pair distribution on the micro-scale for four systems: (i) Denmark buildings, (ii) random positions, (iii) Lennard-Jones disks, and (iv) attractive hard disks. Pair-correlation function $g(r)$ quantifies over- and under-presence of neighbors compared to the ideal gas.
    \textbf{(e)} Pair distribution for cities: (i) random positions, (ii) non-attractive hard disks, and (iii) real data. Weighted distributions replicate macro regime behavior of panel (a).
    \textbf{(f)} Fractal model for building positions.}
    \label{fig:geography}
\end{figure*}

\begin{figure*}[t]
\checkoddpage \ifoddpage \forcerectofloat \else \forceversofloat \fi
    \centering
    \includegraphics[width=\textwidth]{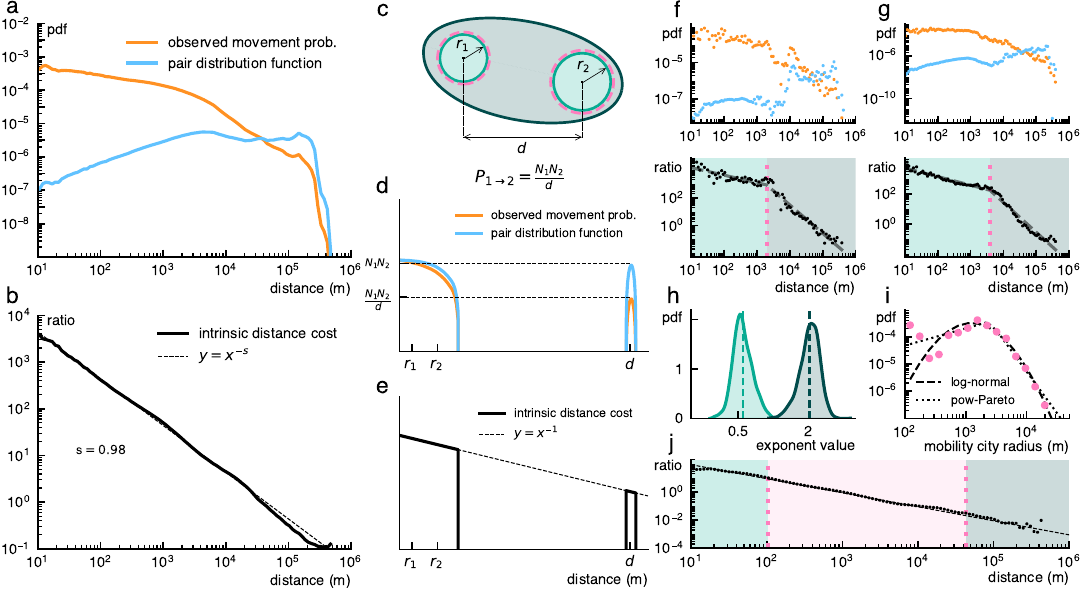}
    \caption[Intrinsic distance cost]{Empirical moving distance distribution and pair distribution in Denmark. The intrinsic distance cost function follows a power law that spans five orders of magnitude.}
    \label{fig:2}
\end{figure*}

By normalizing the residential move distance distribution (Figure \ref{fig:2}a, orange) by the pair distribution function of locations (Figure \ref{fig:2}, blue), we uncover a universal power law spanning five orders of magnitude, from 10 meters to 1000 km (Figure \ref{fig:2}b). We name the power law the intrinsic distance cost, it closely follows a $\pi(r)=1/r$ power law, effectively extending the gravity model of migration into a continuous domain (Figure \ref{fig:2}c-e). This 'geography-free' gravity law is continuous because it operates beyond administrative boundaries, applying at the scale of individual addresses, the finest scale possible. Examination of mobility within Danish cities uncovers piece-wise power laws for intrinsic distance costs (Figure \ref{fig:2}f-g), suggesting a universal two-stage mobility pattern across urban environments (Figure \ref{fig:2}h-i). This pattern integrates into a global model that replicates the empirical intrinsic distance cost (Figure \ref{fig:2}j), confirming the method's accuracy.

\begin{marginfigure}
    \centering
    \includegraphics[width=\textwidth]{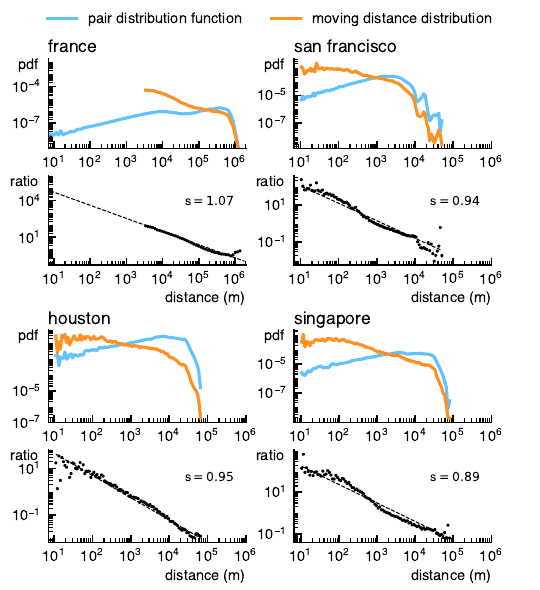}
    \caption[Intrinsic distance cost for day-to-day mobility]{(a) Observed moving distance distribution and pair distribution for residential mobility in France. Observed moving distance distribution and pair distribution for day-to-day mobility between points of interest in San Francisco (b), Houston (c), and Singapore (d). }
    \label{fig:4}
\end{marginfigure}

Our results extend to other regions and for day-to-day mobility. We find consistent patterns in France (Figure \ref{fig:4}a) and across diverse geographies for day-to-day mobility in Houston, Singapore, and San Francisco (\ref{fig:4}b-d). This demonstrates the general applicability of our power law model to various types of mobility and geographical settings, offering a robust framework for understanding human movement. We also explain how the pair distribution function characterizes geography at every scale by modeling locations like particles in a condensed matter system, where the attraction-repulsion interactions between locations reveal structural properties of the urban environment (Figure \ref{fig:geography}a-b). This approach allows us to define the radius of a city by fitting a generalized Gamma distribution to the pair distribution function, providing a meaningful scale for each city based on its spatial layout (Figure \ref{fig:geography}c)

This study explicitly considering the role of geography and the built environment in human mobility. It moves beyond the limitations of previous models that either overlooked geography or reduced it to simplistic density maps. By explicating the relationship between geography and mobility, the findings have implications for urban planning, transportation, and public health. The pair distribution function between provides a powerful framework for integrating geographical constraints into models of human mobility.

\bibliographystyle{unsrt}
\bibliography{bibliography}


\end{document}